\documentclass[hidelinks, 11pt]{article}
\usepackage{amssymb}
\usepackage{amsmath}
\usepackage{amsfonts}
\usepackage{geometry}
\usepackage[doublespacing]{setspace}
\usepackage{hyperref}
\usepackage{graphics}
\usepackage{lscape}
\usepackage{subcaption}
\usepackage{rotating}
\usepackage{multirow}
\usepackage[table]{xcolor}
\usepackage{subcaption}
\usepackage{todonotes}
\usepackage{comment}
\usepackage{threeparttable}
\usepackage[authoryear]{natbib}
\usepackage[toc,page,header]{appendix}

\usepackage{xr}
\DeclareMathOperator*{\argmin}{arg\,min}

\newtheorem{theorem}{Theorem}[section]

\newtheorem{corollary}[theorem]{Corollary}

\newtheorem{lemma}{Lemma}[section]

\newtheorem{remark}{Remark}[section]

\newtheorem{assumption}{Assumption}[section]
\newenvironment{proof}[1][Proof]{\noindent \textbf{#1:} }{\  \rule{0.5em}{0.5em}}

\input{tcilatex}
\begin{document}

	\title{Predictive Quantile Regression with Mixed Roots \\and Increasing Dimensions: The ALQR Approach\thanks{We immensely thank the co-editor, Torben Andersen, the associate editor and three anonymous referees for very constructive comments. We also greatly appreciate helpful comments by Roger Koenker, Xiaofeng Shao, Zhijie Xiao, Viktor Todorov, Kajal Lahiri, the seminar participants at Vanderbilt, Sungkyunkwan, Washington, Syracuse, Northwestern Kellogg, Shanghai University of Finance and Economics, Fudan and Xiamen University. Shin is grateful for financial support by the Social
	Sciences and Humanities Research Council of Canada (SSHRC-435-2018-0275). This research was enabled in part by support provided by Compute Ontario (https://www.computeontario.ca/) and Compute Canada (www.computecanada.ca).}}
\author{\normalsize
	Rui Fan\thanks{\footnotesize\setlength{\baselineskip}{4.4mm} Rui Fan: \href{mailto:	fanr5@rpi.edu}{	fanr5@rpi.edu}. Department of Economics, Rensselaer Polytechnic Institute, Russell Sage Laboratory 4307, 110 8th St., NY 12180, USA\smallskip}
	\and\normalsize	
	Ji Hyung Lee\thanks{\footnotesize\setlength{\baselineskip}{4.4mm} 
		Ji Hyung Lee: \href{mailto:jihyung@illinois.edu}{jihyung@illinois.edu}. Department of Economics, University of Illinois Urbana Champaign, 214 David Kinley Hall, 1407 West Gregory Dr, Urbana, IL 61801, USA\smallskip}
	\and\normalsize
	Youngki Shin\thanks{\footnotesize\setlength{\baselineskip}{4.4mm} Youngki Shin: \href{mailto:shiny11@mcmaster.ca}{shiny11@mcmaster.ca}. Department of Economics, McMaster University, 1280 Main St.\ W., Hamilton, ON, Canada, L8S 4L8}}

\date{Nov 2022}
\maketitle

	\begin{abstract}
		In this paper we propose the adaptive lasso for predictive quantile regression (ALQR). Reflecting empirical findings, we allow predictors to have various degrees of persistence and exhibit different signal strengths. The number of predictors is allowed to grow with the sample size. We study regularity conditions under which stationary, local unit root, and cointegrated predictors are present simultaneously. We next show the convergence rates, model selection consistency, and asymptotic distributions of ALQR.
		We apply the proposed method to the out-of-sample quantile prediction problem of stock returns and find that it outperforms the existing alternatives. We also provide numerical evidence from additional Monte Carlo experiments, supporting the theoretical results.


		\vspace{0.08in}

		
	\end{abstract}
	\section{Introduction}\label{SEC-intro}

	Predictive quantile regression (QR) identifies the impact of predictors on a set of conditional quantiles of a response variable.
	It provides richer information on the heterogeneous distributional prediction.
	For example, the conditional quantile prediction of stock returns receives much attention in finance since the tail quantile information has a crucial role in measuring risk.
	Many economic state variables are employed to predict stock returns and the number of candidate predictors is often large.
	When a large number of predictors are available, researchers encounter the inevitable model selection issue.
	A good model selection can improve forecasting performance but the opposite can also occur.
	Considering the importance of model selection in practice, we need constructive guidance for empirical applications.

\begin{figure}
		\caption{Time-series Plots of Persistent and Stationary Predictors}
		\label{fig_intro}\centering
		\begin{threeparttable}
			\begin{tabular}{cc}
				\includegraphics[scale=0.5]{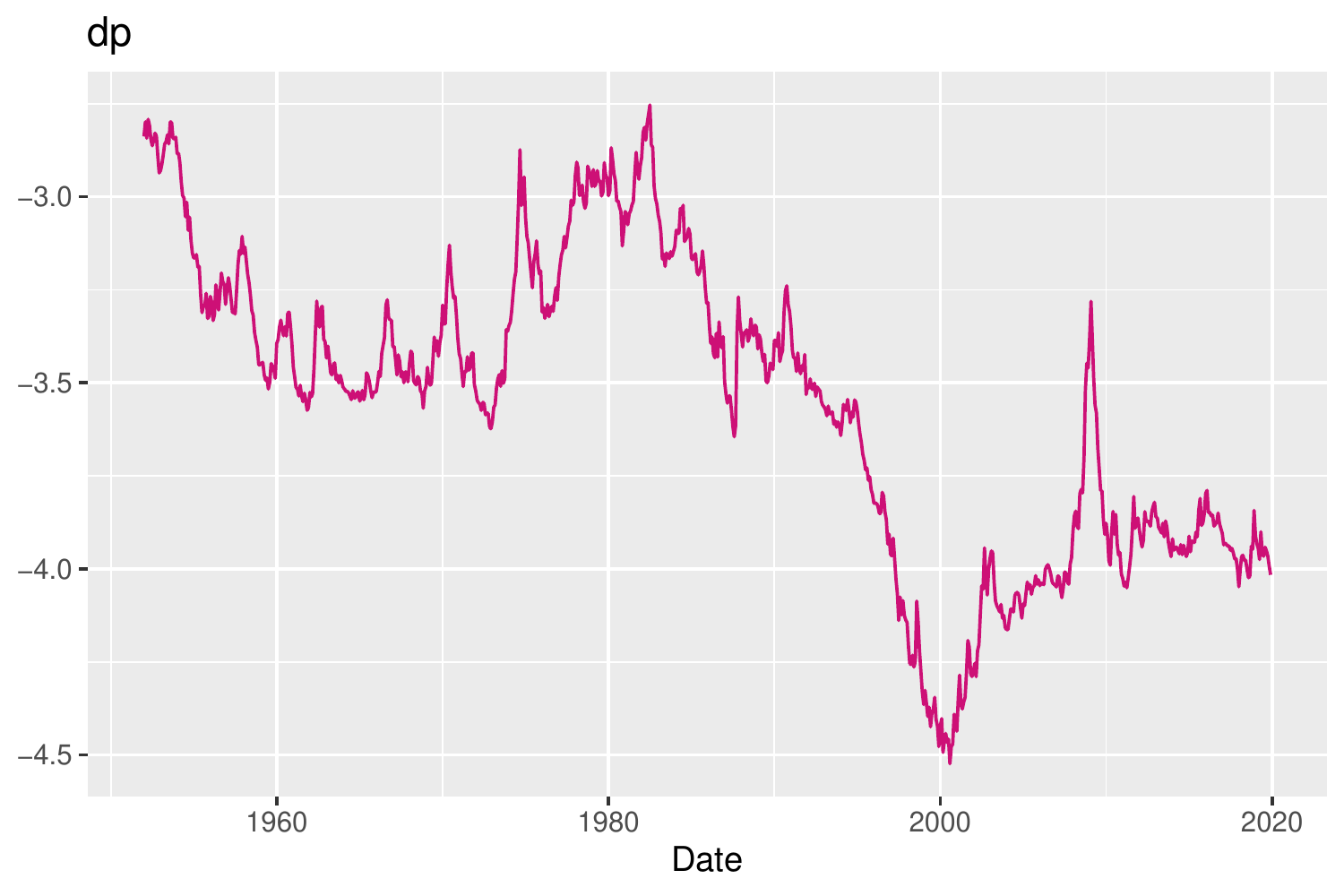} &  \includegraphics[scale=0.5]{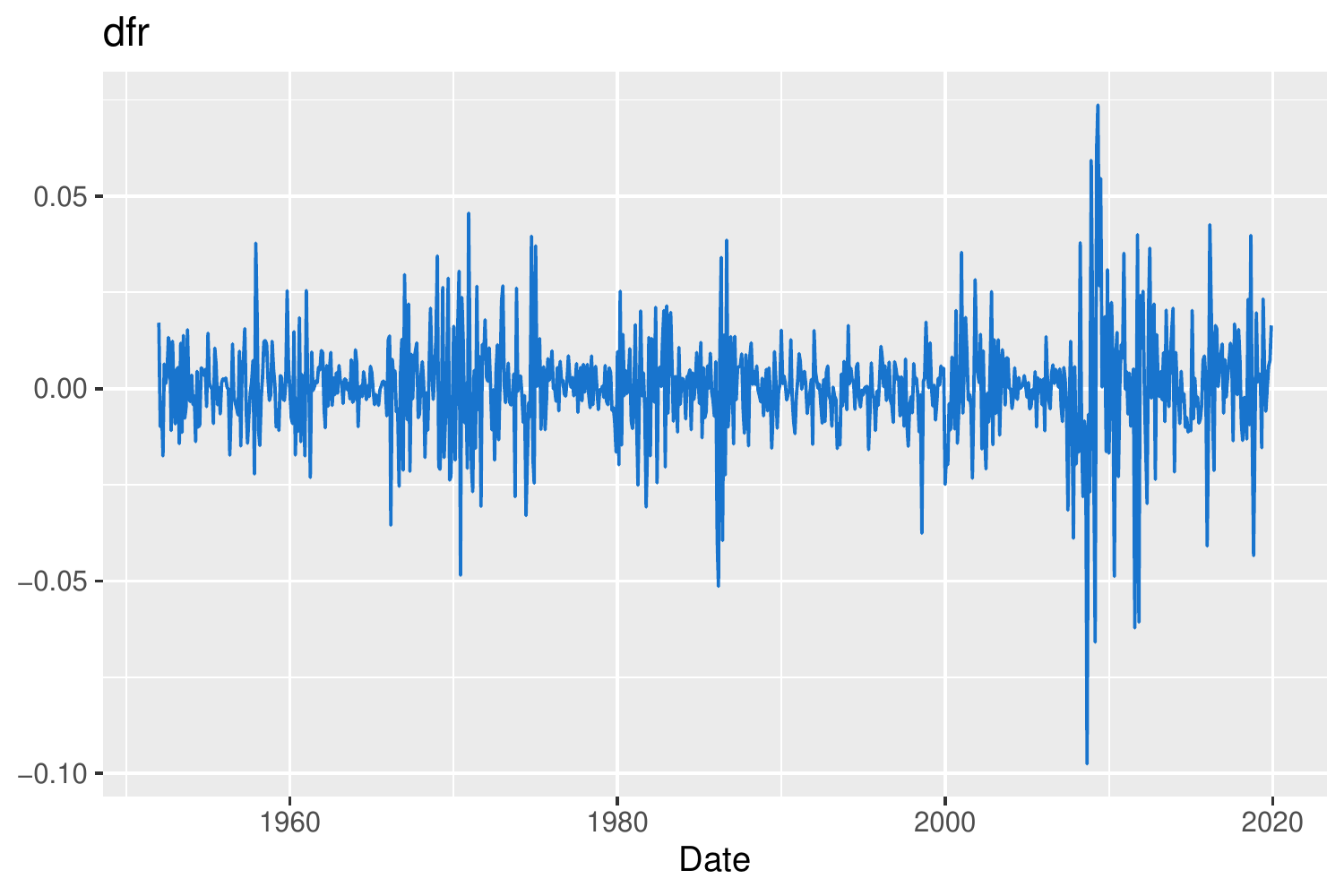} \\
			\end{tabular}
			\begin{tablenotes}[para,flushleft]
				\item {\footnotesize Notes: All plots are based on 816 monthly observations ranging from January 1952 to December 2019. The acronym \emph{dp} denotes the dividend price ratio and \emph{dfr} denotes the default yield spread.}
				\end{tablenotes}
			\end{threeparttable}
		\end{figure}

	In this paper we propose the adaptive lasso for predictive quantile regression (ALQR).
	Although there exists a large volume of literature on predictive mean regression of equity returns (see, e.g.\ \citet{campbell1987stock}, \citet{fama1988dividend}, \citet{hodrick1992dividend}, \citet{cenesizoglu2012return}, \citet{andersen2020pricing} among others), predictive QR is relatively understudied.
	\citet{cenesizoglu2008distribution} is an early paper on predictive QR, and \citet{maynard2011inference}, \citet{lee2016predictive}, \citet{fan2019predictive}, \citet{gungor2019exact}, and \citet{cai2022new} recently develop inference methods in predictive QR with nonstationarity and heteroskedasticity.
	The proposed method is different from these approaches.
	We consider predictive QR with an increasing number of mixed root predictors and address two important problems raised in the stock returns data.
	First, the prediction power of each predictor can vary over different quantiles.
	By adapting the lasso, we allow the model selection based on real data not by the researcher's discretion.
	Second, the predictors widely used in predicting equity returns are composed of stationary, local unit root, and cointegrated processes.
	For example, we plot the time-series of two predictors, dividend price ratio (\emph{dp}) and default yield spread (\emph{dfr}), in Figure \ref{fig_intro}. Even a simple eyeballing test easily confirms their different levels of persistence. (We conduct more informative estimation procedures in Section \ref{sec_application}).
	We provide a unified adaptive lasso framework that allows those mixed root predictors. We show that the estimator converges to the true parameter value at different convergence rates and that the faster convergence rates make the adaptive lasso more efficient in the model selection.

		Naturally, some technical challenges arise.
		Since the seminal paper of \citet{tibshirani1996regression}, the lasso has been intensively studied in various fields of statistical analysis.
		However, most studies have been focusing on the \emph{i.i.d.}\ sample and it has not been a long time since more studies have been conducted with dependent data (see the references in the related literature section below).
		Furthermore, we allow nonstationary predictors, which impose an additional difficulty in the formal analysis of the proposed adaptive lasso.
		We tackle these issues by considering a simple model that only contains unit-root predictors first. Once establishing the desired properties of ALQR in this model, we generalize the model so that it includes all stationary, local unit root, and cointegrated predictors.

		The contributions of this paper are two-fold.
		First, to the best of our knowledge, this is the first paper to study predictive QR with an increasing number of mixed root predictors. We propose the adaptive lasso for predictive quantile regression (ALQR) and derive the convergence rates, model selection consistency and asymptotic distributions of ALQR under some regularity conditions. As a by-product, we also prove that the standard QR estimator is consistent under both mixed roots (including the local unit roots) and the increasing dimension of the predictors, which is new in the literature.
		Second, we conduct an empirical analysis of the stock returns data and find that ALQR can improve prediction performance over existing alternatives across different quantiles. We apply the ALQR method along with the existing alternatives to the data set. The results confirm that ALQR shows better prediction performance across different quantiles, particularly at higher quantiles. As illustrated in Section 3, ALQR can be readily applicable to other applications.

		The rest of the paper is organized as follows.
		This section finishes with a review of the relevant literature.
		In section 2, we introduce predictive QR models formally and define the ALQR estimator.
		In section 3, we investigate the performance of ALQR using the out-of-sample quantile prediction problem of stock returns.
		We study the theoretical properties of ALQR in sections 4 and 5.
		In section 7, we conduct some Monte Carlo simulation experiments.
		Section 7 concludes.
		All technical proofs are relegated to the appendix.

		\paragraph{Related Literature}
		The lasso has been extensively studied for cross-sectional data.
		Recently, there has been development in lasso procedures with dependent data.
		\citet{basu2015regularized} exploit the spectral properties of the
		stationary time series design and investigate the regularity condition of
		the lasso that leads to the non-asymptotic bounds and the consistency
		results.
		\citet{kock2015oracle} investigate the oracle property of the lasso in a
		stationary vector autoregression model.
		\citet{adamek2020lasso} provide an inference procedure based on the
		debiased/desparsified lasso under the near-epoch dependence assumption.
		\citet{chernozhukov2021lasso} propose a penalty selection algorithm with weakly dependent data and the post-selection inference procedure.
		\citet{wu2016performance} and \citet{wong2020lasso} analyze the lasso with non-sub-Gaussian processes that allow a heavy-tail distribution.
		\citet{medeiros2016l1} show the asymptotic properties of the adaptive lasso for stationary high-dimensional time series models.
		For the cointegrated models,
		\citet{kock2016consistent} shows the oracle property of the adaptive lasso
		in the autoregression model. Interestingly, he finds that the unit root test
		can be incorporated into the model selection procedure by adopting the
		Dickey-Fuller form of autoregression.
		\citet{liao2015automated} propose a shrinkage estimator with multiple
		penalty terms to select the rank of cointegration and estimate the
		parameters simultaneously in a vector error correction model.
		\citet{liang2019determination}, \citet{zhang2019identifying}, and \citet{onatski2018alternative} investigate the same model in the high-dimensional setting.

		\citet{koo2020high} recently use lasso
		to improve the prediction of stock returns. It is shown that lasso significantly
		reduces forecasting mean squared errors even with a mixture of stationary,
		unit-root, and cointegrated variables. However, the conventional lasso
		method may not have model selection consistency and the oracle property
		as shown by
		\citet{meinshausen2004consistent}
		and
		\citet{fan2001variable}.
		The adaptive lasso proposed by \citet{zou2006adaptive} improves the performance of the lasso.
		Instead of imposing the same penalty weight on
		all candidate parameters, the adaptive lasso penalizes each parameter proportionally to
		the inverse of its initial estimate. With a proper choice of the tuning
		parameter $\lambda _{n}$, the adaptive penalty weights for the irrelevant
		variables approach infinity, whereas those for the relevant variables
		converge to constants.
		\citet{lee2021lasso} apply the adaptive lasso to a predictive mean regression framework.
		Similar to \citet{koo2020high}, predictors are allowed to have different degrees of persistence and cointegration. %
		\citet{lee2021lasso} find that the adaptive lasso and a newly proposed twin adaptive
		lasso outperform the alternative methods in terms of predictor selection consistency and out-of-sample mean squared errors.

		Some effort has been also made to investigate model selection and model
		estimation in QR under the \emph{i.i.d.}\ samples. The $\ell_1$-penalized method in the QR framework has been studied for high-dimensional data analysis (see, e.g.\ \citet{portnoy1984asymptotic},
		\citet{portnoy1985asymptotic}, \citet{knight2000asymptotics},
		\citet{koenker_2005}, \citet{li20081}, \citet{lee2018oracle} and \citet{belloni2011_L1}).
		To overcome the problem of inconsistent model selection (\citet{fan2014adaptive}; \citet{wang2012quantile}), recent studies have further considered the adaptive lasso in QR.
		\citet{wu2009variable} discuss how to conduct model selection for QR
		models using SCAD and the adaptive lasso method. \citet{zheng2013adaptive} establish
		the oracle property for an adaptive lasso QR model with heterogeneous error
		sequences. \citet{zheng2015globally} study a globally adaptive lasso method for ultra
		high-dimensional QR models.

		\paragraph{Notation} Let $\Vert \cdot \Vert$ and $\Vert \cdot \Vert_0$ denote $\ell_2$-norm and $\ell_0$-norm, respectively. For a matrix $S$, $\Vert S \Vert$ represents the spectral norm.
		Let $F_a(\cdot)$ and $f_a(\cdot)$ denote a cumulative distribution function (CDF) and a probability density function (pdf) of a generic random variable $a$.
		Let $S>0$ denote a generic positive definite matrix $S$.
		Let $\lambda_{\min }\left( S\right) $ and $\lambda _{\max }\left( S\right) $ denote the smallest and largest eigenvalues of $S$. 
		We use $O_p(1)$ and $o_p(1)$ when a sequence is bounded in probability and converges to zero in probability, respectively.
		The $O(1)$ and $o(1)$ denote the non-stochastic counterparts.


		\section{Model and the ALQR Estimator\label{SEC-model}}
		In this section, we introduce the predictive quantile regression model and the adaptive lasso for quantile regression (ALQR).
		We will develop the theory for the ALQR in two steps in Section \ref{SEC:Oracle}. For a better exposition of the theory, we also propose the model and its estimator in two separate cases: (i) unit-root predictors; and (ii) mixed-root predictors.

		\subsection{QR Model with Unit-Root Predictors\label{SEC-model1}}

		Consider a predictive QR model with unit-root predictors:
		\begin{align}
			\label{QR1}
			\begin{split}
				Q_{y_{t}}\left( \tau |\mathcal{F}_{t-1}\right)
				& := \mu _{0\tau}+x_{t-1}^{\prime }\beta _{0\tau }\\
				x_{t} & := x_{t-1}+v_{t}\text{,}
			\end{split}
		\end{align}
		where
		$Q_{y_{t}}\left( \tau |\mathcal{F}_{t-1}\right) $ is the conditional $\tau$-quantile of $y_{t}$,
		$\mathcal{F}_{t}=\left\{ z_{j}\right\} _{j=-\infty
		}^{t}$ with $z_{j}=(y_{j},x_{j}^{\prime })^{\prime }$ is a natural filtration,
		$x_{t}$ is a $p_{n}$-dimensional vector of unit-root predictors with a stationary $O_{p}(1)$ initialization of $v_{0}=\sum_{j=0}^{\infty
		}D_{vj}\epsilon _{-j}$ following the innovation structure below,
		and
		$\beta_{0\tau}$ is the corresponding true parameter vector.

	 The innovation of unit root predictors, $v_{t}$, follows a linear process, which is commonly assumed in the predictive regression literature (see \citet{phillips2013predictive, phillips2016robust}, \citet{cai2022new}, \citet{lee2021lasso} for a few recent papers):
			\begin{equation*}
				v_{t}=\sum_{j=0}^{\infty }D_{vj}\epsilon _{t-j}\text{, }\epsilon _{t}\sim
				mds(0,\Sigma )\text{, }\Sigma >0,
			\end{equation*}%
			\begin{equation*}
				E||\epsilon _{t}||^{2+\kappa }<\infty \text{, }\kappa >0,
			\end{equation*}%
			\begin{equation*}
				D_{v0}=I_{p}\text{, }\sum_{j=0}^{\infty }||D_{vj}||<\infty \text{, }%
				D_{v}(r)=\sum_{j=0}^{\infty }D_{vj}r^{j}\text{, and }D_{v}(1)=\sum_{j=0}^{\infty
				}D_{vj}>0,
			\end{equation*}%
			\begin{equation*}
				\Sigma _{vv}= \sum_{h=-\infty}^{\infty} E(v_t v_{t-h}')=D_{v}(1)\Sigma D_{v}(1)^{\prime },
			\end{equation*}
			where $mds$ denotes martingale difference sequences with respect to the natural filtration.

		We allow the dimension of predictors to increase, i.e.~$p_n\rightarrow \infty$ as $n \rightarrow \infty$. To make notation simple, we omit subscript $n$ and use $p$ unless it may cause any confusion. We define $u_{t\tau }:=y_{t}-Q_{y_{t}}\left( \tau |\mathcal{F}_{t-1}\right)$, the deviation of $y_t$ from the conditional $\tau$-quantile. If the CDF of $u_{t\tau}$ is continuous, we have
		\begin{align*}
			F_{u_{t\tau }}^{-1}\left( \tau | \mathcal{F}_{t-1}\right) = F_{u_{t\tau }}^{-1}\left( \tau \right)  = 0, \quad \text{for all $\tau$}
		\end{align*}
		where $F_{u_{t\tau }}^{-1}$ is the inverse CDF of $u_{t\tau }
		$. It holds by construction that $\Pr \left( u_{t\tau }<0|\mathcal{F}_{t-1}\right) =\tau$, which is equivalent to $F_{u_{t\tau }}^{-1}\left( \tau | \mathcal{F}_{t-1}\right) = 0$.
		To see the equivalence to the unconditional $\tau$-quantile, we note that
		\begin{align*}
			\Pr(u_{t\tau} < 0)
			= E[\mathbf{1}(u_{t\tau} < 0) ] = E[ E [\mathbf{1}(u_{t\tau} < 0) | \mathcal{F}_{t-1}] ]
			= E[ \Pr(u_{t\tau} < 0 |\mathcal{F}_{t-1})]
			= \tau,
		\end{align*}
		where  $\mathbf{1}(\cdot )$ denotes an indicator function.
		The equations above imply that both conditional and unconditional $\tau$-quantiles of $u_{t\tau}$ are zero for any given $\tau$. However, it does not mean that two distribution functions are the same, i.e.~$F_{u_{t\tau_1 }}^{-1}\left( \tau_2 | \mathcal{F}_{t-1}\right) \neq F_{u_{t\tau_1 }}^{-1}\left( \tau_2 \right)$ for $\tau_1 \neq \tau_2$ in general.

		Let $\psi _{\tau}(u_{t\tau }):= \tau - \mathbf{1} \left( u_{t\tau }<0\right)$.
		It is easy to find that $\psi _{\tau }(u_{t\tau })$  is uncorrelated to any predetermined regressor. That is, $\psi _{\tau}(u_{t\tau })$ is uncorrelated with any past innovations, $v_{t-j}$, for $j\geq 1$:
		\begin{align*}
			Cov(\psi _{\tau}(u_{t\tau }), v_{t-j})
			& = E[\psi _{\tau}(u_{t\tau }) \cdot v_{t-j}] - E[\psi _{\tau}(u_{t\tau })] \cdot E[v_{t-j}] \\
			& = E[ E[\psi _{\tau}(u_{t\tau })|\mathcal{F}_{t-1}] \cdot v_{t-j}] - E[E[\psi _{\tau}(u_{t\tau })|\mathcal{F}_{t-1}]] \cdot E[v_{t-j}] \\
			& = 0,
		\end{align*}
		by the law of iterated expectations and the fact that $E\left( \psi_{\tau }(u_{t\tau })|\mathcal{\ F}_{t-1}\right) =\tau -\Pr \left( u_{t\tau}<0|\mathcal{F}_{t-1}\right) =0$. In our settings, however, we allow the QR-induced regression errors $\psi _{\tau}(u_{t\tau })$ to be contemporaneously correlated with the innovations of unit
		root sequences. This is commonly assumed in cointegration and predictive regression literature, inducing a potential second order bias arising from the one-sided correlation, see, e.g., \citet{xiao2009quantile}.

		We now define the ALQR that minimizes the penalized QR objective function as follows:
		\begin{equation}
			(\hat{\mu}_{\tau }^{ALQR},\hat{\beta}_{\tau }^{ALQR })
			:=
			\argmin_{\mu \in \mathbf{R},\beta \in \mathbf{R}^{p}}\sum_{t=1}^{n}%
			\rho _{\tau }(y_{t}-\mu -x_{t-1}^{\prime }\beta )+\sum_{j=1}^{p}\lambda
			_{n,j}|\beta _{j}|,  \label{lassoQR_1}
		\end{equation}%
		where $\rho _{\tau }(u):=u(\tau -\mathbf{1}(u<0))$.
		Following \citet{zou2006adaptive}, we define the tuning parameter of the penalty term as		\begin{equation*}
			\lambda_{n,j}:=\frac{\lambda _{n}}{\omega _{j}} = \frac{\lambda _{n}}{|{\tilde{\beta}_{\tau
						,j}}|^{\gamma }},
		\end{equation*}
		where $\lambda_n$ is a sequence converging to infinity, $\omega _{j}:=|{\tilde{\beta}_{\tau
				,j}}|^{\gamma }$ with a positive constant $\gamma$, and ${\tilde{\beta}_{\tau ,j}}$ is a first-step \textit{consistent} estimator for ${\beta}_{0\tau ,j}$.
		The penalty $\lambda_{n,j}$, which contrasts with the penalty in the standard lasso, is an adaptive weight and it has an inverse relationship with ${\tilde{\beta}_{\tau ,j}}$.
		The idea of using a consistent first-step estimator as an adaptive weight in $\lambda_{n,j}$ is to avoid over-penalizing important predictors (or under-penalizing irrelevant predictors). In this paper, we use the ordinary QR estimator (\citet{koenker1978regression}) below as a first-step estimator:
		\begin{equation}
			(\hat{\mu}_{\tau }^{QR},{\hat{\beta}_{\tau}}^{QR})
			:=
			\argmin_{\mu \in \mathbf{R},\beta \in \mathbf{R}^{p}}\sum_{t=1}^{n}%
			\rho _{\tau }(y_{t}-\mu -x_{t-1}^{\prime }\beta ).  \label{eqn_QR}
		\end{equation}%
		The consistency of the ordinary QR estimator is provided in Section \ref{SEC:Oracle}.

		\subsection{QR Model with Mixed Roots\label{SEC-model2}}

		We now introduce a QR model with mixed-root predictors. We assume the predictors of the model have different degrees of persistence: $I(0)$, local unit roots, and cointegration. This model is the most relevant in practice. The model in the previous section can be seen as a special case of it.

		Let $z_{t}$, $x_{t}^{c}$, and $x_{t}$ be vectors of stationary,\footnote{In this paper, we use covariance (weak) stationarity unless noted otherwise. The linear process assumption below implies covariance stationarity of $z_{t}$. In addition, the first element of $z_{t}$ is $1$ so that $\beta _{0\tau,1 }^{z}$ plays a role of the intercept. } cointegrated, and local unit-root predictors whose lengths are $p_z$, $p_c$, and $p_x$, respectively. Let $p:= p_z+p_c+p_x$. Given a sample of $\{y_{t},z_{t},x_{t}^{c},x_{t}\}_{t=1}^{n}$, the QR model with mixed roots is defined as follows:
		\begin{equation}
			Q_{y_{t}}\left( \tau |\mathcal{F}_{t-1}\right) ={\ z_{t-1}}%
			^{\prime }\beta _{0\tau }^{z}+{x_{t-1}^{c}}^{\prime }\beta _{0\tau }^{c}+{\
				x_{t-1}}^{\prime }\beta _{0\tau }^{x}.  \label{QR_mixed}
		\end{equation}

		The cointegrated system in $x_{t}^{c}$ has the triangular representation by %
		\citet{phillips1991optimal}: for $x_{1t}^c$ and $x_{2t}^c$ whose dimensions are $p_1\times 1$ and $p_2\times 1$, respectively,
		\begingroup\allowdisplaybreaks%
		\begin{align}
			Ax_{t}^{c}& =x_{1t}^{c}-A_{1}x_{2t}^{c}=v_{1t}^{c},  \notag \\
			(I_{p_{2}}-R_{2}L) x_{2t}^{c}& =v_{2t}^{c},  \label{eqn_QR_3}
		\end{align}%
		\endgroup
		where $R_{2}=I_{p_{2}}+c_{2}/n$ with $c_{2}=\mathrm{diag}\left( \check{c}%
		_{1},\ldots,\check{c}_{p_{2}}\right)$, $L$ is the lag operator, and the vector $v_{1t}^{c}$ is the vector of $I$(0) cointegrating residuals.
		The cointegrated regressors $x_{t}^{c}$ are decomposed into $x_{1t}^c$ and $x_{2t}^c$.
		The local-to-unity parameter of $x_{2t}^c$ is assumed to be $\check{c}_{j}\in \left( -\infty ,\infty \right)$ for $j=1,\ldots,p_2$.	Thus, the cointegrated system in $x_t^c$ includes both stationary and nonstationary local unit root regions.
		Using \eqref{eqn_QR_3}, we can easily characterize the cointegration relations in $x_{t}^{c}$ by the matrices $A\equiv(I_{p_{1}},-A_{1})$ and $A_{1}$.
		
		The near integrated process $x_{t}=(x_{t1},...,x_{tp_{x}})$ is defined in a similar way:
		\begin{equation}
			(I_{p_{x}}-R_{x}L)x_{t}=v_{t},\text{ }  \label{eqn_I1}
		\end{equation}%
		where $R_{x}=I_{p_{x}}+c_{x}/n$ with $c_{x}=\mathrm{diag}\left(\tilde{c}%
		_{1},...,\tilde{c}_{p_{x}}\right)$ and $\tilde{c}_{j}\in \left(
		-\infty ,\infty \right) $. Again, the local-to-unity specification includes the
		unit root process as a special case when $\tilde{c}_{j}=0$.

		Let $X_{t}:=(z_{t}^{\prime },x_{t}^{c \prime},x_{t}^{\prime })^{\prime }$ and $\beta ^{\ast }:=(\beta ^{z^{\prime
		}},\beta ^{c^{\prime }},\beta ^{x^{\prime }})^{\prime }$.
		Define a $p_c$-dimensional vector $v_{t}^{c}:=({v_{1t}^{c}}^{\prime },{v_{2t}^{c}}^{\prime })^{\prime }$ and a $p$-dimensional vector $e_{t}:=\left( z_{t}^{\prime },{v_{t}^{c}}^{\prime },v_{t}^{\prime }\right)^{\prime }$.
		Abusing notation on $\epsilon_t$, we assume that $e_t$ follows a linear process:
			\begin{equation*}
				e_{t}=D_{e}(L)\epsilon _{t}=\sum_{j=0}^{\infty }D_{ej}\epsilon _{t-j},
			\end{equation*}%
			\begin{equation*}
				\epsilon _{t}=\left(
				\begin{array}{c}
					\epsilon _{zt} \\
					\epsilon _{ct} \\
					\epsilon _{vt}%
				\end{array}%
				\right) \sim mds\left( 0,\Sigma \right) \text{ with }\Sigma _{p\times
					p}=\left(
				\begin{array}{ccc}
					\Sigma _{zz} & \Sigma _{zc} & \Sigma _{zv} \\
					\Sigma _{zc}^{\prime } & \Sigma _{cc} & \Sigma _{cv} \\
					\Sigma _{zv}^{\prime } & \Sigma _{cv}^{\prime } & \Sigma _{vv}%
				\end{array}%
				\right) >0,
			\end{equation*}%
			\begin{equation*}
				E||\epsilon _{t}||^{2+\kappa }<\infty \text{, }\kappa >0,
			\end{equation*}%
			\begin{equation*}
				D_{e0}=I_{p} \text{, }\sum_{j=0}^{\infty }j||D_{ej}||<\infty
				\text{, }D_{e}(z)=\sum_{j=0}^{\infty }D_{ej}z^{j} \text{, and }D_{e}(1)=\sum_{j=0}^{\infty
				}D_{ej}>0,
			\end{equation*}
			\begin{equation*}
				\Omega _{ee}=\sum_{h=-\infty }^{\infty }E(e_{t}e_{t-h}^{\prime
				})=D_{e}(1)\Sigma D_{e}(1)^{\prime }.
			\end{equation*}
		Note that this assumption implies a covariance
		stationary $O_{p}(1)$ initialization of $e_{0}=\left( z_{0}^{\prime },{%
			v_{0}^{c}}^{\prime },v_{0}^{\prime }\right) ^{\prime }=\sum_{j=0}^{\infty
		}D_{ej}\epsilon _{-j}$.

		Similar to Section \ref{SEC-model1}, the QR-induced regression errors, $\psi _{\tau}(u_{t\tau })$, are uncorrelated to any predetermined regressor but allowed to be contemporaneously correlated with the innovations of unit root sequences $x_{2t}^{c}$ and $x_{t}$, the stationary predictor $z_t$ as well as the cointegrating residuals $v_{2t}^{c}$.

		We define the ALQR for Model \eqref{QR_mixed} as follows:
		\begin{equation}
			\hat{\beta}_{\tau }^{ALQR\ast } =\argmin_{\beta ^{\ast }\in
				\mathbf{R}^{p}}\sum_{t=1}^{n}\rho _{\tau }(y_{t}-X_{t-1}^{\prime }\beta ^{\ast })+\sum_{j=1}^{p}\lambda _{n,j}|\beta
			_{j}^{\ast }|,  \label{lasso_QR_mix}
		\end{equation}%
		where $\lambda _{n,j}={\lambda _{n}}/\omega _{j}$, $\omega _{j}=|{\tilde{\beta}_{\tau ,j}}|^{\gamma }$,  $\gamma $ is a constant, and ${\tilde{\beta}_{\tau ,j}}$ is a first-step \textit{consistent} estimator. We recommend using the ordinary QR estimator for ${\tilde{\beta}_{\tau ,j}}$. The consistency of QR will be discussed in Section \ref{oracle mr}.


		\section{Quantile Prediction of Stock Returns}
		\label{sec_application}

		In this section, we consider the quantile prediction problem of stock returns and illustrate the usefulness of the proposed ALQR method.
		We use an updated version of the data set in \citet{welch2008comprehensive}, which ranges from January 1952 to December 2019.
		It is composed of 816 monthly observations of the US financial and macroeconomic variables.
		Our goal is to predict the quantiles of the excess stock returns ($y_t$) using 12 predictors. The variable names are summarized in Table \ref{tb:v_names}. See also \citet{welch2008comprehensive} for more details on these variables.

		\begin{table}[bhp]
			\centering
			\caption{Variable Names and Persistency}\label{tb:v_names}
			\begin{tabular}{clcc}
				\hline
				Notation  & \multicolumn{1}{c}{Variable Name} & High Persistence & AR(1) Coefficient\\
				\hline
				\emph{y}  & excess stock returns      & No           &  0.102  \\
								dp        & dividend price ratio      & Yes          &  0.996  \\
								dy        & dividend yield ratio      & Yes          &  0.996  \\
								ep        & earnings price ratio      & Yes          &  0.990  \\
								bm        & book-to-market ratio      & Yes          &  0.994  \\
								dfy       & default yield spread      & Yes          &  0.969  \\
								ntis      & net equity expansion      & Yes          &  0.982  \\
								lty       & long term yield           & Yes          &  0.995  \\
								tbl       & treasury bill rates       & Yes          &  0.991  \\
								svar      & stock variance            & No           &  0.475  \\
								dfr       & default return spread     & No           &  -0.078  \\
								ltr       & long term rate of returns & No           &  0.040  \\
								infl      & inflation                 & No           &  0.545  \\
				\hline
			\end{tabular}
		\end{table}

\begin{figure}
			\caption{Time-series Plots of Persistent Predictors}
			\label{fig_persistent}\centering
			\begin{threeparttable}
				\begin{tabular}{cc}
					\includegraphics[scale=0.5]{figures/plot_per_dp} &  \includegraphics[scale=0.5]{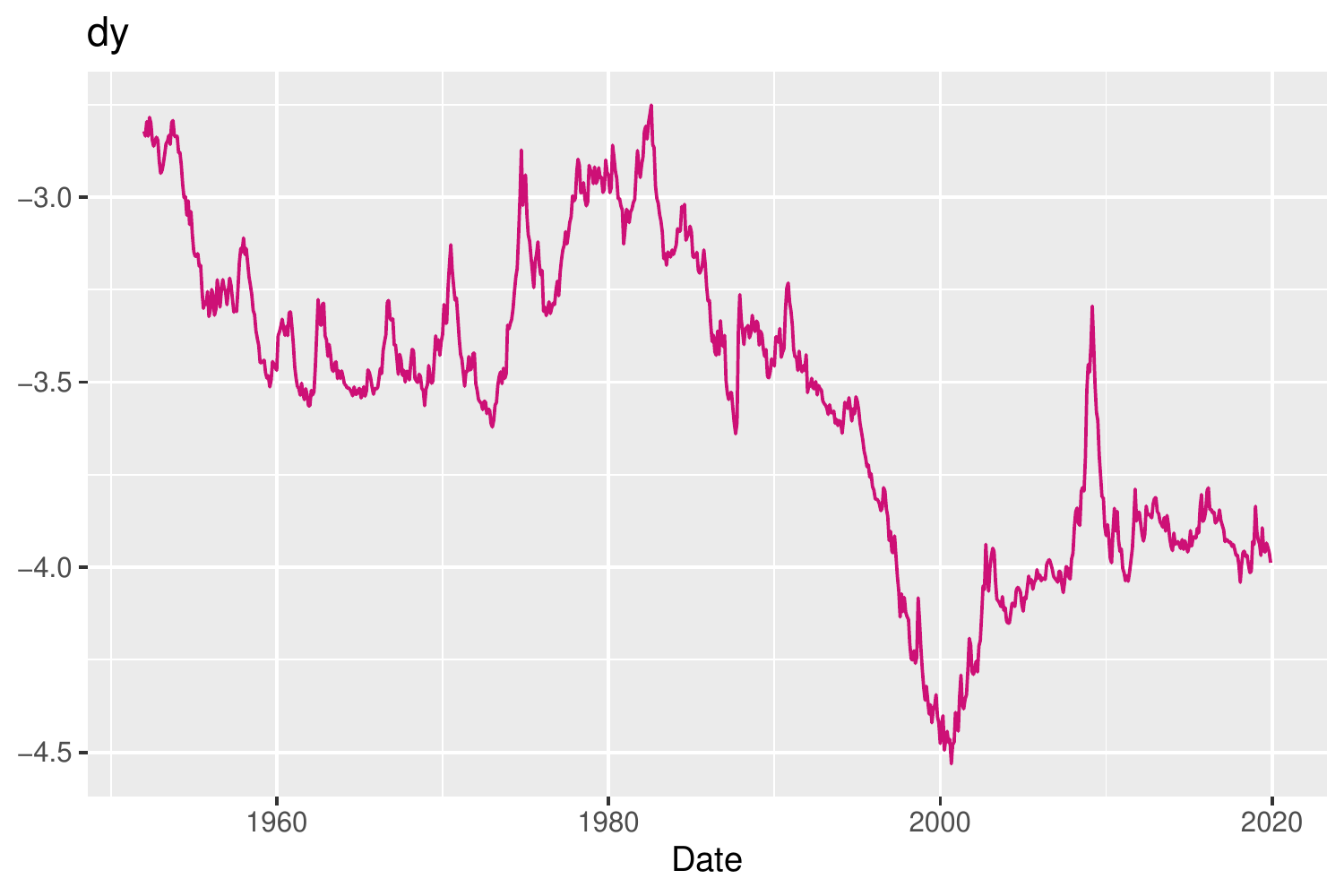} \\
					\includegraphics[scale=0.5]{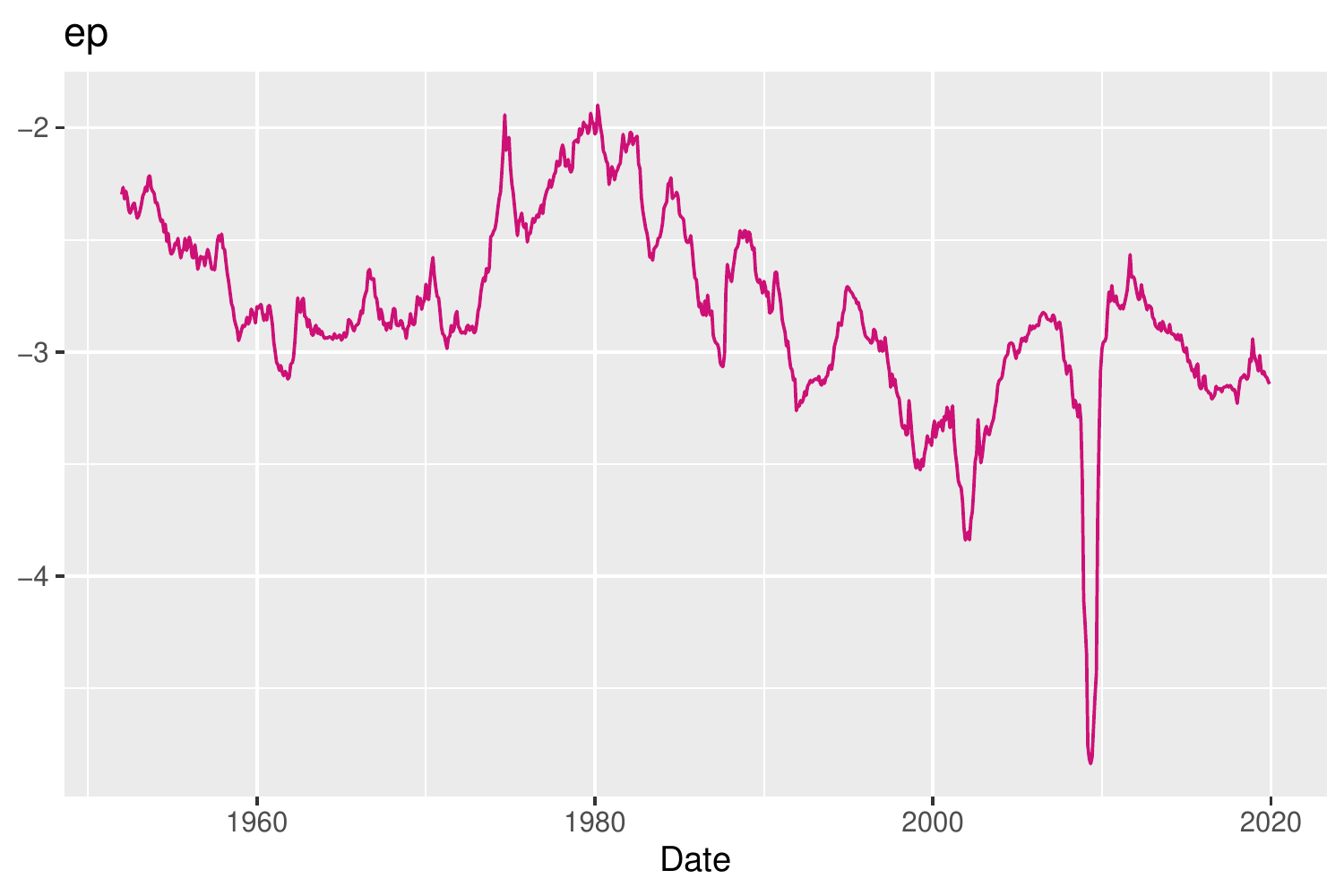} &  \includegraphics[scale=0.5]{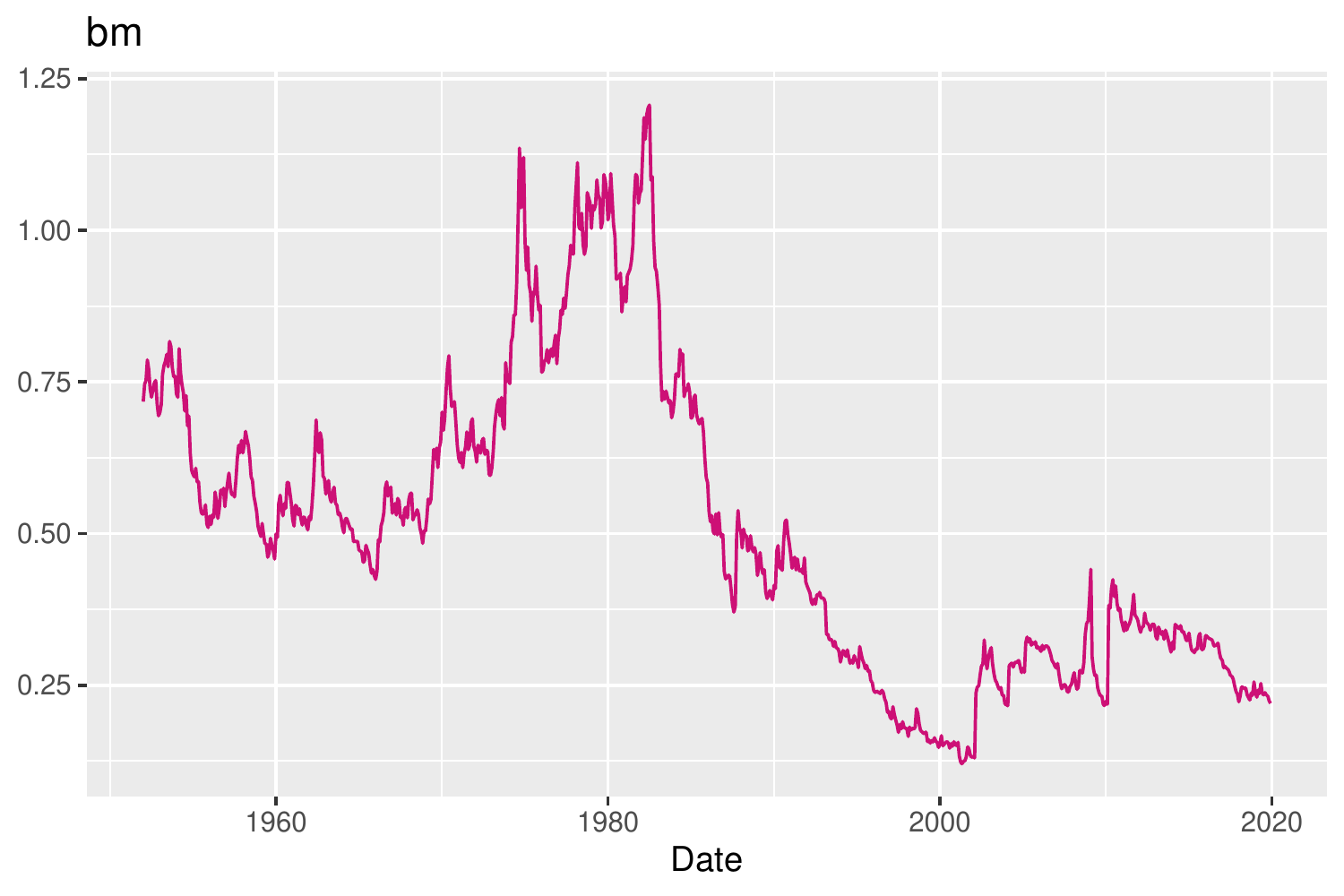} \\
					\includegraphics[scale=0.5]{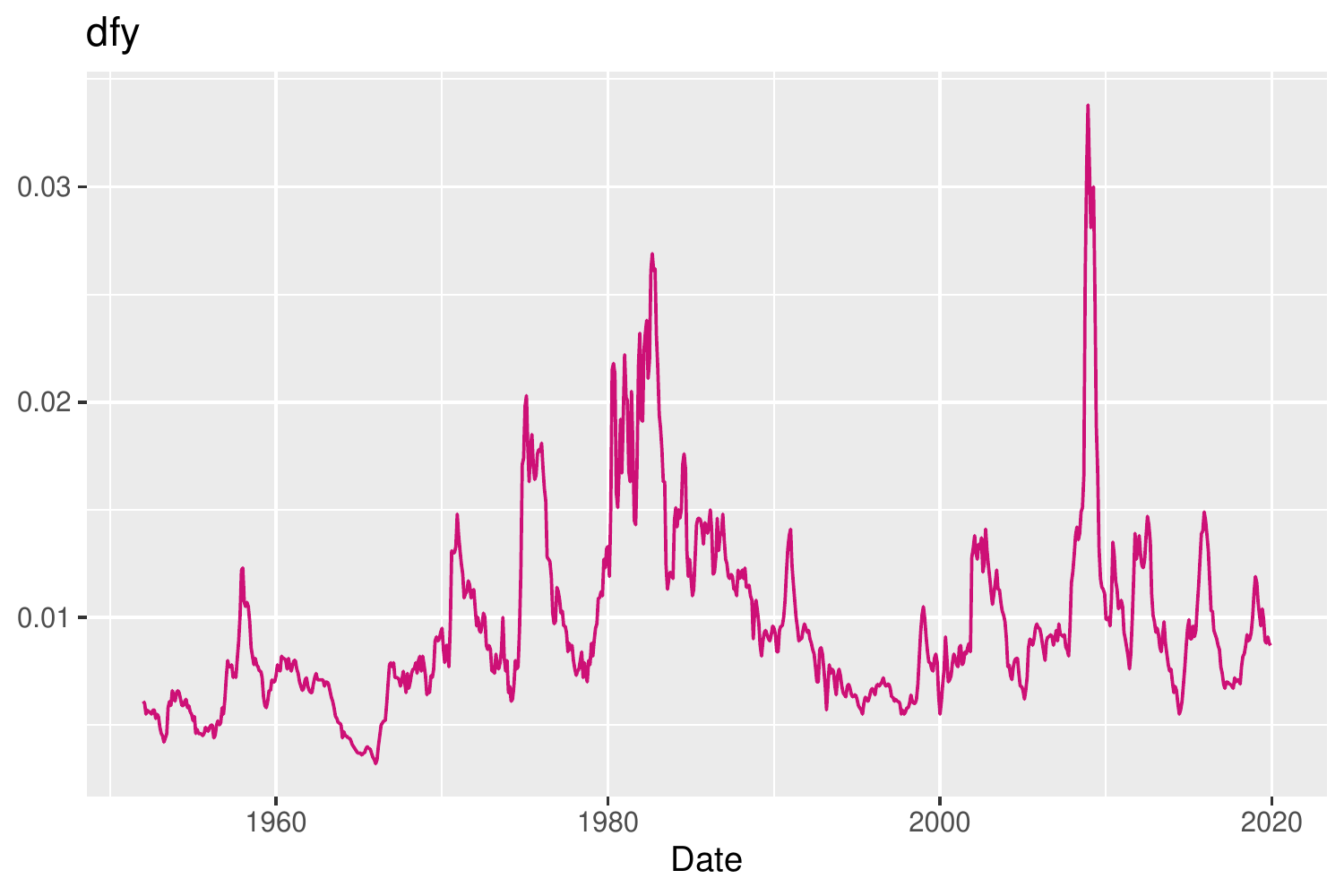} &  \includegraphics[scale=0.5]{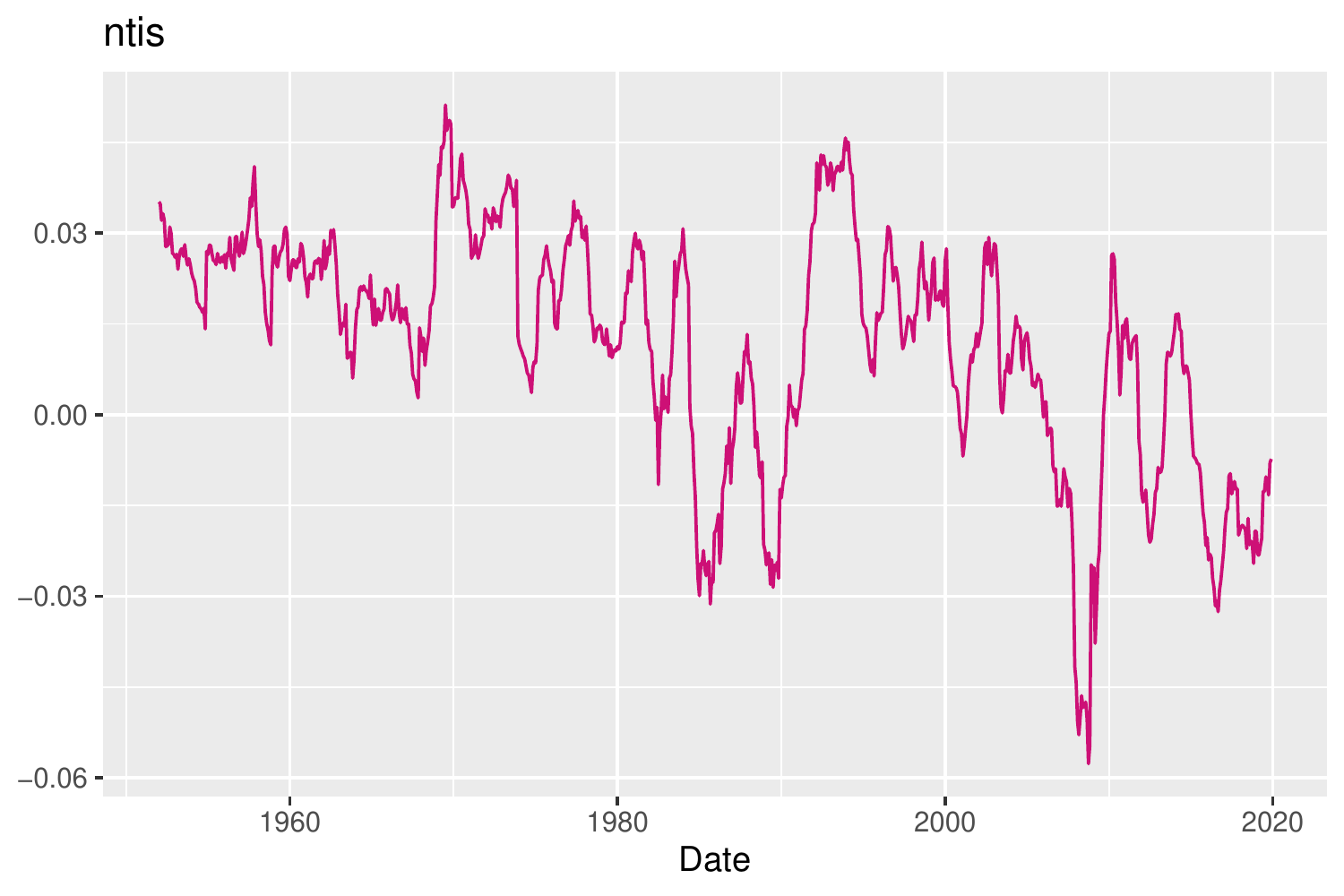} \\
					\includegraphics[scale=0.5]{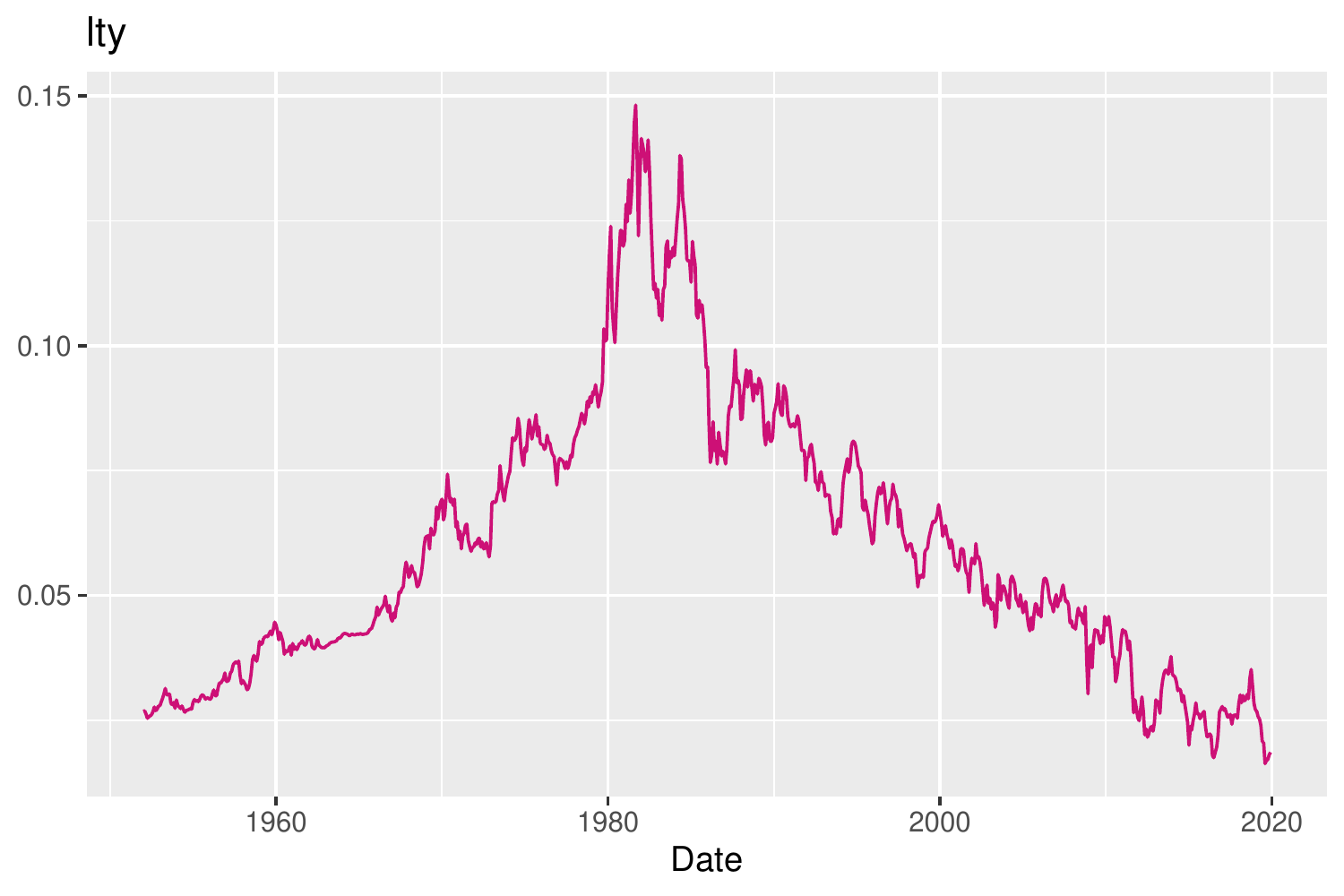} &  \includegraphics[scale=0.5]{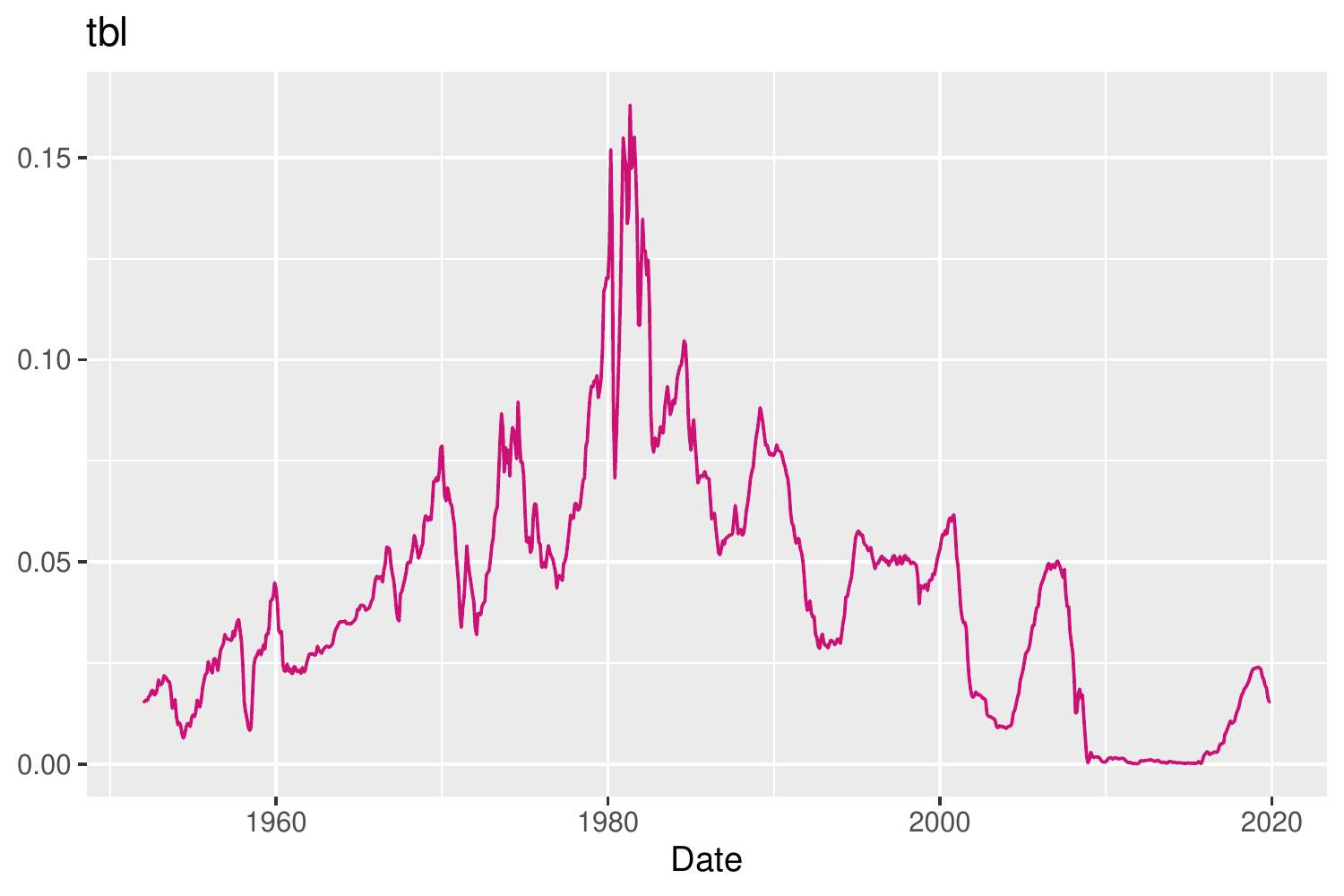} \\
				\end{tabular}
				\begin{tablenotes}[para,flushleft]
					\item {\footnotesize Notes: All plots are based on 816 monthly observations ranging from January 1952 to December 2019. The full predictor names are defined in Table \ref{tb:v_names}.}
				\end{tablenotes}
			\end{threeparttable}
		\end{figure}

\begin{figure}
			\caption{Time-series Plots of Stationary Predictors}
			\label{fig_stationary}\centering
			\begin{threeparttable}
				\begin{tabular}{cc}
					\includegraphics[scale=0.5]{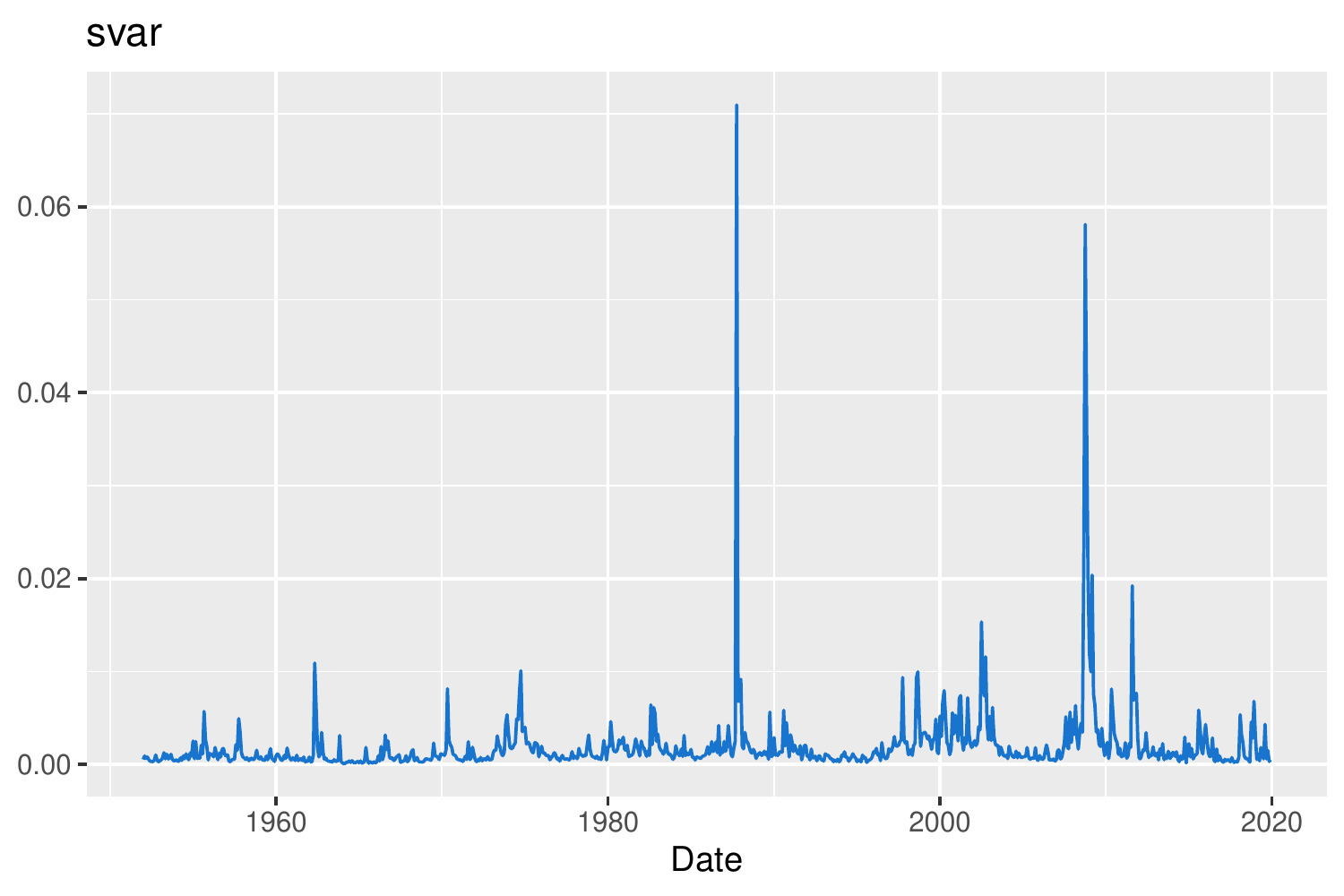} &  \includegraphics[scale=0.5]{figures/plot_sta_dfr} \\
					\includegraphics[scale=0.5]{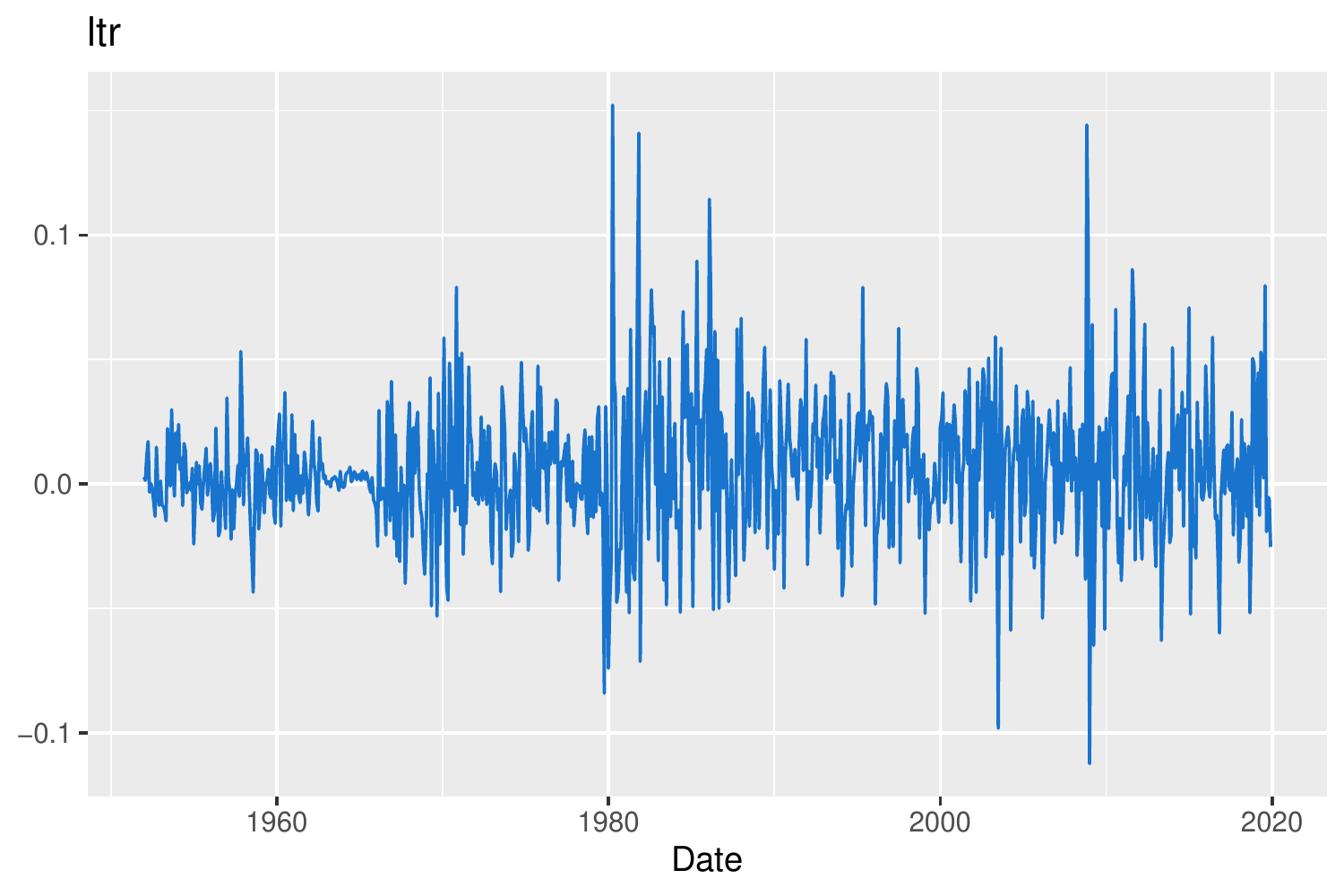} &  \includegraphics[scale=0.5]{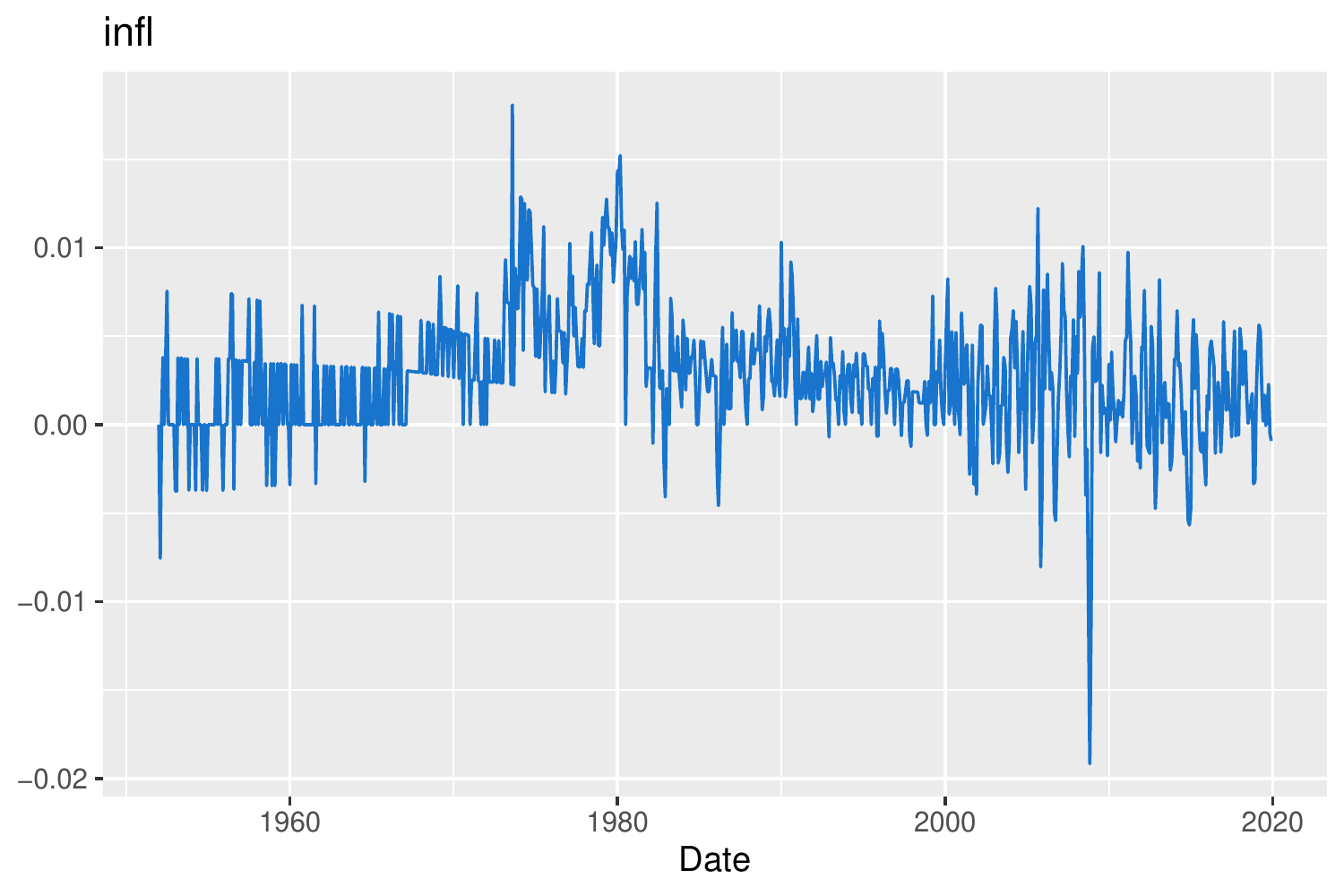} \\
				\end{tabular}
				\begin{tablenotes}[para,flushleft]
					\item {\footnotesize Notes: All plots are based on 816 monthly observations ranging from January 1952 to December 2019. The full predictor names are defined in Table \ref{tb:v_names}.}
				\end{tablenotes}
			\end{threeparttable}
		\end{figure}

		We first check the mixed-root property of the predictors.
		In Figures \ref{fig_persistent}--\ref{fig_stationary}, we plot each predictor using monthly observations. The predictors (dp, dy, ep, bm, dfy, ntis, lty, and tbl) collected in Figure \ref{fig_persistent} are highly persistent and have quite different patterns from the other four predictors (svar, dfr, ltr and infl) in Figure \ref{fig_stationary}.
		The first-order autoregression coefficients reported in Table \ref{tb:v_names} also suggest that predictors have heterogeneous degrees of persistence. We also conduct the Johansen test for cointegration. The results show that the cointegrating rank is $3$ in all of the 804-month rolling windows.\footnote{The Johansen cointegration test results are presented in the appendix, Table \ref{table_Johansen}.} Thus, the data fit into the mixed-root model structure discussed in Section \ref{SEC-model2}.

		We next investigate the performance of ALQR in the quantile prediction problem of stock returns.
		For comparison, we also apply three alternative estimators in the literature: the QR estimator without selecting predictors (QR), the quantile lasso (LASSO), and the unconditional quantile without using any predictor (QUANT).
		For the tuning parameter of LASSO and ALQR, we use both the Bayesian information criteria (BIC) and the generalized information criteria (GIC), which are explained in detail in Section \ref{section_MonteCarlo}.
		We evaluate the performance of quantile prediction using the final prediction error (FPE) and the out-of-sample $R^2$ following \citet{lu2015jackknife}.
		The FPE measures the out-of-sample quantile prediction errors:
		\begin{eqnarray}
			\text{FPE}(\tau) = \frac{1}{S} \sum_{s=1}^{S} \rho_{\tau}(y_s - \hat{y}_s),
			\label{eqn_FPE}
		\end{eqnarray}
		where $\hat{y}_s$ is a prediction for $\tau$-quantile of $y_s$ and $S$ is the number of out-of-sample predictions.
		Note that FPE($\tau$) averages the quantile loss function. It is different from the standard mean squared prediction error.
		At each quantile $\tau$, a smaller FPE implies a better quantile prediction.
		The second measure of the performance is the out-of-sample $R^2$:
		\begin{eqnarray}
			R^2(\tau) = 1- \frac{\sum_{s=1}^{S} \rho_{\tau} (y_{s} - \hat{y}_{s})}{\sum_{s=1}^{S} \rho_{\tau} (y_{s} - \bar{y}_{s})},
			\label{eqn_R2}
		\end{eqnarray}
		where $\bar{y}_{s}$ is the unconditional $\tau$-quantile of $y$ (QUANT) from the training sample. Note that the out-of-sample $R^2$ measures the performance of each method relative to QUANT. For example, a positive $R^2$ indicates that the prediction error is smaller than that of QUANT, and a larger $R^2$ implies a better prediction. By definition, $R^2$ of QUANT is always zero.

		Tables \ref{table_stockreturns_FPE_and_R2_BIC}--\ref{table_stockreturns_FPE_and_R2_BIC_24step} summarize the prediction results with BIC.
		They are based on the one-step-ahead prediction for the last 12 and 24 periods of the sample, respectively.
		The results with GIC are similar and left in the appendix.
		We consider 5 different quantiles, $\tau=0.05, 0.1, 0.5, 0.9$, and $0.95$.
		For each quantile, the best performance results, i.e., the smallest FPE and the highest $R^2$, are marked in bold.
		In addition to performance measures, we also report the average number of selected predictors and the corresponding tuning parameter value $\lambda$ for LASSO and ALQR.

		Overall, ALQR shows quite satisfactory results.
		First, ALQR performs the best in prediction across all quantiles, except for two designs: $\tau=0.95$ at the 12-period experiment and $\tau=0.05$ at the 24-period experiment.
		In the former design, LASSO performs slightly better than ALQR but the difference is negligible. In the latter design, QR performs better than the alternatives.
		Second, the relative performance of ALQR to QUANT improves substantially for higher quantiles. Looking at $R^2$, we find that ALQR predicts 44\% ($\tau=0.9$) and 32\% ($\tau=0.95$) better than QUANT at the 12-period design and 28\% ($\tau=0.9$) and 31\% ($\tau=0.95$) at the 24-period design, respectively. QR
		does not show particularly better prediction performance than QUANT, which is in line with the well-known result in the mean stock return prediction literature (see, e.g., \citet{welch2008comprehensive}).
		Third, ALQR selects fewer predictors than LASSO on average.
		This result is consistent with the report in literature that the standard lasso overselects relevant predictors in practice (See, e.g.,~\citet{wasserman2009high}).
		In Section \ref{SEC:Oracle}, we prove the oracle properties of ALQR including model selection consistency.
		In Section \ref{section_MonteCarlo}, we provide further numerical evidence that the number of selected variables by ALQR is closer to the true sparsity via Monte Carlo simulations.

		\begin{table}[tbp]
			\caption{Prediction Results of Stock Returns: 12 one-period-ahead forecasts, BIC}
			\label{table_stockreturns_FPE_and_R2_BIC}\centering
			\begin{threeparttable}
				\begin{tabular}{ l rrrrr}
					\hline
					& \multicolumn{5}{c}{Quantile ($\tau$)} \\
					\cline{2-6}
					& \multicolumn{1}{c}{0.05} & \multicolumn{1}{c}{0.1} & \multicolumn{1}{c}{0.5} & \multicolumn{1}{c}{0.9} & \multicolumn{1}{c}{0.95} \\
					\hline
					& \multicolumn{5}{c}{\underline{Final Prediction Error (FPE)}}\\
					QR  & 0.0047 & 0.0099 & 0.0134 & 0.0078 & 0.0060 \\
					LASSO & \textbf{0.0045} & 0.0084 & 0.0123 & 0.0033 & \textbf{0.0020} \\
					ALQR & \textbf{0.0045} & \textbf{0.0083}& \textbf{0.0122} & \textbf{0.0031} & \textbf{0.0020} \\
					QUANT & 0.0046 & \textbf{0.0083} & 0.0124 & 0.0055 & 0.0029 \\
					\\
					& \multicolumn{5}{c}{\underline{Out-of-Sample $R^2$}}\\
					QR & -0.0147 & -0.1930 & -0.0831 & -0.4048 & -1.0666 \\
					LASSO & 0.0219 & -0.0045 & 0.0094 & 0.4115 & \textbf{0.3233} \\
					ALQR& \textbf{0.0223} & \textbf{-0.0012} & \textbf{0.0167} & \textbf{0.4421} & 0.3211 \\
					\\
					& \multicolumn{5}{c}{\underline{Average \# of Selected Predictors}}\\
					LASSO & 11.00 & 9.42 & 12.00 & 10.92 & 10.58 \\
					ALQR & 10.00 & 7.83 & 10.00 & 6.08 & 9.25 \\
					\\
					& \multicolumn{5}{c}{\underline{Tuning Parameter ($\lambda$) by BIC}}\\
					LASSO $(\times10^{-4})$ &82.07 & 684.27 & 0.04 & 263.25 & 27.66 \\
					ALQR $(\times10^{-7})$& 5.26 & 89.29 & 1.46 & 283.17 & 0.88 \\
					\hline
				\end{tabular}%
				\begin{tablenotes}[para,flushleft]
					\item {\footnotesize Notes: For each quantile, the best performance result is written in bold font.}
				\end{tablenotes}
			\end{threeparttable}

		\end{table}

		\begin{table}[tbp]
			{}	\caption{Prediction Results of Stock Returns: 24 one-period-ahead forecasts, BIC}
			\label{table_stockreturns_FPE_and_R2_BIC_24step}\centering
			\begin{threeparttable}
				\begin{tabular}{l rrrrr}
					\hline
					& \multicolumn{5}{c}{Quantile ($\tau$)} \\
					\cline{2-6}
					& \multicolumn{1}{c}{0.05} & \multicolumn{1}{c}{0.1} & \multicolumn{1}{c}{0.5} & \multicolumn{1}{c}{0.9} & \multicolumn{1}	{c}{0.95} \\
					\hline
					& \multicolumn{5}{c}{\underline{Final Prediction Error (FPE)}}\\
					QR & \textbf{0.0045} & 0.0102 & 0.0147 & 0.0067 & 0.0050 \\
					LASSO & 0.0058 & 0.0102 & \textbf{0.0140} & \textbf{0.0045} & \textbf{0.0025} \\
					ALQR & 0.0058 & \textbf{0.0097} & \textbf{0.0140} & \textbf{0.0045} & \textbf{0.0025} \\
					QUANT & 0.0055 & 0.0098 & 0.0148 & 0.0063 & 0.0036 \\
					\\
					& \multicolumn{5}{c}{\underline{Out-of-Sample $R^2$}}\\
					QR & \textbf{0.1751} & -0.0388 & 0.0100 & -0.0745 & -0.3938 \\
					LASSO & -0.0590 & -0.0349 & 0.0573 & 0.2767 & 0.2881 \\
					ALQR & -0.0526 & \textbf{0.0103} & \textbf{0.0577} & \textbf{0.2798} & \textbf{0.3088} \\
					\\
					& \multicolumn{5}{c}{\underline{Average \# of Selected Predictors}}\\
					LASSO & 10.88 & 8.96 & 12.00 & 10.63 & 10.71 \\
					ALQR & 9.96 & 8.88 & 10.63 & 10.25 & 9.67 \\
					\\
					& \multicolumn{5}{c}{\underline{Tuning Parameter ($\lambda$) by BIC}}\\
					LASSO $(\times10^{-4})$ &66.33 & 577.58 & 0.30 & 210.13 & 14.76 \\
					ALQR $(\times10^{-7})$& 3.74 & 24.70 & 0.65 & 1.55 & 0.59 \\
					\hline
				\end{tabular}%
				\begin{tablenotes}[para,flushleft]
					\item {\footnotesize Notes: For each quantile, the best performance result is written in bold font.}
				\end{tablenotes}
			\end{threeparttable}
		\end{table}

\begin{figure}[tbph]
			\caption{The ALQR (with BIC) Estimated Coefficients}
			\label{fig_plotcoefalasso}\centering
			\includegraphics[width=1\linewidth]{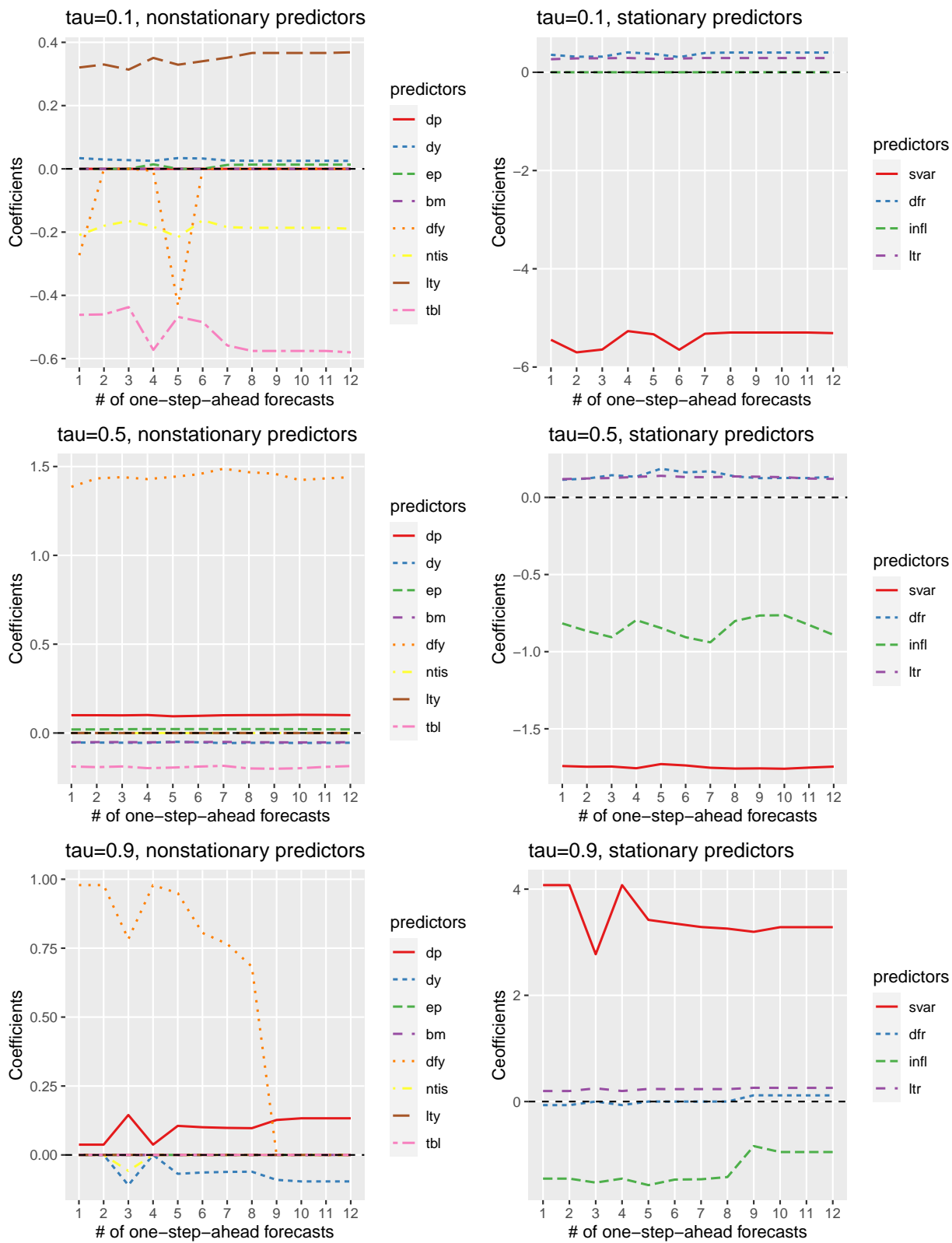}
		\end{figure}

		Finally, we make some remarks from the empirical perspective.
		In Figure \ref{fig_plotcoefalasso}, we plot the coefficient estimates of ALQR at the 12-period experiment.
		To make the graph readable, we divide the predictors into two groups (persistent and stationary) and restrict the quantiles into $\tau=0.1, 0.5$, and $0.9$.
		First, we observe that predictors have heterogeneous effects across quantiles. This provides useful information on predicting the distribution of stock returns.
		For high stock returns ($\tau=0.9$), default yield spread (dfy), dividend price ratio (dp), dividend yield ratio (dy), stock variance (svar), and inflation (infl) show larger effects.
		However, long term yield (lty), net equity expansion (ntis), treasury bill rates (tbl), and stock variance (svar) show large coefficients for low stock returns ($\tau=0.1$).
		Second, the magnitude of the coefficients at the median is relatively smaller than that at both tails.
		This result coincides with the previous findings in the literature (see, e.g.,~\citet{welch2008comprehensive} and \citet{fan2019predictive}) that it is more difficult to predict the center part of the stock return distribution.
		The positive but small $R^{2}$'s of ALQR at $\tau=0.5$ in Tables \ref{table_stockreturns_FPE_and_R2_BIC}--\ref{table_stockreturns_FPE_and_R2_BIC_24step} also reflect such an aspect.
		Third, the coefficient of stock variance (svar) swings from a large positive value at $\tau=0.9$ to a large negative number at $\tau=0.1$. Since the return distribution would spread out when the market is volatile, this result is consistent with the stylized fact in the financial market.

\section{Oracle Properties of ALQR Estimators\label{SEC:Oracle}}

This section investigates the asymptotic properties of ALQR. We show that the oracle properties (\citet{fan2001variable}) of the adaptive lasso remain valid in the QR framework when predictors have both increasing dimensions and various degrees of persistence. This result provides theoretical support for applying the ALQR method to stock returns prediction in Section 3. In Section 4.1, we discuss the case that all predictors are $I(1)$. Section 4.2 then generalizes the results to the case of mixed root predictors.

		\subsection{Unit-Root Predictors\label{oracle ur}}

		Consider the QR model in Section \ref{SEC-model1} with unit-root predictors only. The following assumptions provide regularity conditions.

		\begin{assumption}[Assumption $f$]
			\label{ass: F and f}
			(i) The distribution function of $u_{t\tau }$, $F(\cdot
			)$, has a continuous density $f(\cdot )$ with $f(a)>0$ on $\{a:0<F(a)<1\}$.
			(ii) The derivative of conditional distribution function $F_{t-1}(a)=Pr[u_{t \tau }<a|\mathcal{F}_{t-1}]$, which we denote $f_{t-1}(\cdot )$, is
			continuous and uniformly bounded above by a finite constant $c_{f}$. (iii) For any sequence $\zeta _{n}\rightarrow F^{-1}(\tau )$, $%
			f_{t-1}(\zeta _{n})$ is uniformly integrable, and $E[f_{t-1}^{1+\eta
			}(F^{-1}(\tau ))]<\infty $ for some $\eta >0$.
		\end{assumption}

For the bound and rate conditions, we define $A_{(t,p)}:=E\left[f_{t-1}(0)x_{t-1}x_{t-1}^{\prime }\right]$ and $B_{(t,p)}:=E\left[ \psi _{\tau}(u_{t\tau })^{2}x_{t-1}x_{t-1}^{\prime }\right]$. 

		\begin{assumption}[Assumption $L1$]
			\label{assumption A1} 
			\label{assumption L1} 
			For each $t$ and $p$, there exist some constants $
			\underline{c}_{A(t,p)}$ and $\underline{c}_{B(t,p)}$ such that $0<\underline{c}_{A(t,p)}\leq \lambda _{\min }(\frac{A_{(t,p)}}{t})$ and $0<\underline{c}_{B(t,p)}\leq \lambda _{\min }(\frac{B_{(t,p)}}{t})$.
		\end{assumption}

Note that, for each $t$ and $p$, $\lambda_{\max}(A_{(t,p)}/p)$ and $\lambda_{\max}(B_{(t,p)}/p)$ are bounded above since $|f_{t-1}(\cdot)|$ and $|\psi_{\tau}(\cdot)|$ are uniformly bounded and the variance of $x_{t-1}$ is finite by the design of $v_{t}$. 
Let $\overline{c}_{A(t,p)}<\infty$ and $\overline{c}_{B(t,p)}<\infty$ be these finite bounds. 
We next define the following uniform bounds over $t$ for each $n$: $\underline{c}_{A(n,p)}:=\min_{1\leq t\leq n}\underline{c}_{A(t,p)}$, $\overline{c}_{A(n,p)}:=\max_{1\leq t\leq n}\overline{c}_{A(t,p)}$, and $\overline{c}_{B(n,p)}:=\max_{1\leq t\leq n}\overline{c}_{B(t,p)}$.


		\begin{assumption}[Assumption $U1$]
			\label{assumption B1}
			\label{assumption U1} 
			For some $\alpha>0$, (i) $p=n^{\zeta }$ with $0<\alpha \zeta <1$;
			(ii) $\frac{\overline{c}_{A(n,p)}^{1/2}n^{1/2}}{p^{\alpha }}\vee \frac{\overline{c}%
				_{B(n,p)}^{1/2}}{p^{\alpha }}=o(\underline{c}_{A(n,p)})$;
			 (iii)$\frac{%
				p^{3/2}}{n^{2+\alpha \zeta}\overline{c}_{A(n,p)}^{1/2}}=\frac{n^{(3/2)\zeta
			}}{n^{2+\alpha \zeta }\overline{c}_{A(n,p)}^{1/2}}=o(1).$
		\end{assumption}


		\begin{assumption}[Assumption $\protect\lambda 1$]
			\label{assumption lambda} $\lambda _{n}$ satisfies that (i) $\frac{\lambda _{n}n^{\frac{1}{2}\zeta }}{n^{1+\alpha \zeta }\underline{c}_{A(n,p)}}\rightarrow 0$; (ii) $\frac{\lambda _{n}n^{(1-\alpha
					\zeta )\gamma }}{n^{2+\alpha \zeta }\overline{c}_{A(n,p)}^{1/2}}\rightarrow
			\infty $, with $\gamma >0.$
		\end{assumption}
	
		\begin{assumption}
		\label{5.1}
		Let $q_{n}=\left\Vert \mathcal{A}_{n}\right\Vert_0 $ be the size of the
		true active set $\mathcal{A}_{n}$. There exist positive constants $c_{q}$, $%
		c_{\beta }$ and $c_0$ such that (i) $q_{n}=O(n^{c_{q}})$ with $c_{q}<1/3;$(ii) 
		$2c_{q}<c_{\beta }<1$ and (iii) 
		\begin{equation*}
			n^{(1-c_{\beta })/2}\cdot \min_{1\leq j\leq q_{n}}\left\vert \beta
			_{0j}\right\vert \geq c_0.
		\end{equation*}
	\end{assumption}

	We make some remarks on these assumptions.
	Assumption $f$ is a standard restriction in the QR literature on the conditional density of regression residuals. 
	Assumptions $L1$ and $U1$ modify the conventional conditions from stationary QR
		literature to allow for serial dependence in predictors and regression errors. In particular, we modify Assumption $A.2$ of \cite{lu2015jackknife}
		to accommodate nonstationary $x_t$. Based on the model settings in Section \ref{SEC-model1}, we find that 
		\begin{equation*}
			E\left[ x_{t-1}x_{t-1}^{\prime }\right] =t\cdot \Sigma _{vv}+o(t),
		\end{equation*}
	  for $t=\left\lfloor rn\right\rfloor $ with $r\in \left( 0,1\right) $, as $%
	  n\rightarrow \infty$. Therefore, we impose a similar set of restricted eigenvalue
		conditions for $\frac{E\left[ x_{t-1}x_{t-1}^{\prime }\right] }{t}$. This is a natural extension of the existing conditions in the \emph{i.i.d.}\ or stationary QR setup to a nonstationary time series model with an increasing dimension. 
		We allow the upper bounds $\overline{c}_{A(t,p)}$, $\overline{c}_{B(t,p)}$ and the lower bounds $\underline{c}_{A(t,p)}$, $\underline{c}_{B(t,p)}$ to depend on the dimension of predictors, hence the upper bounds can diverge to infinity and the lower bounds can converge to zero when $p\rightarrow\infty$ as $n\rightarrow \infty$.
		However, Assumption $U1$ (ii) imposes further restrictions on the upper bounds: The rate of convergence in $p$ is slow enough so that $(n/p^\alpha)$-consistency of the ALQR estimator can be achieved.
		Specifically, Assumption $L1$ prevents high correlation between predictor innovations. Note that $\lambda_{\min}(E[x_{t-1}x_{t-1}]/t)=0$ if there exits perfect multicollinearity between predictor innovations.
		Also, Assumption $U1$ restricts the strength of the contemporaneous correlation in $v_t$ by limiting nonzero off-diagonal terms in $\Sigma_{vv}$.\footnote{
		In the i.i.d.\ mean regression model with highly correlated regressors, it is well known that the lasso performs worse than the ridge estimator.
		To accommodate the lasso with highly correlated regressors, researchers have also developed variants of the lasso such as the elastic net \citep{zou2005regularization} and the group lasso \citep{yuan2006model}.
		As shown in Figure \ref{fig_heatmap} in the appendix, however, most predictor innovations in our empirical application are not highly correlated to each other and satisfy this assumption.
		Furthermore, we investigate the performance of the proposed ALQR estimator when predictor innovations are highly correlated in the simulation experiments.
		}
		In Assumption $U1$ (i), the condition $0<\alpha \zeta <1$ imposes $a_{n}=\frac{p^{\alpha }}{n}=\frac{n^{\alpha \zeta }}{n}\rightarrow 0$, indicating the restriction on the growth rate of the number of parameters. In the appendix,
		we discuss the technicality of these restrictions on $p$ and $\lambda $ in detail.
		Assumption $\lambda 1$ collects a set of rate conditions on $\lambda_n$ to show the asymptotic properties of the ALQR estimators. Assumption $\lambda 1$ (i) is comparable to the condition of $\lambda_n/\sqrt{n}\rightarrow 0$ in a stationary time series model with fixed dimensions. This condition allows $(n/p^\alpha)$-rate under the lasso variable selection. Assumption $\lambda 1$ (ii) presents a necessary condition for model selection consistency of ALQR, which depends on the consistency of the initial estimator for $\beta_{0\tau}$ (as shown in the appendix, proof of Theorem \ref{thm_sparsity}). Note that this condition is developed given that the ordinary QR estimator defined in (\ref{eqn_QR}) is used as an initial estimator for $\beta _{0\tau }$. It can be generalized further since we do not require a certain rate of consistency for the initial estimator. If another consistent estimator is used, condition (ii) may be adjusted to the corresponding rate of consistency. Assumption \ref{5.1} is for the minimum signal strength of the coefficients in the true active set, which we adopted from \citet*{sherwood2016partially}.


		Next we show that the ordinary QR estimator is consistent and can be a qualified initial estimator for ALQR.

		\begin{lemma}
			\label{initial estimator}\textbf{(Consistency of QR Estimator with unit-root
				predictors)} Under Assumptions \ref{ass: F and f}--\ref{assumption U1}, the QR
			estimator defined in (\ref{eqn_QR}) satisfies
			\begin{equation*}
				\left\Vert \hat{\beta}_{\tau }^{QR}-\beta _{0\tau }\right\Vert =O_{p}\left(
				\frac{p^{\alpha }}{n}\right) .
			\end{equation*}
		\end{lemma}

		It is well-known that the rate of convergence is $\frac{1}{n^{1/2}}$ with
		fixed $p$ and $\frac{p^{1/2}}{n^{1/2}}$ with increasing $p$ when predictors are $I$(0). This indicates that the information loss (or the degrees of freedom) to estimate the increasing number of parameters is $p^{1/2}$.
		It is also known that when the predictors are $I$(1), the rate of convergence becomes $\frac{1}{n}$ with fixed $p$. The super-consistency is caused by the stronger signal of $X^{\prime }X$, where $X$ is a $n\times p$ matrix of predictors. To allow for increasing $p$, Lemma \ref{initial estimator} extends the existing result for nonstationary QR and finds that the rate becomes $\frac{p^{\alpha }
		}{n}=\frac{p^{1/2}\cdot p^{\alpha-1/2 }}{n}$, where the additional rate
		loss $p^{\alpha -1/2}$ comes from the increasing singularity of $E\left[
		x_{t}x_{t}^{\prime }\right] $ as summarized in Assumptions $L1$ and $U1$.
		In contrast, we have $0<\left[ \lambda _{\min }\left( E\left[
		x_{t}x_{t}^{\prime }\right] \right) \right] <\left[ \lambda _{\max }\left( E\left[
		x_{t}x_{t}^{\prime }\right] \right) \right] <\infty $ for $I$(0) predictors with $p/n\rightarrow 0$.

		Taking the QR estimator of Lemma \ref{initial estimator} as an initial estimator for $\beta _{0\tau }$, we show the consistency of ALQR and model selection consistency under unit roots.
		\begin{theorem}
			\label{thm_consistency}{\textbf{(Consistency of ALQR under Unit Roots)}} Under Assumptions \ref{ass: F and f}--\ref{assumption lambda}, the ALQR
			estimator defined in (\ref{lassoQR_1}) satisfies
			\begin{equation*}
				\left\Vert \hat{\beta}_{\tau }^{ALQR}-\beta _{0\tau }\right\Vert
				=O_{p}\left( \frac{p^{\alpha }}{n}\right) .
			\end{equation*}
		\end{theorem}


		Theorem \ref{thm_consistency} confirms that both the ALQR estimator and the QR estimator of Lemma \ref{initial estimator} are $(n/p^\alpha)$-consistent. With a proper choice of the penalty parameter $\lambda_n$, the rate of consistency for the ALQR estimate is not affected. As we show in the proof of Theorem \ref{thm_consistency}, this estimation rate is possible under Assumption $\lambda 1$.

		\begin{theorem}
			\label{thm_sparsity}{\textbf{(Model Selection Consistency under Unit Roots)}} Let $\hat{\mathcal{A}}_{n}:=\{j:\hat{\beta}_{\tau ,j}^{ALQR}\neq 0\}$ and $\mathcal{A}_{0}:=\{j:\beta _{0\tau ,j}\neq0\}$.
			Under Assumptions \ref{ass: F and f}--\ref{5.1},
			it holds that
			\begin{equation*}
				\Pr (j\in \hat{\mathcal{A}}_{n})\longrightarrow 0,\quad \text{for $j\notin
					\mathcal{A}_{0}$},
			\end{equation*}%
			as $n\rightarrow \infty $.
		\end{theorem}

		Theorem \ref{thm_sparsity} shows that the ALQR procedure is an oracle procedure. When the predictors with diverging dimensions are all unit root, the adaptive lasso is at least as good as other ``oracle" procedures in terms of variable selection. This is a major difference from the standard lasso, which does not enjoy the oracle properties. As in \citet{zou2006adaptive}, with a proper choice of $\lambda_n$, the adaptive lasso procedure can simultaneously estimate relevant coefficients and remove unimportant ones with probability approaching unity. In this study, we further develop the result of \citet{zou2006adaptive} and generalize the adaptive lasso procedure to a nonstationary QR model. Note that the consistency in model selection can be achieved only under Assumption $\lambda1$.


		\subsection{Mixed-Root Predictors\label{oracle mr}}

		In this section, we consider a general QR model discussed in Section \ref{SEC-model2}:
			\begin{align} \label{untransformed QR}
		\begin{split}
		Q_{y_{t}}\left( \tau |\mathcal{F}_{t-1}\right)
		&={\ z_{t-1}}%
		^{\prime }\beta _{0\tau }^{z}+{x_{1t-1}^{c}}^{\prime }\beta _{0,1\tau }^{c}+{%
			x_{2t-1}^{c}}^{\prime }\beta _{0,2\tau }^{c}+{x_{t-1}}^{\prime }\beta
		_{0\tau }^{x} \\
		&=(z_{t-1}^{\prime },{\ x_{1t-1}^{c}}^{\prime },{x_{2t-1}^{c}}%
		^{\prime })\beta _{0\tau }^{(0)}+{x_{t-1}}^{\prime }\beta _{0\tau }^{(1)} \\
		&=X_{t-1}^{\prime }\beta _{\tau }
		\end{split}
		\end{align}%
	where $X_{t}=(z_{t}^{\prime },{\ x_{1t}^{c}}^{\prime },{x_{2t}^{c}}^{\prime
	},x_{t}^{\prime })^{\prime }$ and $\beta _{0\tau }=(\beta _{0\tau
	}^{z^{\prime }},\beta _{0,1\tau }^{c^{\prime }},\beta _{0,2\tau }^{c^{\prime
	}},\beta _{0\tau }^{x^{\prime }})^{\prime }$ are $p$-dimensional vectors.
		 This model allows predictors to have heterogeneous degrees of persistence as well as increasing dimensions. The mixed-root predictors in $X_t$ include stationary ($z_{t}$), local unit root ($x_{t}$), and cointegrated ($x_{1t}^c$, $x_{2t}^c$) processes.

		To capture the cointegration relation between $x_{1t}^{c}$ and $x_{2t}^{c}$, we define a transformation matrix $H$ as follows:
		\begin{equation*}
			H_{(p\times p)}:=\left(
			\begin{array}{cccc}
				I_{p_{z}} & 0 & 0 & 0 \\
				0 & I_{p_{1}} & 0 & 0 \\
				0 & A_{1}^{\prime } & I_{p_{2}} & 0 \\
				0 & 0 & 0 & I_{p_{x}}%
			\end{array}
			\right).
		\end{equation*}%
		Using $H$, the original model can be rewritten as:
		\begin{align} \label{transformed QR}
			\begin{split}
				Q_{y_{t}}\left( \tau |\mathcal{F}_{t-1}\right)
				&=X_{t-1}^{\prime }\beta _{0\tau }=\left( X_{t-1}^{\prime
				}H^{-1}\right) H\beta _{0\tau } \\
				&\equiv \tilde{x}_{t-1}^{\prime }\tilde{\beta}_{0\tau },
			\end{split}
		\end{align}%
		where $\tilde{\beta}_{0\tau }:=H\beta _{0\tau }=(\beta _{0\tau }^{z \prime},\beta _{0,1\tau }^{c \prime}, (A_{1}^{\prime }\beta _{0,1\tau }^{c}+\beta _{0,2\tau }^{c})', \beta _{0\tau }^{x \prime})'$ and $\tilde{x}_{t-1}^{\prime }:=X_{t-1}^{\prime }H^{-1}=(z_{t}^{\prime },{\ v_{1t}^{c}}^{\prime },{x_{2t}^{c}}^{\prime},x_{t}^{\prime })\equiv\left( w_{t}^{(0)\prime },w_{t}^{(1)\prime }\right)$.
		The $w_{t}^{(0)}=(z_{t}^{\prime },{v_{1t}^{c}}^{\prime })^{\prime }\ $ is a $(p_z+p_1)$-dimensional vector that collects all $I(0)$ processes, and $\tilde{\beta}_{0\tau }^{(0)}$ is the corresponding $(p_{z}+p_{1})
		$-dimensional QR regression coefficient. The $w_{t}^{(1)}=({x_{2t}^{c}}^{\prime },x_{t}^{\prime })^{\prime }$ is a $(p_2+p_x)$-dimensional vector of all $I(1)$ processes, $\tilde{\beta}_{0\tau }^{(1)}$ is the corresponding $(p_{2}+p_{x})$%
		-dimensional QR regression coefficient.
		Since the cointegration rank of this system is $p_1$, we let $r:=p_{z}+p_{1}$ be the number of $I(0)$ predictors in $\tilde{x}_{t-1}$. Note that the discussion in Section \ref{oracle ur} can be considered as a special case of the mixed root model with $r=0$. In this section, we generalize the results in Section \ref{oracle ur} and examine a more practical model with $r>0$.

		The following assumptions provide additional regularity conditions for mixed roots. Abusing notation slightly, we will use the same notation for the sequences of bounds. Let $M_{n}$ be a diagonal matrix of the form
			\begin{equation*}
				M_{n}:=\left(
				\begin{array}{cc}
					\sqrt{n}I_{r} & 0 \\
					0 & I_{p-r}%
				\end{array}%
				\right).
			\end{equation*}

		\begin{assumption}[Assumption$~L2$]
			\label{assumption A2}
			\label{assumption L2}
			There exist $\underline{c}_{A(n,p)}$ and $\underline{c}_{B(n,p)}$ such that
			\begin{align*}
				0<\underline{c}_{A(n,p)} & \leq \lambda _{\min }\left( \frac{M_{n}^{\prime}\sum_{t=1}^{n}E\left[ f_{t-1}(0)\tilde{x}_{t-1}\tilde{x}_{t-1}^{\prime }\right] M_{n}}{n^{2}}\right) 
			\end{align*}and 
			\begin{align*}
				0<\underline{c}_{B(n,p)} & \leq \lambda _{\min }\left( \frac{M_{n}^{\prime}\sum_{t=1}^{n}E\left[ \psi _{\tau }(u_{t\tau })^{2}\tilde{x}_{t-1}\tilde{x}_{t-1}^{\prime }\right] M_{n}}{n^{2}}\right) 
			\end{align*}%
			$\,$ as $n\rightarrow \infty $.
		\end{assumption}

		\begin{assumption}[Assumption$~U2$]
			\label{assumption B2}
			\label{assumption U2}
			There exist $\overline{c}_{A(n,p)}$ and $\overline{c}_{B(n,p)}$ such that
			\begin{align*}
				\lambda _{\max }\left( \frac{M_{n}^{\prime}\sum_{t=1}^{n}E\left[ f_{t-1}(0)\tilde{x}_{t-1}\tilde{x}_{t-1}^{\prime }\right] M_{n}}{n^{2}}\right)  \leq
				\overline{c}_{A(n,p)}<\infty
			\end{align*}%
			and
			\begin{align*}
				\lambda _{\max }\left(\frac{M_{n}^{\prime }\sum_{t=1}^{n}E\left[ \psi _{\tau }(u_{t\tau })^{2}\tilde{x}_{t-1}\tilde{x}_{t-1}^{\prime }\right] M_{n}}{n^{2}}\right) \leq
				\overline{c}_{B(n,p)}<\infty
			\end{align*}%
			$\,$ as $n\rightarrow \infty. $\newline
		\end{assumption}

		\begin{assumption}[Assumption$~\protect\lambda 2$]
				\label{assumption lambda2}
			When  $\lambda _{n}\rightarrow \infty$ as $n \rightarrow \infty$,  (i) $\lambda_{n}\frac{n^{\frac{1}{2}\zeta }}{n^{\frac{1}{2}+\alpha \zeta }\underline{c}_{A(n,p)}}\rightarrow 0$; (ii) $\lambda_{n} \frac{n^{\left( 1/2-\alpha \zeta \right) \gamma }}{n^{2+\alpha \zeta }\overline{c}_{A(n,p)}^{1/2}}\rightarrow \infty .$
		\end{assumption}


		Assumption $L2$, $U2$, and $\lambda2$ are analogies to Assumption $L1$, $U1$ and $\lambda1$ in Section \ref{SEC:Oracle}.1. They are essential regularity conditions used to show ALQR's consistency in parameter estimation and model selection under mixed roots. Compared with Assumption $\lambda 1$, Assumption $\lambda 2$ are more restrictive on the choice of $\lambda_n$, because $\lambda_n$ of Assumption $\lambda 2$ accommodates both stationary and local unit root predictors.

		We first show the consistency of QR for the transformed model in \eqref{transformed QR}.
		\begin{lemma}
			\label{initial estimator_transformed}
			\textbf{(Consistency of the transformed QR Estimator under Mixed Roots)}
			Under Assumptions
			\ref{ass: F and f},
			\ref{assumption U1},
			\ref{assumption L2} and \ref{assumption U2},
			we have
			\begin{align*}
				\left\Vert \hat{\beta}_{\tau }^{(0),QR}-\tilde{\beta}_{0\tau
				}^{(0)}\right\Vert =O_{p}\left( \frac{p^{\alpha }}{\sqrt{n}}\right) \text{,
					and }\left\Vert \hat{\beta}_{\tau }^{(1),QR}-\tilde{\beta}_{0\tau
				}^{(1)}\right\Vert =O_{p}\left( \frac{p^{\alpha }}{n}\right).
			\end{align*}
		\end{lemma}

		In a mixed root model, Lemma \ref{initial estimator_transformed} shows that the QR estimators for the $I(0)$ predictors and $I(1)$ predictors have different convergence rates. As we discussed in the previous section, the faster convergence rate for the $I(1)$ predictors comes from the stronger signal of $E\left[w_{t}^{(1)}w_{t}^{(1) \prime}\right]$. For the later use of these rates, we define $a_{n}^{(0)}:=(p^{\alpha }/\sqrt{n})$ and $a_{n}^{(1)}:=(p^{\alpha }/n)$.

		Based on the definition of the transformed model, we know that Lemma \ref{initial estimator_transformed} can further imply consistency of the QR estimator in the original model \eqref{QR_mixed}. Since all parameters are identical in model \eqref{QR_mixed} and \eqref{transformed QR} except $\tilde{\beta}_{2\tau }^{c}=A_{1}^{\prime }\hat{\beta}_{1\tau }^{c}+\hat{\beta}_{2\tau }^{c}$, the consistency of QR estimator in \eqref{QR_mixed} can be verified if we can show that $\hat{\beta}_{2\tau }^{c} $ is an $a_n^{(0)}$-consistent estimator for $\beta_{0,2\tau}^{c}$.  To see the convergence rate of $\hat{\beta}_{2\tau }^{c}$, we have
		\begin{eqnarray*}
			\hat{\beta}_{2\tau }^{c} &=&\left( \tilde{\beta}_{2\tau }^{c}\right)
			^{QR}-A_{1}^{\prime }\hat{\beta}_{1\tau }^{c} \\
			&=&\left( \left( \tilde{\beta}_{2\tau }^{c}\right) ^{QR}-\tilde{\beta}%
			_{0,2\tau }^{c}\right) -A_{1}^{\prime }\left( \hat{\beta}_{1\tau }^{c}-\beta
			_{0,1\tau }^{c}\right) +\left( \tilde{\beta}_{0,2\tau }^{c}-A_{1}^{\prime }\beta
			_{0,1\tau }^{c}\right) \\
			&=&O_{p}\left( a_{n}^{(1)}\right) +O_{p}\left( a_{n}^{(0)}\right) +\beta
			_{0,2\tau }^{c} \\
			&=&\beta _{0,2\tau }^{c}+O_{p}\left( a_{n}^{(0)}\right).
		\end{eqnarray*}%
		Thus, the $I$(0) rate dominates, and $\left\Vert \hat{\beta}_{2\tau}^{c}-\beta _{0,2\tau }^{c}\right\Vert =O_{p}\left( a_{n}^{(0)}\right) $.

				These results are summarized in the following corollary.
		\begin{corollary}
			\label{corollary consistency mix}
			\textbf{(Consistency of the QR Estimator under Mixed Roots) }
			Under Assumptions
			\ref{ass: F and f},
			\ref{assumption U1},
				\ref{assumption A2} and
				\ref{assumption B2},we have
			\begin{align*}
			\left\Vert \hat{\beta}_{\tau }^{(0),QR\ast }-\beta _{0\tau
			}^{(0)}\right\Vert =O_{p}\left(  a_{n}^{(0)} \right) \text{,
				and }\left\Vert \hat{\beta}_{\tau }^{(1),QR\ast }-\beta _{0\tau
			}^{(1)}\right\Vert =O_{p}\left(  a_{n}^{(1)}\right).
			\end{align*}

		\end{corollary}

		Note that ${\beta}_{0\tau }^{(0)}$ is the $(p_{z}+p_{1}+p_{2})
		$-dimensional QR regression coefficient, and ${\beta}_{0\tau }^{(1)}$ is the $p_{x}$%
		-dimensional QR regression coefficient, as given in Equation (\ref{untransformed QR}).
		We now investigate the oracle properties of ALQR with mixed roots.

		\begin{theorem}
			\label{thm_consistency_mix}
			{\textbf{(Consistency of ALQR under Mixed Roots)}}
			Under Assumptions
			\ref{ass: F and f},
			\ref{assumption U1},
				\ref{assumption A2}-\ref{assumption lambda2},
			we have
			\begin{align*}
			\left\Vert \hat{\beta}_{\tau }^{(0),ALQR\ast }-\beta _{0\tau
			}^{(0)}\right\Vert =O_{p}\left(  a_{n}^{(0)} \right) \text{,
				and }\left\Vert \hat{\beta}_{\tau }^{(1),ALQR\ast }-\beta _{0\tau
			}^{(1)}\right\Vert =O_{p}\left(  a_{n}^{(1)}\right).
			\end{align*}
		\end{theorem}

		\begin{theorem}
			\label{thm_sparsity_mix}
			{\textbf{(Model Selection Consistency under Mixed Roots)}}
			Under Assumptions
			\ref{ass: F and f},
			\ref{assumption U1},
				\ref{5.1}-\ref{assumption lambda2}, 
			it holds that
			\begin{equation*}
				\Pr (j\in \hat{\mathcal{A}}_{n})\longrightarrow 0,\quad \text{for $j\notin
					\mathcal{A}_{0}$},
			\end{equation*}%
			where $\hat{\mathcal{A}}_{n}=\{j:\hat{\beta}_{\tau ,j}^{ALQR\ast }\neq 0\}$
			and $\mathcal{A}_{0}=\{j:\beta _{0\tau ,j}\neq 0\}$.
		\end{theorem}

		It is worth emphasizing that the ALQR procedure does not require practitioners to conduct any pretest to identify different types of predictors. With the proper choice of $\lambda_n$ along with consistent initial estimates of the parameters, ALQR can simultaneously estimate relevant regression coefficients and provide the consistent variable selection for the mixed root QR model.

\section{Asymptotic Distributions of ALQR Estimators}\label{section_LimitTheory}
In this section, we derive the asymptotic distributions of the proposed ALQR
estimators. In QR literature, the Convexity Lemma (\citet{pollard1991asymptotics}) typically
provides a convenient limit theory, bypassing the stochastic equicontinuity
argument, see Section 4 of \citet{koenker_2005}. Unfortunately, for the increasing
dimension, we cannot use the Convexity Lemma which is defined in $\mathbb{R}%
^{p}$ with a fixed $p<\infty $. Thus we modify the classical chaining
argument of \citet{ruppert1980trimmed} to our ALQR\ framework. \citet{lu2015jackknife} recently modified the proof of \citet{ruppert1980trimmed} to show the
asymptotic distribution of Jackknife Model Averaging QR with \emph{iid} regressors of
increasing dimensions. We follow a similar proof strategy but substantially
refine the results to prove the distributional limit theory with weakly dependent
regressors of increasing dimensions, along with the adaptive lasso penalty
functions. For the asymptotic distribution with non-cointegrated local unit root predictors, we assume that the number of predictors ($p_{x}$) is
fixed. Note that $p=p_{z}+p_{1}+p_{2}+p_{x}$ is
still allowed to increase, even with this additional restriction of $%
p_{x}<\infty $. 

We now state the additional assumptions we need to prove the asymptotic distributions of
the ALQR estimators.

\begin{assumption}
	\label{5.2}
 The $p_{z}\times1$ vector of stationary predictors $%
z_{t}=\sum_{j=0}^{\infty }D_{zj}\epsilon _{t-j}$ satisfies the following
condition from Corollary 2 of \citet{withers1981conditions}%
\begin{equation*}
	\Vert D_{zj} \Vert=O\left( e^{-vj}\right) \text{ with }v>0
\end{equation*}%
along with the conditions (1), (2), (5) of \citet{withers1981conditions}, provided in the
appendix for brevity, see (\ref{withers1})-(\ref{withers5}) in Section \ref{appen_proof_section5} of the appendix.
Then $z_{t}$ is strong mixing with the strong mixing number $\alpha
(j)=O(e^{-v\lambda j})$ with another constant $\lambda >0$.
\end{assumption}

\begin{assumption} \label{5.3}
	$\sup_{j\geq 1}E(z_{t-1,j}^{8})\leq c_{z}$ for some $c_{z}\leq \infty $%
	.
\end{assumption}

\begin{assumption} \label{5.4}
	As $n\rightarrow \infty $, $\underline{c}_{B}^{-1/2}\overline{c}%
	_{A}^{1/2}p^{2}n^{-1/2}\rightarrow 0$, and $n^{-3}p^{12}\underline{c}%
	_{B}^{-4}\rightarrow 0$.
\end{assumption}

\begin{assumption} \label{5.5}
	We require $p_{x}$ (the number of non-cointegrated local unit root predictors) to be
	finite: 
	$p_{x}<\infty$.
\end{assumption}

 Assumption \ref{5.2} is a strong mixing condition with the specific mixing rates to use the Bernstein Inequality of \citet{merlevede2009bernstein}. Assumption \ref{5.3} is the moment condition we adopt from \citet{lu2015jackknife}. The moment condition is stronger than the typical assumptions when the number of the regressors is fixed. Assumption \ref{5.4} is another set of combined rate conditions required to prove the asymptotic theories of ALQR estimators. Finally, Assumption \ref{5.5} restricts the number of non-cointegrated local unit root predictors, which is also assumed in \cite{koo2020high} who study the mean regressions with increasing dimensions. The distributional QR limit theory with the increasing number of
(non-cointegrated) unit-root regressors is an open question, and we leave it
for a future research.

We are now ready to provide the asymptotic distributions of ALQR estimators for the coefficients in the true active set, which we denote $\hat{\beta}_{\tau }^{(i),ALQR\ast }(1)$ with $i=1$ and $2$ for $I(0)$ and $I(1)$ predictors, respectively. Let $R$ be a generic constant $l\times p_{i\text{ }}$matrix, where $l$ is
fixed and $p_{i}$ is defined conformably below. Notation $\Longrightarrow $
indicates the convergence in distribution.

	\begin{theorem}
	\label{thm_AsymDist_mix}
	{\textbf{(Asymptotic Distributions of ALQR Estimators)}}
	When Assumptions 	
	\ref{ass: F and f},
	\ref{assumption U1},
	\ref{5.1}-\ref{assumption lambda2},
	\ref{5.2}-\ref{5.5} hold,  
	
(i) For I(0) \emph{active} predictors (the first $p_{z}+p_{1}$
predictors; hence $p_{i}=p_{z}+p_{1}$), the first $q_{n}$ non-zero elements
has the following limit theory:\ 
\begin{equation*}
	\sqrt{n}R\left[ \hat{\beta}_{\tau }^{(0),ALQR\ast }(1)-\beta _{0\tau
	}^{(0)}(1)\right] \Longrightarrow N\left(
	0,RA_{(t,p)}^{-1}B_{(t,p)}A_{(t,p)}^{-1}R^{\prime }\right)
\end{equation*}

(ii) For the middle $p_{2}$ predictors (the parameters with respect to
the \emph{active} cointegrated predictors; hence $p_{i}=p_{2}$), the first $%
q_{n}$ non-zero elements has the following limit theory: 
\begin{equation*}
	\sqrt{n}R\left[ \hat{\beta}_{\tau }^{(1),ALQR\ast }(1)-\beta _{0\tau
	}^{(1)}(1)\right] \Longrightarrow N\left( 0,RA_{1}^{\prime
	}A_{(t,p)}^{-1}B_{(t,p)}A_{(t,p)}^{-1}A_{1}R^{\prime }\right)
\end{equation*}%

(iii) For the last $p_{x}$ (non-cointegrated) \emph{active} local unit
root predictors, with $p_{x}<\infty $, the first $q_{n}$ non-zero elements
has the following limit theory:\textbf{\ } 
\begin{equation*}
	n\left[ \hat{\beta}_{\tau }^{(1),ALQR\ast }(1)-\beta _{0\tau }^{(1)}(1)%
	\right] \Longrightarrow \left( M_{\beta _{\tau }}^{x}\right) ^{-1}G_{\tau
	}^{x},
\end{equation*}%
where 
\begin{equation*}
	G_{\tau }^{x}=\int J_{x}^{c}(r)dB_{\psi _{\tau }}\text{, and }M_{\beta
		_{\tau }}^{x}=f_{u}(0)^{-1}\int J_{x}^{c}(r)J_{x}^{c}(r)^{\prime }dr
\end{equation*}%
and $B_{\psi _{\tau }}:=BM(\tau (1-\tau ))$ is a Brownian motion and $%
J_{x}^{c}(r)=\int_{0}^{r}e^{(r-s)C}dB_{x}(s)$ is an Ornstein-Uhlenbeck (OU)
process.
\end{theorem}

The result of Theorem \ref{thm_AsymDist_mix}-(i) is in line with Theorem 2 of \citet{medeiros2016l1} who studied the asymptotic distributions of adaptive lasso estimators in mean regressions. It is also interesting to see that there is an additional nuisance parameter $A_{1}$ (the cointegrating
vector) in the asymptotic distribution of ALQR with the cointegrated predictors. 

\section{Monte Carlo Simulations}\label{section_MonteCarlo}

		In this section, we conduct a set of Monte Carlo simulation studies to
		evaluate the forecasting performance of the proposed ALQR method. As is motivated by the data patterns of the stock returns application in Section \ref{sec_application}, we construct a simulation environment that has 12 predictors. As seen in Table \ref{tb:v_names}, we allow for mix-root predictors and let 8 of them be persistent.

		Based on the monthly observations of the stationary predictors (\emph{svar}, \emph{dfr}, \emph{infl} and \emph{ltr}), we calibrate the structure of $z_t$ by estimating a VAR($p$) model with a Bayesian information criterion (BIC). An estimated VAR(2)
		model is obtained as follows:
		\begin{eqnarray*}
			z_{t} &=&
			\begin{pmatrix}
				0.427 & -0.059 & 0 & -0.017 \\
				0 & 0 & 0 & 0.074 \\
				0 & 0.022 & 0.538 & 0 \\
				1 & 0 & 0 & 0
			\end{pmatrix}
			z_{t-1} +
			\begin{pmatrix}
				0.208 & 0 & 0 & 0 \\
				0.421 & -0.089 & -0.312 & 0 \\
				0.092 & 0.023 & 0.239 & 0 \\
				0 & 0 & 1.038 & 0
			\end{pmatrix}
			z_{t-2} \\
			&& +u_{zt},
		\end{eqnarray*}
		where $u_{zt}\sim N(0,1)$.
		To identify the cointegrating relations among persistent predictors, we apply the Johansen test. The estimated cointegrating rank is $3$.
		Thus, we generate a set of cointegrating predictors from the following model:
		\begin{equation*}
			(x_{1t}^{c},x_{2t}^{c})^{\prime }=\left(
			\begin{array}{cccc}
				0.14 & 0.8495 & 0.0039 & -0.1545 \\
				0.19 & 0.8084 & 0.0033 & -0.0507 \\
				-1.24 & 1.2438 & 0.9788 & 0.3206 \\
				0.02 & -0.0205 & 0.0009 & 0.9835%
			\end{array}
			\right)  (x_{1,t-1}^{c},x_{2,t-1}^{c})^{\prime }+v_{t}^{c},
		\end{equation*}
		where $v_{t}^{c}\sim N(0,1)$. For unit-root predictors, we use $i.i.d.$\ $N(0,1)$ innovations with stationary initializations.

		We consider the following scenarios with different numbers of
		zero coefficients:\newline
		(1) 6 non-zero coefficients:
		\begin{equation*}
			\beta^z = (-0.4, \mathbf{0}, \mathbf{0}, 0.1), \quad ({\beta^c}^{\prime }, {%
				\beta^x}^{\prime }) = (0.2, -0.4, 0.1, 0.5, \mathbf{0}, \mathbf{0}, \mathbf{0%
			}, \mathbf{0});
		\end{equation*}
		(2) 8 non-zero coefficients:
		\begin{equation*}
			\beta^z = (-0.4, \mathbf{0}, \mathbf{0}, 0.1), ({\beta^c}^{\prime }, {\beta^x%
			}^{\prime }) = (0.2, -0.4, 0.1, 0.5, 0.2, 0.15, \mathbf{0}, \mathbf{0});
		\end{equation*}
		(3) 12 non-zero coefficients (no zero coefficient):
		\begin{equation*}
			\beta^z =(-0.4, 0.2, 0.15, 0.1), ({\beta^c}^{\prime }, {\beta^x}^{\prime })
			= (0.2, -0.4, 0.1, 0.5, 0.2, 0.15, 0.1, -0.05).
		\end{equation*}
		The sample size is set to be $n=1000$. The last $12$ periods are used for out-of-sample prediction evaluation and the remaining periods are used for in-sample estimation. The quantiles of interest are $0.05$, $0.1$, $0.5$, $0.9$, $0.95$.
		The prediction performance is measured by the final prediction error (FPE) and the out-of-sample $R^2$ defined in \eqref{eqn_FPE} and \eqref{eqn_R2}, respectively. The performance measures are computed by averaging over $1000$ replications.

		\begin{table}[tbhp]
			\caption{Simulation Results: Scenario 1 (6 non-zero coefficients)}
			\label{table_simulation_S1}\centering
			\begin{tabular}{l rrrrr}
				\hline
				& \multicolumn{5}{c}{Quantile ($\tau$)} \\
				\cline{2-6}
				& \multicolumn{1}{c}{0.05} & \multicolumn{1}{c}{0.1} & \multicolumn{1}{c}{0.5} & \multicolumn{1}{c}{0.9} & \multicolumn{1}{c}{0.95} \\
				\hline
				\\
				\multicolumn{5}{l}{\underline{Tuning parameter selection: BIC}} \\
				\\
				& \multicolumn{5}{c}{\underline{Final Prediction Error (FPE)}}\\
				QR    & 0.1821 & 0.2835 & 0.5807 & 0.2817 & 0.1795 \\
				LASSO & 0.1091 & 0.1813 & 0.4048 & 0.1807 & 0.1084 \\
				ALQR  & \textbf{0.1068} & \textbf{0.1795} & \textbf{0.4016} & \textbf{0.1793} & \textbf{0.1064} \\
				QUANT & 2.7366 & 4.9691 & 12.1063 & 4.8815 & 2.6289 \\
				\\
				& \multicolumn{5}{c}{\underline{Out-of-Sample $R^2$}}\\
				QR    & 0.9335 & 0.9430 & 0.9520 & 0.9423 & 0.9317 \\
				LASSO & 0.9601 & 0.9635 & 0.9666 & 0.9630 & 0.9588 \\
				ALQR  & \textbf{0.9610} & \textbf{0.9639} & \textbf{0.9668} & \textbf{0.9633} & \textbf{0.9595} \\
				\\
				& \multicolumn{5}{c}{\underline{Average \# of Selected Predictors}}\\
				LASSO & 9.80 & 9.39 & 8.99 & 9.43 & 9.74 \\
				ALQR  & 6.24 & 6.06 & 5.97 & 6.05 & 6.26 \\
				\\
				& \multicolumn{5}{c}{\underline{Tuning Parameter ($\lambda$)}}\\
				LASSO $(\times10^{0})$  & 8.27 & 15.68 & 36.46 & 15.17 & 8.56 \\
				ALQR $(\times10^{-4})$  & 1.71 & 2.38 & 3.99 & 2.34 & 1.69 \\
				\hline
				\\
				\multicolumn{5}{l}{\underline{Tuning parameter selection: GIC}} \\
				\\
				& \multicolumn{5}{c}{\underline{Final Prediction Error (FPE)}}\\
				QR    & 0.1815 & 0.2834 & 0.5805 & 0.2817 & 0.1793 \\
				LASSO & 0.1084 & 0.1817 & 0.4045 & 0.1805 & 0.1084 \\
				ALQR  & \textbf{0.1065} & \textbf{0.1800} & \textbf{0.4017} & \textbf{0.1791} & \textbf{0.1068} \\
				QUANT & 2.7285 & 4.9162 & 12.0961 & 4.8751 & 2.6290 \\
				\\
				& \multicolumn{5}{c}{\underline{Out-of-Sample $R^2$}}\\
				QR    & 0.9335 & 0.9424 & 0.9520 & 0.9422 & 0.9318 \\
				LASSO & 0.9603 & 0.9630 & 0.9666 & 0.9630 & 0.9588 \\
				ALQR  & \textbf{0.9610} & \textbf{0.9634} & \textbf{0.9668} & \textbf{0.9633} & \textbf{0.9594} \\
				\\
				& \multicolumn{5}{c}{\underline{Average \# of Selected Predictors}}\\
				LASSO & 10.14 & 9.79 & 9.35 & 9.80 & 10.13 \\
				ALQR  & 6.80 & 6.36 & 6.11 & 6.35 & 6.76 \\
				\\
				& \multicolumn{5}{c}{\underline{Tuning Parameter ($\lambda$)}}\\
				LASSO $(\times10^{0})$  & 6.21 & 11.66 & 28.39 & 11.54 & 6.32 \\
				ALQR $(\times10^{-4})$  & 1.22 & 1.93 & 3.55 & 1.94 & 1.24 \\
				\hline
			\end{tabular}%
		\end{table}

		\begin{table}[tbhp]
			\caption{Simulation Results: Scenario 2 (8 non-zero coefficients)}
			\label{table_simulation_S2}\centering
			\begin{tabular}{l rrrrr}
				\hline
				& \multicolumn{5}{c}{Quantile ($\tau$)} \\
				\cline{2-6}
				& \multicolumn{1}{c}{0.05} & \multicolumn{1}{c}{0.1} & \multicolumn{1}{c}{0.5} & \multicolumn{1}{c}{0.9} & \multicolumn{1}{c}{0.95} \\
				\hline
				\\
				\multicolumn{5}{l}{\underline{Tuning parameter selection: BIC}} \\
				\\
				& \multicolumn{5}{c}{\underline{Final Prediction Error (FPE)}}\\
				QR    & 0.1845 & 0.2864 & 0.5882 & 0.2878 & 0.1848 \\
				LASSO & 0.1085 & 0.1818 & 0.4056 & 0.1800 & 0.1078 \\
				ALQR  & \textbf{0.1074} & \textbf{0.1805} & \textbf{0.4033} & \textbf{0.1794} & \textbf{0.1069} \\
				QUANT & 2.6803 & 4.8142 & 12.1974 & 4.8994 & 2.5749 \\
				\\
				& \multicolumn{5}{c}{\underline{Out-of-Sample $R^2$}}\\
				QR    & 0.9312 & 0.9405 & 0.9518 & 0.9413 & 0.9282 \\
				LASSO & 0.9595 & 0.9622 & 0.9667 & 0.9633 & 0.9581 \\
				ALQR  & \textbf{0.9599} & \textbf{0.9625} & \textbf{0.9669} & \textbf{0.9634} & \textbf{0.9585} \\
				\\
				& \multicolumn{5}{c}{\underline{Average \# of Selected Predictors}}\\
				LASSO & 10.33 & 10.09 & 9.74 & 10.09 & 10.27 \\
				ALQR  & 8.23 & 8.06 & 7.99 & 8.04 & 8.20 \\
				\\
				& \multicolumn{5}{c}{\underline{Tuning Parameter ($\lambda$)}}\\
				LASSO $(\times10^{0})$ & 6.75 & 12.56 & 29.98 & 12.22 & 7.18 \\
				ALQR $(\times10^{-4})$  & 1.17 & 1.60 & 2.54 & 1.61 & 1.18 \\
				\hline
				\\
				\multicolumn{5}{l}{\underline{Tuning parameter selection: GIC}} \\
				\\
				& \multicolumn{5}{c}{\underline{Final Prediction Error (FPE)}}\\
				QR    & 0.1852 & 0.2876 & 0.5880 & 0.2874 & 0.1848 \\
				LASSO & 0.1090 & 0.1825 & 0.4035 & 0.1799 & 0.1076 \\
				ALQR  & \textbf{0.1077} & \textbf{0.1810} & \textbf{0.4021} & \textbf{0.1792} & \textbf{0.1071} \\
				QUANT & 2.7245 & 4.8459 & 12.2354 & 4.8893 & 2.5796 \\
				\\
				& \multicolumn{5}{c}{\underline{Out-of-Sample $R^2$}}\\
				QR    & 0.9320 & 0.9407 & 0.9519 & 0.9412 & 0.9283 \\
				LASSO & 0.9600 & 0.9623 & 0.9670 & 0.9632 & 0.9583 \\
				ALQR  & \textbf{0.9605} & \textbf{0.9627} & \textbf{0.9671} & \textbf{0.9633} & \textbf{0.9585} \\
				\\
				& \multicolumn{5}{c}{\underline{Average \# of Selected Predictors}}\\
				LASSO & 10.58 & 10.37 & 9.98 & 10.34 & 10.56 \\
				ALQR  & 8.58 & 8.28 & 8.09 & 8.29 & 8.55 \\
				\\
				& \multicolumn{5}{c}{\underline{Tuning Parameter ($\lambda$)}}\\
				LASSO $(\times10^{0})$ & 5.18 & 9.33 & 23.82 & 9.46 & 5.27 \\
				ALQR $(\times10^{-4})$  & 0.85 & 1.30 & 2.30 & 1.29 & 0.87 \\
				\hline
			\end{tabular}%
		\end{table}

		\begin{table}[tbhp]
			\caption{Simulation Results: Scenario 3 (12 non-zero coefficients)}
			\label{table_simulation_S3}\centering
			\begin{tabular}{l rrrrr}
				\hline
				& \multicolumn{5}{c}{Quantile ($\tau$)} \\
				\cline{2-6}
				& \multicolumn{1}{c}{0.05} & \multicolumn{1}{c}{0.1} & \multicolumn{1}{c}{0.5} & \multicolumn{1}{c}{0.9} & \multicolumn{1}{c}{0.95} \\
				\hline
				\\
				\multicolumn{5}{l}{\underline{Tuning parameter selection: BIC}} \\
				\\
				& \multicolumn{5}{c}{\underline{Final Prediction Error (FPE)}}\\
				QR    & 0.1927 & 0.2968 & 0.5992 & 0.2964 & 0.1906 \\
				LASSO & \textbf{0.1093} & \textbf{0.1820} & 0.4046 & \textbf{0.1802} & \textbf{0.1091} \\
				ALQR  & 0.1099 & 0.1825 & \textbf{0.4045} & 0.1807 & 0.1092 \\
				QUANT & 2.6958 & 4.9349 & 12.1341 & 4.8949 & 2.6794 \\
				\\
				& \multicolumn{5}{c}{\underline{Out-of-Sample $R^2$}}\\
				QR    & 0.9285 & 0.9398 & 0.9506 & 0.9395 & 0.9289 \\
				LASSO & \textbf{0.9595} & \textbf{0.9631} & \textbf{0.9667} & \textbf{0.9632} & \textbf{0.9593} \\
				ALQR  & 0.9592 & 0.9630 & \textbf{0.9667} & 0.9631 & 0.9592 \\
				\\
				& \multicolumn{5}{c}{\underline{Average \# of Selected Predictors}}\\
				LASSO & 11.97 & 11.99 & 12.00 & 11.99 & 11.98 \\
				ALQR  & 11.77 & 11.86 & 11.95 & 11.86 & 11.78 \\
				\\
				& \multicolumn{5}{c}{\underline{Tuning Parameter ($\lambda$)}}\\
				LASSO $(\times10^{-1})$ & 2.01 & 1.59 & 0.646 & 1.12 & 1.97 \\
				ALQR $(\times10^{-5})$  & 1.75 & 1.88 & 1.87 & 1.89 & 1.75 \\
				\hline
				\\
				\multicolumn{5}{l}{\underline{Tuning parameter selection: GIC}} \\
				\\
				& \multicolumn{5}{c}{\underline{Final Prediction Error (FPE)}}\\
				QR    & 0.1945 & 0.2972 & 0.6002 & 0.2964 & 0.1906 \\
				LASSO & \textbf{0.1100} & \textbf{0.1818} & 0.4048 & \textbf{0.1802} & \textbf{0.1091} \\
				ALQR  & 0.1105 & 0.1820 & \textbf{0.4047} & 0.1803 & 0.1092 \\
				QUANT & 2.6718 & 4.9361 & 12.2134 & 4.8949 & 2.6794 \\
				\\
				& \multicolumn{5}{c}{\underline{Out-of-Sample $R^2$}}\\
				QR    & 0.9272 & 0.9398 & 0.9509 & 0.9395 & 0.9289 \\
				LASSO & \textbf{0.9588} & \textbf{0.9632} & \textbf{0.9669} & \textbf{0.9632} & \textbf{0.9593} \\
				ALQR  & 0.9586 & 0.9631 & \textbf{0.9669} & \textbf{0.9632} & 0.9592 \\
				\\
				& \multicolumn{5}{c}{\underline{Average \# of Selected Predictors}}\\
				LASSO & 11.98 & 11.99 & 12.00 & 12.00 & 11.98 \\
				ALQR  & 11.86 & 11.92 & 11.98 & 11.93 & 11.85 \\
				\\
				& \multicolumn{5}{c}{\underline{Tuning Parameter ($\lambda$)}}\\
				LASSO $(\times10^{-1})$ & 1.42 & 1.24 & 0.64 & 0.90 & 1.45 \\
				ALQR $(\times10^{-5})$  & 1.34 & 1.39 & 1.27 & 1.36 & 1.40 \\
				\hline
			\end{tabular}%
		\end{table}

		We briefly discuss the choice of tuning parameter $\lambda$.
		For ALQR, we recommend using the Bayesian information criterion (BIC) proposed by \citet{wang2007unified} or the generalized information criterion (GIC) proposed by \citet{fan2013tuning} and \citet{zheng2015globally}.
		The objective function for choosing $\lambda$ is defined as
		\begin{equation}  \label{eqn{BIC_GIC}}
			\log\left( \frac{1}{n}\sum_{t=1}^{n}\rho _{\tau }(y_{t}-x_{t-1}^{\prime }\hat{\beta}%
			_{\tau}(\lambda))\right) +\Gamma _{n}\cdot \hat{S}(\lambda),
		\end{equation}
		where $\hat{\beta}_{\tau}(\lambda)$ is ALQR given $\lambda$, $\Gamma _{n}$ is a
		positive sequence converging to $0$, and $\hat{S}(\lambda):=\Vert\hat{\mathcal{A}}_{n} \Vert_0$ is the number of active predictors selected by $\hat{\beta}_{\tau}(\lambda)$.
		We set $\Gamma _{n}={\log(n)}/{n}$ for BIC, and $\Gamma_{n}={\log(p)} \cdot \log(\log(n))/{n}$ for GIC and select $\hat{\lambda}$ that minimizes the information criterion (objective) function.
		We use the same procedures for LASSO.
		Different approaches like the $k$-fold cross-validation and the Akaike information criterion (AIC) are available but it is known that they sometimes fail to effectively identify the true model (see, e.g.~\citet{shao1997asymptotic}; \citet{wang2007tuning}; \citet{zhang2010regularization}).
		\citet{zheng2013adaptive} also discussed that the statistical properties of the $k$-fold CV have not been well understood for high-dimensional regression with heavy-tailed errors, where QR is often applied.
		On the contrary, \citet{wang2007robust} show the model selection consistency of BIC for the fixed dimension and \citet{wang2009shrinkage} do for the increasing dimension with $p<n$.
		Also, \citet{fan2013tuning} and \citet{zheng2015globally} show that GIC can identify the underlying true model consistently with probability approaching $1$.

		Overall, the performance of ALQR is satisfactory and confirms the theory developed in the previous section.
		Tables \ref{table_simulation_S1}--\ref{table_simulation_S3} summarize the simulation results from Scenarios 1--3.
		The upper and lower panels of each table contain the results of using BIC and GIC, respectively.
		First, when there are zero coefficients in the model (Scenarios 1 and 2), ALQR performs better than its alternatives in terms of FPE (or $R^2$).
		As predicted by existing theory in the literature, both LASSO and ALQR show better performance than QR. In the meantime, ALQR performs uniformly, though slightly, better than LASSO.
		When all predictors have non-zero coefficients (Scenario 3), LASSO performs slightly better than ALQR over all quantiles except $\tau=0.5$.
		Second, the simulation results confirm the model selection consistency of ALQR derived in Theorem \ref{thm_sparsity_mix}.
		When we take a look at the average number of selected predictors in each table, the results of ALQR are quite close to the number of the true non-zero coefficients (6, 8, and 12 in each scenario).
		As known in the literature, LASSO mostly overselects them in Scenarios 1--2.
		Even in Scenario 3 where there are no zero predictors, ALQR selects most of them successfully.
		These results show the robustness of ALQR in terms of model selection.
		Finally, both BIC and GIC perform well under different designs and we do not see much difference between these two methods.

		To investigate the robustness of the simulation results above, we further conduct simulation experiments with predictors with highly correlated innovations. In section \ref{SEC:Oracle}, the oracle properties of the ALQR estimator are developed under the assumption that most predictor innovations are not highly correlated to each other. Although our empirical application satisfies this assumption, we provide additional simulation results in Tables \ref{table_more_simulation_S1_rho010}-\ref{table_more_simulation_S3_rho090} of the appendix  to demonstrate the robustness of ALQR in various settings of correlated predictor innovations. Specifically, we generate the $I$(1) predictors in Scenario 1-3 with correlated innovations. We consider correlation $\rho=$0.1, 0.5, and 0.9. We find that when the true DGP includes zero coefficients, ALQR is better than other alternative methods in most of the settings. Only when DGP has no zero coefficient ALQR is at risk of underperforming across quantiles, though the performance of the other methods is only slightly better than ALQR. An interesting finding is that ridge regression, commonly recommended when many predictors are highly correlated, has a mixed performance across quantiles. Based on the simulation results, we recommend using ALQR, especially in the case of sparse data. But ridge regression and other methods may be used if sparsity is not found in the data and the quantile of interest is around the center of the distribution.

		In sum, the numerical experiments confirm that ALQR can provide satisfactory prediction performance in finite samples.
		The results are robust over different quantiles and simulation designs.
		Therefore, we can expect similar results in other quantile prediction applications when the predictors have mixed roots and are composed of $I(0)$, $I(1)$, and cointegrated processes.


		\section{Conclusion}

		\label{section_Conclusion}

		In this paper, we show that the adaptive lasso for quantile regression (ALQR) is attractive in forecasting with stationary and nonstationary predictors as well as cointegrated predictors.
		The framework is general enough to include mixed roots but ALQR does not require any researchers' knowledge on the specific structure of each predictor nor the order of integration.
		In this general framework, we show that ALQR preserves the oracle properties.
		These advantages offer substantial convenience and robustness to empirical researchers working with quantile prediction using time series data.

		We have focused on the case where the number of covariates, $p$, is allowed to grow as sample size increases, although $p$ is smaller than $n$. 
		This framework justifies a wide range of practical applications in economics, such as the stock return quantile prediction in this paper.
		It would be an interesting future research to allow $p$ to be even larger than $n$, which has not been studied in a general time series framework with mixed roots.

\clearpage


		\begin{appendix}

			\section*{Appendix}

			\setcounter{equation}{0}
		\renewcommand{\theequation}{\thesection.\arabic{equation}}
			
         \setcounter{table}{0}
         \renewcommand{\thetable}{\thesection.\arabic{table}}
         
          \setcounter{figure}{0}
         \renewcommand\thefigure{\thesection.\arabic{figure}}

			%



			\section{Proofs for Section \protect\ref{oracle ur}}
For simplicity, we remove the intercept terms in Model \eqref{QR1} and define the dequantiled dependent variable (\citet{lee2016predictive}) as
\begin{equation*}
	y_{t\tau }:=y_{t}-\hat{\mu}_{\tau }^{QR}.
\end{equation*}%
Note that, in a mixed-root model, the intercept term is included as one of $I(0)$ predictors (see Footnote 1), so there is no need to dequantile the dependent variable in Section 4.2.

			\begin{proof}[Proof of Lemma \protect\ref{initial estimator}]
				Let $a_n = p^{\alpha}/n $ and $c \in \mathbb{R}^{p}$ such that $||c|| =
				C $, where $C$ is a finite constant. Denote the (unpenalized) quantile
				objective function as $Q_n^{QR}(\beta_{\tau})$.

				To show the result of consistency, it suffices to show that for any $%
				\epsilon>0$ , there exists a sufficiently large $C$ such that
				\begin{equation}  \label{eqn_obj_consis_QR}
					P\left\{ \inf_{||c|| = C} Q_n^{QR}(\beta_{0\tau}+a_n c) >
					Q_n^{QR}(\beta_{0\tau}) \right\} \geq 1-\epsilon.
				\end{equation}
				This inequality implies that with probability at least $1-\epsilon$, there
				is a local minimizer $\tilde{ \beta}_{\tau}$ in the shrinking ball $%
				\{\beta_{0\tau}+a_n c, ||c||\leq C\}$ such that $||\tilde{ \beta}
				_{\tau}-\beta_{0\tau}||=O_p(a_n)$. Thus, the proof is completed if we show
				that the following term is positive:
				\begin{align}
					Q_n^{QR}(\beta_{0\tau}+a_n c) - Q_n^{QR}(\beta_{0\tau}) = \sum_{t=1}^{n}\rho
					_{\tau }(u_{t\tau} - x_{t-1}^{\prime }a_nc) - \sum_{t=1}^{n}\rho _{\tau
					}(u_{t\tau})  \label{eqn_diffQ_QR}
				\end{align}

				By Knight's Identity,
				\begin{align*}
					\sum_{t=1}^{n} & \left[\rho _{\tau }(u_{t\tau} - x_{t-1}^{\prime}a_nc) -
					\rho _{\tau }(u_{t\tau})\right] \\
					& = - a_n\sum_{t=1}^{n} x_{t-1}^{\prime}c \cdot \psi_{\tau}(u_{t\tau}) +
					\sum_{t=1}^{n} \int_{0}^{x_{t-1}^{\prime }a_nc} \left( \mathbf{1}
					(u_{t\tau}\leq s) - \mathbf{1}(u_{t\tau}\leq 0) \right) ds \\
					&= - a_n\sum_{t=1}^{n} x_{t-1}^{\prime}c \cdot \psi_{\tau}(u_{t\tau}) +
					\sum_{t=1}^{n} E\left[ \int_{0}^{x_{t-1}^{\prime }a_nc} \left( \mathbf{1}
					(u_{t\tau}\leq s) - \mathbf{1}(u_{t\tau}\leq 0) \right) ds \right] \\
					& \hskip15pt +\sum_{t=1}^{n} \left\{\int_{0}^{x_{t-1}^{\prime }a_nc} \left(
					\mathbf{1} (u_{t\tau}\leq s) - \mathbf{1}(u_{t\tau}\leq 0) \right) ds - E%
					\left[ \int_{0}^{x_{t-1}^{\prime }a_nc} \left( \mathbf{1} (u_{t\tau}\leq s)
					- \mathbf{1}(u_{t\tau}\leq 0) \right) ds \right] \right\} \\
					&\equiv I_1 + I_2 + I_3.
				\end{align*}%
				We will show that $I_1$ and $I_3$ are dominated by $I_2$ and that $I_2>0$.

				First, we derive the upper bound of $I_{1}$.
				\begin{align*}
					E|I_{1}|^{2}& =a_{n}^{2}\sum_{t=1}^{n}c^{\prime }E\left[ \psi _{\tau
					}(u_{t\tau })^{2}x_{t-1}x_{t-1}^{\prime }\right] c+2\sum_{t=2}^{n}%
					\sum_{k=1}^{t-1}c^{\prime }E\left[ \psi _{\tau }(u_{t\tau })\psi _{\tau
					}(u_{k\tau })x_{t-1}x_{k-1}^{\prime }\right] c \\
					& =a_{n}^{2}\sum_{t=1}^{n}c^{\prime }E\left[ \psi _{\tau }(u_{t\tau
					})^{2}x_{t-1}x_{t-1}^{\prime }\right] c \\
					& \leq a_{n}^{2}\sum_{t=1}^{n}t\overline{c}_{B(t,p)}C^{2} \\
					& \leq C^{2}a_{n}^{2}\overline{c}_{B(n,p)}\frac{n(n+1)}{2}.
				\end{align*}%
				The second equality holds since, for $k\leq t-1$,
				\begin{equation*}
					E\left[ \psi _{\tau }(u_{t\tau })^{2}x_{t-1}x_{t-1}^{\prime }\right] =E\left[
					E_{t-1}\left[ \psi _{\tau }(u_{t\tau })\right] \psi _{\tau }(u_{k\tau
					})x_{t-1}x_{k-1}^{\prime }\right] =0.
				\end{equation*}%
				The third inequality holds by the definition of $\overline{c}_{B(t,p)}$. Therefore, the Chebyshev's
				inequality implies that
				\begin{equation*}
					I_{1}=O_{p}(a_{n}\overline{c}_{B(n,p)}^{1/2}n)=O_{p}(a_{n}^{2}\overline{c}%
					_{B(n,p)}^{1/2}n^{2}p^{-\alpha }).
				\end{equation*}

				Next, we derive the lower bound of $I_{2}$.
				\begin{align*}
					I_{2}& =\sum_{t=1}^{n}E\int_{0}^{a_{n}x_{t-1}^{\prime }c}\left(
					F_{t-1}(s)-F_{t-1}(0)\right) ds \\
					& =\sum_{t=1}^{n}E\int_{0}^{a_{n}x_{t-1}^{\prime }c}\left( f_{t-1}(0)\cdot
					s\right) ds\left\{ 1+o_{p}(1)\right\} \\
					& =\frac{1}{2}a_{n}^{2}\sum_{t=1}^{n}c^{\prime }E\left[
					f_{t-1}(0)x_{t-1}x_{t-1}^{\prime }\right] c\left\{ 1+o(1)\right\} \\
					& \geq \frac{1}{4}C^{2}a_{n}^{2}\underline{c}_{A(n,p)}\frac{n(n+1)}{2} \\
					& =O(a_{n}^{2}n^{2}\underline{c}_{A(n,p)}).
				\end{align*}%
				The first equality holds by the law of iterated expectations and the second
				does by the Taylor expansion. The inequality holds under Assumption $L1$.

				Finally, we derive the upper bound of $I_{3}$.
				\begin{align*}
					Var(I_{3})& =Var\left( \sum_{t=1}^{n}\int_{0}^{x_{t-1}^{\prime
						}a_{n}c}\left( \mathbf{1}(u_{t\tau }\leq s)-\mathbf{1}(u_{t\tau }\leq
					0)\right) ds\right) \\
					& \leq E\left[ \left( \sum_{t=1}^{n}\int_{0}^{x_{t-1}^{\prime }a_{n}c}\left(
					\mathbf{1}(u_{t\tau }\leq s)-\mathbf{1}(u_{t\tau }\leq 0)\right) ds\right)
					^{2}\right] \\
					& =E\Bigg[\sum_{t=1}^{n}\left( \int_{0}^{x_{t-1}^{\prime }a_{n}c}\left(
					\mathbf{1}(u_{t\tau }\leq s)-\mathbf{1}(u_{t\tau }\leq 0)\right) ds\right)
					^{2} \\
					& \hskip15pt+2\sum_{t=2}^{n}\sum_{k=1}^{t-1}\left( \int_{0}^{x_{t-1}^{\prime
						}a_{n}c}\left( \mathbf{1}(u_{t\tau }\leq s)-\mathbf{1}(u_{t\tau }\leq
					0)\right) ds\right) \left( \int_{0}^{x_{k-1}^{\prime }a_{n}c}\left( \mathbf{1%
					}(u_{k\tau }\leq s)-\mathbf{1}(u_{k\tau }\leq 0)\right) ds\right) \Bigg] \\
					& \leq E\Bigg[\sum_{t=1}^{n}\left( x_{t-1}^{\prime }a_{n}c\right)
					^{2}+2\sum_{t=2}^{n}\sum_{k=1}^{t-1}\left\vert x_{t-1}^{\prime
					}a_{n}c\right\vert \left\vert x_{k-1}^{\prime }a_{n}c\right\vert \Bigg] \\
					& =a_{n}^{2}\sum_{t=1}^{n}c^{\prime }E\left[ x_{t-1}x_{t-1}^{\prime }\right]
					c+2a_{n}^{2}\sum_{t=2}^{n}\sum_{k=1}^{t-1}E\left[ \left\vert x_{t-1}^{\prime
					}c\right\vert \left\vert x_{k-1}^{\prime }c\right\vert \right] \\
					& \equiv V_{3,1}+V_{3,2}
				\end{align*}%
				Using the similar arguments in $I_{1}$, we have
				\begin{equation*}
					V_{3,1}\leq a_{n}^{2}C^{2}\frac{\overline{c}_{A(n,p)}}{c_{f}}\frac{n(n+1)}{2}%
					=O(a_{n}^{2}\overline{c}_{A(n,p)}n^{2}).
				\end{equation*}%
				By the Cauchy-Schwarz inequality, the definition of $\overline{c}_{A(n,p)}$, and $t>k$, we have
				\begin{align*}
					V_{3,2}& \leq 2a_{n}^{2}\sum_{t=2}^{n}\sum_{k=1}^{t-1}\sqrt{E\left[
						(x_{t-1}^{\prime }x_{t-1})\right] }\sqrt{E\left[ (x_{t-1}^{\prime }x_{t-1})%
						\right] } \\
					& \leq 2a_{n}^{2}\sum_{t=2}^{n}\sum_{k=1}^{t-1}\sqrt{C^{2}t\frac{\overline{c}%
							_{A(t,p)}}{c_{f}}}\sqrt{C^{2}t\frac{\overline{c}_{A(t,p)}}{c_{f}}} \\
					& \leq 2a_{n}^{2}C^{2}\frac{\overline{c}_{A(n,p)}}{c_{f}}\sum_{t=2}^{n}%
					\sum_{k=1}^{t-1}t \\
					& =O(a_{n}^{2}\overline{c}_{A(n,p)}n^{3})
				\end{align*}%
				Therefore, $Var(I_{3})=O(a_{n}^{2}\overline{c}_{A}n^{3})$, and Chebyshev's
				inequality implies that
				\begin{equation*}
					I_{3}=O_{p}(\overline{c}_{A(n,p)}^{1/2}a_{n}n^{3/2})=O_{p}\left( \overline{c}%
					_{A(n,p)}^{1/2}a_{n}^{2}n^{5/2}p^{-\alpha }\right) .
				\end{equation*}

				By Assumptions $U1$ and $U2$, we establish the desired result.
			\end{proof}


			\begin{proof}[Proof of Theorem \protect\ref{thm_consistency}]
				For simplicity, in this proof, we use $\hat{ \beta}_{\tau}$ to represent the
				ALQR estimator $\hat{ \beta}_{\tau}^{ALQR}$. Without loss of generality, let
				the values of $\beta_{0\tau,1}, \beta_{0\tau,2},..., \beta_{0\tau,q_n}$ be
				nonzero and $\beta_{0\tau,q_n+1},$ $\beta_{0\tau,q_n+2},...,
				\beta_{0\tau,p}$ be zero. Let $E_{t-1}(\cdot)\equiv E(\cdot | \mathcal{F}%
				_{t-1})$.

				To show the result of consistency, it suffices to show that for any $%
				\epsilon>0$ , there exists a sufficiently large $C$ such that
				\begin{equation}  \label{eqn_obj_consis}
					P\left\{ \inf_{||c|| = C} Q_n(\beta_{0\tau}+a_n c) > Q_n(\beta_{0\tau})
					\right\} \geq 1-\epsilon.
				\end{equation}
				This inequality implies that with probability at least $1-\epsilon$, there
				is a local minimizer $\hat{ \beta}_{\tau}$ in the shrinking ball $%
				\{\beta_{0\tau}+a_n c, ||c||\leq C\}$ such that $||\hat{ \beta}
				_{\tau}-\beta_{0\tau}||=O_p(a_n)$.

				Since 
				\begingroup
				\allowdisplaybreaks
				\begin{align}
					& Q_n(\beta_{0\tau}+a_n c) - Q_n(\beta_{0\tau})  \label{eqn_diffQ} \\
					&= \left(\sum_{t=1}^{n}\rho _{\tau }(u_{t\tau} - x_{t-1}^{\prime }a_nc) -
					\sum_{t=1}^{n}\rho _{\tau }(u_{t\tau})\right) +
					\left(\sum_{j=1}^{p}\lambda _{n,j}|\beta _{0\tau,j}+a_n c_j| -
					\sum_{j=1}^{p}\lambda _{n,j}|\beta _{0\tau,j}|\right)  \notag \\
					&= \left(\sum_{t=1}^{n}\rho _{\tau }(u_{t\tau} - x_{t-1}^{\prime }a_nc) -
					\sum_{t=1}^{n}\rho _{\tau }(u_{t\tau})\right)  \notag \\
					&+\left(\sum_{j=1}^{q_n}\lambda _{n,j}(|\beta _{0\tau,j}+a_n c_j| - |\beta
					_{0\tau,j}|) + \sum_{j=q_n+1}^{p}\lambda _{n,j}(|\beta _{0\tau,j}+a_n c_j|
					- |\beta _{0\tau,j}|)\right)  \notag \\
					&= \left(\sum_{t=1}^{n}\rho _{\tau }(u_{t\tau} - x_{t-1}^{\prime }a_nc) -
					\sum_{t=1}^{n}\rho _{\tau }(u_{t\tau})\right)  \notag \\
					&+\left(\sum_{j=1}^{q_n}\lambda _{n,j}(|\beta _{0\tau,j}+a_n c_j| - |\beta
					_{0\tau,j}|) + \sum_{j=q_n+1}^{p}\lambda _{n,j}(|a_n c_j|)\right)  \notag
					\\
					&\geq \left(\sum_{t=1}^{n}\rho _{\tau }(u_{t\tau} - x_{t-1}^{\prime }a_nc) -
					\sum_{t=1}^{n}\rho _{\tau }(u_{t\tau})\right) +\sum_{j=1}^{q_n}\lambda
					_{n,j}(|\beta _{0\tau,j}+a_n c_j| - |\beta _{0\tau,j}|)  \notag \\
					&\equiv D_1 + D_2,  \notag
				\end{align}
				\endgroup
				we need to show that $D_1+D_2$ is positive.

				For $D_{2}$, we know $\left\vert |\beta _{0\tau ,j}+a_{n}c_{j}|-|\beta
				_{0\tau ,j}|\right\vert \leq \left\vert a_{n}c_{j}\right\vert .$ Given that $%
				\tilde{\beta}_{\tau }$ is a $(a_{n}^{-1})$-consistent estimate of $\beta
				_{0\tau }$, we have $a_{n}^{-1}(\tilde{\beta}_{\tau ,j}-\beta _{0\tau
					,j})=O_{p}(1)$, and then $\tilde{\beta}_{\tau ,j}=O_{p}(a_{n})+\beta _{0\tau
					,j}=o_{p}(1)+\beta _{0\tau ,j}$. Thus, \begingroup\allowdisplaybreaks%
				\begin{align}
					|D_{2}|\leq & \sum_{j=1}^{q_{n}}\lambda _{n,j}|a_{n}c_{j}|  \notag \\
					=& \lambda _{n}a_{n}\sum_{j=1}^{q_{n}}\left\vert \frac{1}{|{\tilde{\beta}%
							_{\tau ,j}}|^{\gamma }}c_{j}\right\vert  \notag \\
					\leq &\lambda _{n}a_{n}\left( \sum_{j=1}^{q_{n}}\frac{1}{(\tilde{\beta}%
						_{\tau ,j})^{2\gamma }}\right) ^{1/2}\cdot ||c|| \notag \\
					\leq & C \lambda _{n}a_{n}\left( \sum_{j=1}^{q_{n}}\frac{1}{(o_{p}(1)+\beta
						_{0\tau ,j})^{2\gamma }}\right) ^{1/2}  \notag \\
					=& C\lambda _{n}a_{n}\left( O_{p}(1)\sum_{j=1}^{q_{n}}1\right) ^{1/2}
					\notag \\
					= & O_{p}(\lambda _{n}a_{n}q_{n}^{1/2})  \notag \\
					= & O_{p}(\lambda _{n}a_{n}p^{1/2}).  \label{eqn_D2}
				\end{align}%
				\endgroup

				We next consider $D_{1}$. Following the proof of Lemma \ref{initial
					estimator}, we have that the dominating term of $D_{1}$ is $O_{p}(a_{n}^{2}n^{2}\underline{c}_{A(n,p)})$ and that it is positive with probability approaching $1$. Under
				Assumption $\lambda 1$ (i), $D_{1}$ dominates $D_{2}$. We complete the proof of
				Theorem \ref{thm_consistency}.
			\end{proof}


			\begin{proof}[Proof of Theorem \protect\ref{thm_sparsity}]
				From Theorem \ref{thm_consistency} for a sufficiently large constant $C$, $%
				\hat{ \beta}_{\tau}^{ALQR}$ is a local minimizer lies in the ball $%
				\{\beta_{0\tau} + a_n c, ||c||\leq C\}$ with probability approaching $1$ and $%
				a_n=p^{\alpha}/n$. For simplicity, in this proof, we use $\hat{ \beta}%
				_{\tau}$ to represent the ALQR estimator $\hat{ \beta}_{\tau}^{ALQR}$.

				First, note that the subgradient of the unpenalized objective function, $%
				s(\beta_{\tau})=(s_1(\beta_{\tau}),...,s_{p}(\beta_{\tau}))^{\prime }$ is
				given by (\citet{sherwood2016partially}, page 298 and Lemma 1):
				\begin{eqnarray}
					s_j(\beta_{\tau}) = -\sum_{t=1}^{n} x_{t-1,j} \psi_{\tau} \left(y_{t\tau} -
					x_{t-1}^{\prime }{\beta}_{\tau}\right) + \sum_{t=1}^{n} x_{t-1,j} k_t \quad
					\text{for $1\leq j \leq p$},  \notag
				\end{eqnarray}
				where $k_t=0$ if $y_{t\tau} - x_{t-1}^{\prime }{\beta}_{\tau} \neq 0$ and $%
				k_t \in [-\tau, 1-\tau]$ if $y_{t\tau} - x_{t-1}^{\prime }{\beta}_{\tau}=0$.

				Let $\mathcal{D} = \{ t: y_{t\tau} - x_{t-1}^{\prime }\hat{\beta}_{\tau}=0
				\} $, then 
				\begin{eqnarray}
					s_j(\hat\beta_{\tau}) &=& -\sum_{t=1}^{n} x_{t-1,j} \psi_{\tau}
					\left(y_{t\tau} - x_{t-1}^{\prime }\hat{\beta}_{\tau}\right) + \sum_{t\in
						\mathcal{D}} x_{t-1,j} (k_t^*+(1-\tau)),  \notag \\
					&=& -\sum_{t=1}^{n} x_{t-1,j} \psi_{\tau} \left(y_{t\tau} - x_{t-1}^{\prime }%
					\hat{\beta}_{\tau}\right) + h_n,  \notag
				\end{eqnarray}
				where $h_n :=\sum_{t\in \mathcal{D}} x_{t-1,j} (k_t^*+(1-\tau)) $ and
				\begin{equation*}
					k_t^* =
					\begin{cases}
						0, & \text{ if } y_{t\tau} - x_{t-1}^{\prime }\hat{\beta}_{\tau} \neq 0 \\
						\in [-\tau, 1-\tau], & \text{ if } y_{t\tau} - x_{t-1}^{\prime }\hat{\beta}%
						_{\tau}=0.%
					\end{cases}%
				\end{equation*}
				With probability one (\citet[Section 2.2]{koenker_2005}), $|\mathcal{D}| =
				p$. 
				Thus $h_n = O_p(p^{3/2})=O_p(n^{(3/2)\zeta}). $

				Next, define the subgradient of the penalized objective function as $%
				S_j(\beta_{\tau})$:
				\begin{eqnarray}
					S_j(\beta_{\tau}) &=& -\sum_{t=1}^{n} x_{t-1,j} \psi_{\tau} \left(y_{t\tau}
					- x_{t-1}^{\prime }{\beta}_{\tau}\right) + h_n + \frac{\lambda_n}{|\tilde{%
							\beta}_{j,\tau}|^{\gamma}} sgn(\beta_{j,\tau})  \notag \\
					&=& -\sum_{t=1}^{n} x_{t-1,j} \psi_{\tau} \left(u_{t\tau} - x_{t-1}^{\prime
					}\delta_{\tau} \right) + h_n + \frac{\lambda_n}{|\tilde{\beta}%
						_{j,\tau}|^{\gamma}} sgn(\beta_{j,\tau}),  \notag
				\end{eqnarray}
				where $u_{t\tau}=u_{0t} - F_{u_0}^{-1}(\tau)$ and $\delta_{\tau} =
				\beta_{\tau}-\beta_{0\tau}$. The subgradient condition requires that at the
				optimum, $\hat{ \beta}_{\tau}$,
				\begin{eqnarray}
					0 \in S_j(\hat\beta_{\tau}).  \notag
				\end{eqnarray}
				That is,
				\begin{eqnarray}
					\sum_{t=1}^{n} x_{t-1,j} \psi_{\tau} \left(u_{t\tau} - x_{t-1}^{\prime
					}\hat\delta_{\tau} \right) - h_n= \frac{\lambda_n}{|\tilde{\beta}%
						_{j,\tau}|^{\gamma}} sgn(\hat\beta_{j,\tau}).  \label{eqn_subgradient}
				\end{eqnarray}

				If $j=1,..,q_n$, it implies that $|sgn(\hat\beta_{j,\tau})|=1$. Then we can
				write the subgradient condition (\ref{eqn_subgradient}) as:
				\begin{eqnarray}  \label{eqn_KKT_pn}
					\left|\sum_{t=1}^{n} x_{t-1,j} \psi_{\tau} \left(u_{t\tau} -
					\hat\delta_{\tau}^{\prime }x_{t-1}\right) - h_n \right| &=& \frac{\lambda_n}{%
						|\tilde{ \beta}_{j,\tau}|^{\gamma}}.  \label{eqn_foc}
				\end{eqnarray}
				In the following, we show that this subgradient condition does not hold for $%
				j \notin \mathcal{A}_0$, i.e., $j=q_n+1,...,p$. It suffices to show:

				\begin{enumerate}
					\item[(a)] $(n^{\alpha \zeta +2}\overline{c}_{A(n,p)}^{1/2})^{-1}\frac{%
						\lambda _{n}}{|\tilde{\beta}_{j,\tau }|^{\gamma }}\rightarrow \infty $;

					\item[(b)] $\sum_{t=1}^{n}x_{t-1,j}\psi _{\tau }\left( u_{t\tau
					}-x_{t-1}^{\prime }\delta _{\tau }\right) -h_{n}\leq O_{p}(n^{\alpha \zeta
						+2}\overline{c}_{A(n,p)}^{1/2})$.
				\end{enumerate}

				For (b), we first prove that the first term on the left-hand side of (\ref%
				{eqn_KKT_pn}) is dominated by $O_{p}(n^{\alpha \zeta +2}\overline{c}%
				_{A(n,p)}^{1/2})$:
				\begin{eqnarray}
					&&\sum_{t=1}^{n}x_{t-1,j}\psi _{\tau }\left( u_{t\tau }-\delta _{\tau
					}^{\prime }x_{t-1}\right)  \notag \\
					&=&\sum_{t=1}^{n}x_{t-1,j}\Big[\psi _{\tau }\left( u_{t\tau }-\delta _{\tau
					}^{\prime }x_{t-1}\right) -E_{t-1}\psi _{\tau }\left( u_{t\tau }-\delta
					_{\tau }^{\prime }x_{t-1}\right) -\psi _{\tau }\left( u_{t\tau }\right)
					+E_{t-1}\psi _{\tau }\left( u_{t\tau }\right) \Big]  \notag \\
					&&+\sum_{t=1}^{n}x_{t-1,j}E_{t-1}\psi _{\tau }\left( u_{t\tau }-\delta
					_{\tau }^{\prime }x_{t-1}\right) +\sum_{t=1}^{n}x_{t-1,j}\psi _{\tau }\left(
					u_{t\tau }\right)  \notag \\
					&\equiv &I_{1}+I_{2}+I_{3}.  \label{eqn_taylor_pn}
				\end{eqnarray}

				For $I_{1}$, let $I_1=A+B$, where
				\begin{eqnarray}
					A &\equiv &\sum_{t=1}^{n}x_{t-1,j}\Big[\psi _{\tau }\left( u_{t\tau }-\delta
					_{\tau }^{\prime }x_{t-1}\right) -E_{t-1}\psi _{\tau }\left( u_{t\tau
					}-\delta _{\tau }^{\prime }x_{t-1}\right) \Big],  \notag \\
					B &\equiv &\sum_{t=1}^{n}x_{t-1,j}\Big[\psi _{\tau }\left( u_{t\tau }\right)
					-E_{t-1}\psi _{\tau }\left( u_{t\tau }\right) \Big].  \notag
				\end{eqnarray}

				Note that
				\begin{eqnarray}
					{E_{t-1}\psi _{\tau }\left( u_{t\tau }-\delta _{\tau }^{\prime
						}x_{t-1}\right) } &=& E_{t-1}\psi _{\tau }\left( u_{t\tau }\right) +\left.
					\frac{\partial E_{t-1}\psi _{\tau }\left( u_{t\tau }-\delta _{\tau }^{\prime
						}x_{t-1}\right) }{\partial \delta _{\tau }^{\prime }}\right\vert _{\delta
						_{\tau }=\mathbf{0}}\delta _{\tau }+o_{p}(a _{n})  \notag \\
					&=& -x_{t-1}^{\prime }f_{t-1}(0)\delta _{\tau }+o_{p}(a _{n}),
					\label{eqn_psi_mid}
				\end{eqnarray}
				\begin{eqnarray}
					{E_{t-1}\psi _{\tau }^{2}\left( u_{t\tau }- \delta _{\tau }^{\prime
						}x_{t-1}\right) } &=& E_{t-1}\psi_{\tau }^2 \left( u_{t\tau }\right) +\left.
					\frac{\partial E_{t-1}\psi^2_{\tau }\left( u_{t\tau }-\delta _{\tau
						}^{\prime }x_{t-1}\right) }{\partial \delta _{\tau }^{\prime }}\right\vert
					_{\delta _{\tau }=\mathbf{0}}\delta _{\tau }+o_{p}(a _{n})  \notag \\
					&=& \tau(1-\tau) -x_{t-1}^{\prime }f_{t-1}^2(0)\delta _{\tau }+o_{p}(a_n).
					\label{eqn_psi2}
				\end{eqnarray}

				Then, we have
				\begin{eqnarray}
					E(A) = EE_{t-1}(A) &=&E\left\{\sum_{t=1}^{n}x_{t-1,j}\Big[E_{t-1}\psi _{\tau
					}\left( u_{t\tau }-\delta _{\tau }^{\prime }x_{t-1}\right) -E_{t-1}\psi
					_{\tau }\left( u_{t\tau }-\delta _{\tau }^{\prime }x_{t-1}\right) \Big]%
					\right\}=0,  \notag
				\end{eqnarray}
				and \begingroup
				\allowdisplaybreaks
				\begin{align}
					E_{t-1}(A^{2}) =&\sum_{t=1}^{n}x_{t-1,j}^{2}E_{t-1}\Big[\psi _{\tau }\left(
					u_{t\tau }-\delta _{\tau }^{\prime }x_{t-1}\right) -E_{t-1}\psi _{\tau
					}\left( u_{t\tau }-\delta _{\tau }^{\prime }x_{t-1}\right) \Big]^{2}+0
					\notag \\
					=&\sum_{t=1}^{n}x_{t-1,j}^{2}\Big[E_{t-1}\psi _{\tau }^{2}\left( u_{t\tau
					}-\delta _{\tau }^{\prime }x_{t-1}\right) -\left( E_{t-1}\psi _{\tau }\left(
					u_{t\tau }-\delta _{\tau }^{\prime }x_{t-1}\right) \right) ^{2}\Big]  \notag
					\\
					=& \sum_{t=1}^{n}x_{t-1,j}^{2} \Big( \tau(1-\tau) - x_{t-1}^{\prime
					}f_{t-1}^2(0)\delta _{\tau } + o_{p}(a_n) \Big)  \notag \\
					& - \sum_{t=1}^{n}x_{t-1,j}^{2} \Big( -x_{t-1}^{\prime }f_{t-1}(0)\delta
					_{\tau }+o_{p}(\gamma _{n}) \Big)^2 \quad \text{(by (\ref{eqn_psi_mid}) and (%
						\ref{eqn_psi2}))}  \notag \\
					=& O_p(n^2) - \sum_{t=1}^{n}x_{t-1,j}^{2} x_{t-1}^{\prime }cf_{t-1}^2(0)
					O_p(a_n) + O_p(n^2) o_{p}(a_n)  \notag \\
					& - \sum_{t=1}^{n}x_{t-1,j}^{2} \Big( c^{\prime}x_{t-1} x_{t-1}^{\prime }c
					f_{t-1}^2(0)O_p(a_n^2) - 2 x_{t-1}^{\prime }cf_{t-1}(0) O_p(a_n)o_{p}(a_n) +
					o_p(a_n^2) \Big).  \notag \\
					& \text{(since $\delta_{\tau} = \beta_{\tau}-\beta_{0\tau} = O_p(a_n)c$ and $%
						x_{t-1}^{\prime}\delta _{\tau} = x_{t-1}^{\prime}cO_p(a_n)$)}
					\label{eqn_EAsq}
				\end{align}
				\endgroup

				Under Assumption $f$ and the definition of $\overline{c}_{A(n,p)}$, we have:
				\begin{equation*}
					E\left[ \sum_{t=1}^{n}c^{\prime }f_{t-1}(0)x_{t-1}x_{t-1}^{\prime }c\right]
					^{1/2}\leq \left[ \sum_{t=1}^{n}Ec^{\prime }f_{t-1}(0)x_{t-1}x_{t-1}^{\prime
					}c\right] ^{1/2}\leq O(\overline{c}_{A(n,p)}^{1/2}n),\quad \text{(by
						Jensen's inequality)}
				\end{equation*}%
				\begin{eqnarray}
					E(\max_{1\leq t\leq n}|x_{t-1}^{\prime }cf_{t-1}(0)|)^{2} &\leq
					&E(\sum_{t=1}^{n}|x_{t-1}^{\prime }cf_{t-1}(0)|)^{2}  \notag \\
					&=&E\left[ \sum_{t=1}^{n}(x_{t-1}^{\prime
					}cf_{t-1}(0))^{2}+2\sum_{t=2}^{n}\sum_{k=1}^{t-1}|x_{t-1}^{\prime
					}cf_{t-1}(0)|\cdot |x_{k-1}^{\prime }cf_{k-1}(0)|\right]  \notag \\
					&\leq &c_{f}\sum_{t=1}^{n}E(c^{\prime }x_{t-1}x_{t-1}^{\prime
					}cf_{t-1}(0))+2E\left[ \sum_{t=2}^{n}\sum_{k=1}^{t-1}|x_{t-1}^{\prime
					}cf_{t-1}(0)|\cdot |x_{k-1}^{\prime }cf_{k-1}(0)|\right]  \notag \\
					&\leq &O(c_{f}\overline{c}_{A(n,p)}n^{2})+2\sum_{t=2}^{n}\sum_{k=1}^{t-1}%
					\sqrt{E\left[ (x_{t-1}^{\prime }cf_{t-1}(0))^{2}\right] }\cdot \sqrt{E\left[
						(x_{k-1}^{\prime }cf_{k-1}(0))^{2}\right] }  \notag \\
					&\leq &O(c_{f}\overline{c}_{A(n,p)}n^{2})+O(c_{f}\overline{c}%
					_{A(n,p)}n^{3})=O(c_{f}\overline{c}_{A(n,p)}n^{3}),\text{(by
						Cauchy-Schwarz inequality)}  \notag
				\end{eqnarray}%
				and
				\begin{equation*}
					E(\max_{1\leq t\leq n}|x_{t-1}^{\prime }cf_{t-1}(0)|)\leq \left[
					E(\max_{1\leq t\leq n}|x_{t-1}^{\prime }cf_{t-1}(0)|)^{2}\right] ^{1/2}\leq %
					\left[ O(c_{f}\overline{c}_{A(n,p)}n^{3})\right] ^{1/2}=O(c_{f}^{1/2}%
					\overline{c}_{A(n,p)}^{1/2}n^{3/2}).
				\end{equation*}%
				Using the above results and the Cauchy-Schwarz inequality, we can show that: %
				\begingroup\allowdisplaybreaks%
				\begin{align}
					EE_{t-1}(A^{2})\leq & \quad
					O(n^{2})-O(a_{n})E[\sum_{t=1}^{n}x_{t-1,j}^{2}x_{t-1}^{\prime
					}cf_{t-1}^{2}(0)]+o(n^{2}a_{n})-O(a_{n}^{2})E[\sum_{t=1}^{n}x_{t-1,j}^{2}c^{%
						\prime }x_{t-1}x_{t-1}^{\prime }cf_{t-1}^{2}(0)]  \notag \\
					& +o(a_{n}^{2})E[\sum_{t=1}^{n}x_{t-1,j}^{2}x_{t-1}^{\prime
					}cf_{t-1}(0)]-o(a_{n}^{2})O(n^{2})  \notag \\
					\leq & \quad O(n^{2})+O(a_{n})E\left( \max_{1\leq t\leq n}|x_{t-1}^{\prime
					}cf_{t-1}^{2}(0)|\sum_{t=1}^{n}x_{t-1,j}^{2}\right) +o(n^{2}a_{n})  \notag \\
					& +O(a_{n}^{2})E\left( (\max_{1\leq t\leq n}|x_{t-1}^{\prime
					}cf_{t-1}(0)|)^{2}\sum_{t=1}^{n}x_{t-1,j}^{2}\right) +o(a_{n}^{2})E\left(
					\max_{1\leq t\leq n}|x_{t-1}^{\prime
					}cf_{t-1}(0)|\sum_{t=1}^{n}x_{t-1,j}^{2}\right)  \notag \\
					& +o(n^{2}a_{n}^{2})  \notag \\
					\leq & \quad O(n^{2})+O(a_{n})O(n^{2})O(c_{f}c_{f}^{1/2}\overline{c}%
					_{A(n,p)}^{1/2}n^{3/2})+o(n^{2}a_{n})+O(a_{n}^{2})O(n^{2})O(c_{f}\overline{c}%
					_{A(n,p)}n^{3})  \notag \\
					& +o(a_{n}^{2})O(n^{2})O(c_{f}^{1/2}\overline{c}%
					_{A(n,p)}^{1/2}n^{3/2})+o(n^{2}a_{n}^{2})  \notag \\
					=& \quad O(n^{2})+O(a_{n}n^{7/2}\overline{c}%
					_{A(n,p)}^{1/2})+o(a_{n}n^{2})+O(a_{n}^{2}n^{5}\overline{c}%
					_{A(n,p)})+o(a_{n}^{2}n^{7/2}\overline{c}_{A(n,p)}^{1/2})+o(a_{n}^{2}n^{2})
					\notag \\
					=& \quad O(n^{2})+O(n^{\alpha \zeta +(5/2)}\overline{c}%
					_{A(n,p)}^{1/2})+o(n^{\alpha \zeta +1}) \notag\\
					& +O(n^{2\alpha \zeta +3}\overline{c}%
					_{A(n,p)})+o(n^{2\alpha \zeta +(3/2)}\overline{c}_{A(n,p)}^{1/2})+o(n^{2%
						\alpha \zeta }).
				\end{align}%
				\endgroup Given the conditions: $0<\alpha \zeta <1$, it is easy to verify
				that the fourth term, $O(n^{2\alpha \zeta +3}\overline{c}_{A(n,p)})$, dominates the other terms. Thus, $E(A^{2})\leq O(n^{2\alpha \zeta
					+3}\overline{c}_{A(n,p)})$. The Chebyshev's inequality implies that
				\begin{equation}
					A\leq O_{p}(n^{\alpha \zeta +(3/2)}\overline{c}_{A(n,p)}^{1/2}).
					\label{eqn_A}
				\end{equation}

				For B, it is easy to obtain that
				\begin{equation}
					E_{t-1}[\psi _{\tau }\left( u_{t\tau }\right)] =0,\quad E_{t-1}[\psi _{\tau
					}^{2}\left( u_{t\tau }\right)] =\tau (1-\tau ).
				\end{equation}
				Then we can show:\begingroup
				\allowdisplaybreaks
				\begin{align}
					E_{t-1}(B) &=\sum_{t=1}^{n}x_{t-1,j}\Big[E_{t-1}\psi _{\tau }\left( u_{t\tau
					}\right) -E_{t-1}\psi _{\tau }\left( u_{t\tau }\right) \Big]=0,  \notag \\
					E_{t-1}(B^{2}) &=\sum_{t=1}^{n}x_{t-1,j}^{2}E_{t-1}\Big[\psi _{\tau }\left(
					u_{t\tau }\right) -E_{t-1}\psi _{\tau }\left( u_{t\tau }\right) \Big] %
					^{2}=\sum_{t=1}^{n}x_{t-1,j}^{2}E_{t-1}\psi _{\tau }^{2}\left( u_{t\tau
					}\right) = O_{p}(n^{2}),  \notag
				\end{align}
				\endgroup
				which implies that
				\begin{equation}
					B \leq O_{p}(n).  \label{eqn_B}
				\end{equation}

				By the results of (\ref{eqn_A}) and (\ref{eqn_B}), we have
				\begin{equation}\label{eqn_pf_I1_ur}
					I_{1}\leq O_{p}(n^{\alpha \zeta +(3/2)}\overline{c}_{A(n,p)}^{1/2}).
				\end{equation}

				For $I_{2}$, by (\ref{eqn_psi_mid}),
				\begin{equation*}
					\sum_{t=1}^{n}x_{t-1,j}E_{t-1}\psi _{\tau }\left( u_{t\tau }-\delta _{\tau
					}^{\prime }x_{t-1}\right) =-\sum_{t=1}^{n}x_{t-1,j}x_{t-1}^{\prime
					}f_{t-1}(0)\delta _{\tau }+o_{p}(a_{n})\sum_{t=1}^{n}x_{t-1,j}.
				\end{equation*}%
				Thus,
				\begin{eqnarray}
					&&E\left[ \sum_{t=1}^{n}x_{t-1,j}E_{t-1}\psi _{\tau }\left( u_{t\tau
					}-\delta _{\tau }^{\prime }x_{t-1}\right) \right]  \notag \\
					&=&-E\sum_{t=1}^{n}[x_{t-1,j}x_{t-1}^{\prime
					}cf_{t-1}(0)O_{p}(a_{n})]+E[o_{p}(a_{n}n^{3/2})]  \notag \\
					&\leq &O(a_{n})E\left[ \sum_{t=1}^{n}x_{t-1,j}x_{t-1}^{\prime }cf_{t-1}(0)%
					\right] +o(a_{n}n^{3/2})  \notag \\
					&\leq &O(a_{n})E\left( \max_{1\leq t\leq n}|x_{t-1}^{\prime
					}cf_{t-1}(0)|\sum_{t=1}^{n}x_{t-1,j}\right) +o(a_{n}n^{3/2})  \notag \\
					&\leq &O(a_{n})O(n^{3/2})O(c_{f}^{1/2}\overline{c}%
					_{A(n,p)}^{1/2}n^{3/2})+o(a_{n}n^{3/2})  \notag \\
					&=&O(a_{n}n^{3}\overline{c}_{A(n,p)}^{1/2})+o(a_{n}n^{3/2})  \notag \\
					&=&O(n^{\alpha \zeta +2}\overline{c}_{A(n,p)}^{1/2}),  \label{eqn_I2_a}
				\end{eqnarray}%
				and \begingroup\allowdisplaybreaks%
				\begin{align}
					& E\left[ \left( \sum_{t=1}^{n}x_{t-1,j}E_{t-1}\psi _{\tau }\left( u_{t\tau
					}-\delta _{\tau }^{\prime }x_{t-1}\right) \right) ^{2}\right]  \notag \\
					=& \quad E\left[ \left( -\sum_{t=1}^{n}x_{t-1,j}x_{t-1}^{\prime
					}f_{t-1}(0)\delta _{\tau }+o_{p}(a_{n})\sum_{t=1}^{n}x_{t-1,j}\right) ^{2}%
					\right]  \notag \\
					=& \quad E\bigg[\left( \sum_{t=1}^{n}x_{t-1,j}x_{t-1}^{\prime
					}f_{t-1}(0)\delta _{\tau }\right) ^{2}+\left(
					o_{p}(a_{n})\sum_{t=1}^{n}x_{t-1,j}\right) ^{2}  \notag \\
					& \quad -2\left( \sum_{t=1}^{n}x_{t-1,j}x_{t-1}^{\prime }f_{t-1}(0)\delta
					_{\tau }\right) \cdot \left( o_{p}(a_{n})\sum_{t=1}^{n}x_{t-1,j}\right) %
					\bigg]  \notag \\
					\leq & \quad O(a_{n}^{2})E\left[ \sum_{t=1}^{n}\left(
					x_{t-1,j}x_{t-1}^{\prime }cf_{t-1}(0)\right)
					^{2}+2\sum_{t=2}^{n}\sum_{k=1}^{t-1}x_{t-1,j}x_{k-1,j}x_{t-1}^{\prime
					}cf_{t-1}(0)\cdot x_{k-1}^{\prime }cf_{k-1}(0)\right]  \notag \\
					& \quad +o(a_{n}^{2})E\left[ \sum_{t=1}^{n}x_{t-1,j}^{2}+2\sum_{t=2}^{n}%
					\sum_{k=1}^{t-1}x_{t-1,j}x_{k-1,j}\right]  \notag \\
					& \quad +o(a_{n}^{2})E\left[ \left( \sum_{t=1}^{n}x_{t-1,j}x_{t-1}^{\prime
					}cf_{t-1}(0)\right) \cdot \left( \sum_{t=1}^{n}x_{t-1,j}\right) \right]
					\notag \\
					\leq & \quad \quad O(a_{n}^{2})E\left[ \sum_{t=1}^{n}x_{t-1,j}^{2}c^{\prime
					}x_{t-1}x_{t-1}^{\prime
					}cf_{t-1}^{2}(0)+2\sum_{t=2}^{n}\sum_{k=1}^{t-1}x_{t-1,j}x_{k-1,j}x_{t-1}^{%
						\prime }cf_{t-1}(0)\cdot x_{k-1}^{\prime }cf_{k-1}(0)\right]  \notag \\
					& \quad +o(a_{n}^{2})\left[ O(n^{2})+O(n^{3})\right] +o(a_{n}^{2})E\left[
					\left( \max_{1\leq t\leq n}|x_{t-1}^{\prime
					}cf_{t-1}(0)|\sum_{t=1}^{n}x_{t-1,j}\right) \cdot O_{p}(n^{3/2})\right]
					\notag \\
					\leq & \quad O(a_{n}^{2})E\left( (\max_{1\leq t\leq n}|x_{t-1}^{\prime
					}cf_{t-1}(0)|)^{2}\sum_{t=1}^{n}x_{t-1,j}^{2}\right) +O(a_{n}^{2})E\left[
					(\max_{1\leq t\leq n}|x_{t-1}^{\prime
					}cf_{t-1}(0)|)^{2}\sum_{t=2}^{n}\sum_{k=1}^{t-1}x_{t-1,j}x_{k-1,j}\right]
					\notag \\
					& \quad +o(a_{n}^{2}n^{3})+o(a_{n}^{2})E\left[ \left( \max_{1\leq t\leq
						n}|x_{t-1}^{\prime }cf_{t-1}(0)|\sum_{t=1}^{n}x_{t-1,j}\right) \cdot
					O_{p}(n^{3/2})\right]  \notag \\
					\leq & \quad O(a_{n}^{2})O(n^{2})O(c_{f}\overline{c}%
					_{A(n,p)}n^{3})+O(a_{n}^{2})O(n^{3})O(c_{f}\overline{c}%
					_{A(n,p)}n^{3})+o(a_{n}^{2}n^{3})+o(a_{n}^{2})O(n^{3/2})O(n^{3/2})O(c_{f}^{1/2}%
					\overline{c}_{A(n,p)}^{1/2}n^{3/2})  \notag \\
					=& \quad O(a_{n}^{2}n^{5}\overline{c}_{A(n,p)})+O(a_{n}^{2}n^{6}\overline{c}%
					_{A(n,p)})+o(a_{n}^{2}n^{3})+o(a_{n}^{2}n^{9/2}\overline{c}_{A(n,p)}^{1/2})
					\notag \\
					=& \quad O(a_{n}^{2}n^{6}\overline{c}_{A(n,p)})=O(n^{2\alpha \zeta +4}%
					\overline{c}_{A(n,p)}),  \label{eqn_I2_b}
				\end{align}%
				\endgroup where the second inequality holds using Cauchy-Schwarz inequality
				and the following result:
				\begin{eqnarray}
					\sum_{t=2}^{n}\sum_{k=1}^{t-1}x_{t-1,j}x_{k-1,j} &\leq &\sum_{t=2}^{n}\left[
					x_{t-1,j}\left( \sum_{k=1}^{t-1}x_{k-1,j}\right) \right]  \notag \\
					&\leq &\left( \sum_{t=2}^{n}x_{t-1,j}^{2}\right) ^{1/2}\cdot \left[
					\sum_{t=2}^{n}\left( \sum_{k=1}^{t-1}x_{k-1,j}\right) ^{2}\right] ^{1/2}
					\notag \\
					&=&\left( O_{p}(n^{2})\right) ^{1/2}\cdot \left[ \sum_{t=2}^{n}O_{p}(t^{3})%
					\right] ^{1/2}  \notag \\
					&=&O_{p}(n)\cdot O_{p}(1)\left[ \sum_{t=2}^{n}t^{3}\right] ^{1/2}  \notag \\
					&=&O_{p}(n)\cdot O_{p}(1)\left[ O(n^{4})\right] ^{1/2}  \notag \\
					&=&O_{p}(n^{3}).  \notag
				\end{eqnarray}

				By the results of (\ref{eqn_I2_a}) and (\ref{eqn_I2_b}) , we apply
				Chebyshev's inequality and obtain that:
				\begin{equation}\label{eqn_pf_I2_ur}
					I_{2}\leq O_{p}(n^{\alpha \zeta +2}\overline{c}_{A(n,p)}^{1/2}).
				\end{equation}

				For $I_3$, it can be shown that
				\begin{eqnarray}
					E_{t-1} \sum_{t=1}^{n} x_{t-1,j} \psi_{\tau} \left(u_{t\tau}\right) &=&
					\sum_{t=1}^{n} x_{t-1,j} E_{t-1} \psi_{\tau} \left(u_{t\tau}\right) = 0 ,
					\notag \\
					E_{t-1} \left[\sum_{t=1}^{n} x_{t-1,j} \psi_{\tau} \left(u_{t\tau}\right) %
					\right]^2 &=& \sum_{t=1}^{n} x_{t-1,j}^2 E_{t-1} \psi_{\tau}^2
					\left(u_{t\tau}\right) + 0 = \tau(1-\tau)O_p(n^2).  \notag
				\end{eqnarray}
				This implies that
				\begin{eqnarray}
					I_3 \leq O_p(n). \label{eqn_pf_I3_ur}
				\end{eqnarray}

				By the results of (\ref{eqn_pf_I1_ur}), (\ref{eqn_pf_I2_ur}) and (\ref{eqn_pf_I3_ur}), we find
				the upper bound of the first term in (b):
				\begin{equation*}
					\sum_{t=1}^{n}x_{t-1,j}\psi _{\tau }\left( u_{t\tau }-\delta _{\tau
					}^{\prime }x_{t-1}\right) \leq O_{p}(n^{\alpha \zeta +2}\overline{c}%
					_{A(n,p)}^{1/2}).
				\end{equation*}

				For the second term in (b), under Assumption $U1$ (iii), we have
				\begin{equation*}
					\frac{h_{n}}{n^{\alpha \zeta +2}\overline{c}_{A(n,p)}^{1/2}}=\frac{%
						O_{p}(p^{3/2})}{n^{\alpha \zeta +2}\overline{c}_{A(n,p)}^{1/2}}%
					=O_{p}(1)o(1)=o_{p}(1).
				\end{equation*}%
				Thus the result in (b) is shown:
				\begin{equation}
					\sum_{t=1}^{n}x_{t-1,j}\psi _{\tau }\left( u_{t\tau }-x_{t-1}^{\prime }\hat{%
						\delta}_{\tau }\right) -h_{n}\leq O_{p}(n^{\alpha \zeta +2}\overline{c}%
					_{A(n,p)}^{1/2}).  \label{eqn_xpsi}
				\end{equation}

				For (a), under Assumption $\lambda1$ (ii) that ${\frac{\lambda _{n}n^{(1-\alpha
							\zeta )\gamma }}{n^{\alpha \zeta +2}\overline{c}_{A(n,p)}^{1/2}}\rightarrow
					\infty }$, we obtain:
				\begin{equation}\label{eqn_subgradient_penalty}
					(n^{\alpha \zeta +2}\overline{c}_{A(n,p)}^{1/2})^{-1}\frac{\lambda _{n}}{|%
						\tilde{\beta}_{j,\tau }|^{\gamma }}=\frac{\lambda _{n}n^{(1-\alpha \zeta
							)\gamma }}{n^{\alpha \zeta +2}\overline{c}_{A(n,p)}^{1/2}}\frac{1}{%
						|(n/p^{\alpha })\tilde{\beta}_{j,\tau }|^{\gamma }}=\frac{\lambda
						_{n}n^{(1-\alpha \zeta )\gamma }}{n^{\alpha \zeta +2}\overline{c}%
						_{A(n,p)}^{1/2}}\frac{1}{O_{p}(1)}\rightarrow \infty .
				\end{equation}

				By (\ref{eqn_foc}), (\ref{eqn_xpsi}), and (\ref{eqn_subgradient_penalty}),
				we can show that, for any $j\notin \mathcal{A}_{0}$,
				\begin{eqnarray}
					&&\Pr \left( j\in \hat{\mathcal{A}}_{n}\right)  \notag \\
					&\leq &\Pr \left( \left\vert \sum_{t=1}^{n}x_{t-1,j}\psi _{\tau }\left(
					u_{t\tau }-\delta _{\tau }^{\prime }x_{t-1}\right) -h_{n}\right\vert =\frac{%
						\lambda _{n}}{|\tilde{\beta}_{j,\tau }|^{\gamma }}\right)  \notag \\
					&=&\Pr \left( \frac{1}{n^{\alpha \zeta +2}\overline{c}_{A(n,p)}^{1/2}}%
					\left\vert \sum_{t=1}^{n}x_{t-1,j}\psi _{\tau }\left( u_{t\tau }-\delta
					_{\tau }^{\prime }x_{t-1}\right) -h_{n}\right\vert =\frac{\lambda
						_{n}n^{(1-\alpha \zeta )\gamma }}{n^{\alpha \zeta +2}\overline{c}%
						_{A(n,p)}^{1/2}}\frac{1}{|(n/p^{\alpha })\tilde{\beta}_{j,\tau
						}|^{\gamma }}\right)  \notag \\
					&\longrightarrow &0.  \notag
				\end{eqnarray}
			\end{proof}


			\section{Proofs for Section \protect\ref{oracle mr}}

			\begin{proof}[Proof of Lemma \protect\ref{initial estimator_transformed}]
				Let $Q_{n}^{QR}(\beta _{\tau })$ be the (unpenalized) QR
				objective function. To show the consistency of QR estimator, it suffices to show
				that for any $\epsilon >0$ , there exists a sufficiently large $C$ such that
				\begin{equation}
					P\left\{ \inf_{||c||=C}Q_{n}^{QR}(\tilde{\beta}_{\tau
					}+a_{n}M_{n}c)>Q_{n}^{QR}(\tilde{\beta}_{\tau })\right\} \geq 1-\epsilon,
				\end{equation}
				where $a_{n}M_{n}=\left(
				\begin{array}{cc}
					a_{n}^{(0)}I_{r} & 0 \\
					0 & a_{n}^{(1)}I_{p-r}%
				\end{array}%
				\right) $ so that $a_{n}M_{n}c=\left(
				a_{n}^{(0)}c_{1},...,a_{n}^{(0)}c_{r},a_{n}^{(1)}c_{r+1},...a_{n}^{(1)}c_{p}\right) ^{\prime }
				$.

				As is shown in the proof of Lemma \ref{initial estimator}, this proof
				can be completed if we show that the following term is positive:
				\begin{equation*}
					Q_{n}^{QR}(\tilde{\beta}_{\tau }+a_{n}M_{n}c)-Q_{n}^{QR}(\tilde{\beta}_{\tau
					})=\sum_{t=1}^{n}\rho _{\tau }(u_{t\tau }-\tilde{x}_{t-1}^{\prime
					}a_{n}M_{n}c)-\sum_{t=1}^{n}\rho _{\tau }(u_{t\tau }).
				\end{equation*}%
				By Knight's Identity,
				\begin{align*}
					\sum_{t=1}^{n}& \left[ \rho _{\tau }(u_{t\tau }-\tilde{x}_{t-1}^{\prime
					}a_{n}M_{n}c)-\rho _{\tau }(u_{t\tau })\right] \\
					& =-a_{n}\sum_{t=1}^{n}\tilde{x}_{t-1}^{\prime }M_{n}c\cdot \psi _{\tau
					}(u_{t\tau })+\sum_{t=1}^{n}\int_{0}^{\tilde{x}_{t-1}^{\prime
						}a_{n}M_{n}c}\left( \mathbf{1}(u_{t\tau }\leq s)-\mathbf{1}(u_{t\tau }\leq
					0)\right) ds \\
					& =-a_{n}\sum_{t=1}^{n}\tilde{x}_{t-1}^{\prime }M_{n}c\cdot \psi _{\tau
					}(u_{t\tau })+\sum_{t=1}^{n}E\left[ \int_{0}^{\tilde{x}_{t-1}^{\prime
						}a_{n}M_{n}c}\left( \mathbf{1}(u_{t\tau }\leq s)-\mathbf{1}(u_{t\tau }\leq
					0)\right) ds\right] \\
					& +\sum_{t=1}^{n}\left\{ \int_{0}^{\tilde{x}_{t-1}^{\prime
						}a_{n}M_{n}c}\left( \mathbf{1}(u_{t\tau }\leq s)-\mathbf{1}(u_{t\tau }\leq
					0)\right) ds-E\left[ \int_{0}^{\tilde{x}_{t-1}^{\prime }a_{n}M_{n}c}\left(
					\mathbf{1}(u_{t\tau }\leq s)-\mathbf{1}(u_{t\tau }\leq 0)\right) ds\right]
					\right\} \\
					& \equiv I_{1}+I_{2}+I_{3}.
				\end{align*}%
				We will show that $I_{1}$ and $I_{2}$ are dominated by $I_{2}$ and that $%
				I_{2}>0$.

				First, we derive the upper bound of $I_{1}$. For a large $n$,
				\begin{align*}
					E|I_{1}|^{2}& =a_{n}^{2}\sum_{t=1}^{n}c^{\prime }M_{n}^{\prime }E\left[ \psi
					_{\tau }(u_{t\tau })^{2}\tilde{x}_{t-1}\tilde{x}_{t-1}^{\prime }\right]
					M_{n}c+2a_{n}^{2}\sum_{t=2}^{n}\sum_{k=1}^{t-1}c^{\prime }M_{n}^{\prime }E%
					\left[ \psi _{\tau }(u_{t\tau })\psi _{\tau }(u_{k\tau })\tilde{x}_{t-1}%
					\tilde{x}_{k-1}^{\prime }\right] M_{n}c \\
					& =a_{n}^{2}\sum_{t=1}^{n}c^{\prime }M_{n}^{\prime }E\left[ \psi _{\tau
					}(u_{t\tau })^{2}\tilde{x}_{t-1}\tilde{x}_{t-1}^{\prime }\right] M_{n}c \\
					& =a_{n}^{2}c^{\prime }M_{n}^{\prime }\sum_{t=1}^{n}E\left[ \psi _{\tau
					}(u_{t\tau })^{2}\tilde{x}_{t-1}\tilde{x}_{t-1}^{\prime }\right] M_{n}c\text{
						\ \ \ \ \textbf{\ \ }} \\
					& \leq a_{n}^{2}n^{2}\overline{c}_{B(n,p)}C^{2}\text{ , \quad (from Assumption U2)}
				\end{align*}%
				so that
				\begin{equation*}
					I_{1}=O\left( a_{n}n\overline{c}_{B(n,p)}^{1/2}\right).
				\end{equation*}%
				For $I_{2}$, for a large $n$,
				\begin{align*}
					I_{2}& =\sum_{t=1}^{n}E\int_{0}^{\tilde{x}_{t-1}^{\prime }a_{n}M_{n}c}\left(
					\mathbf{1}(u_{t\tau }\leq s)-\mathbf{1}(u_{t\tau }\leq 0)\right) ds \\
					& =\sum_{t=1}^{n}E\int_{0}^{\tilde{x}_{t-1}^{\prime }a_{n}M_{n}c}\left(
					f_{t-1}(0)\cdot s\right) ds\left\{ 1+o_{p}(1)\right\} \\
					& =\frac{1}{2}a_{n}^{2}\sum_{t=1}^{n}c^{\prime }M_{n}^{\prime }E\left[
					f_{t-1}(0)\tilde{x}_{t-1}\tilde{x}_{t-1}^{\prime }\right] M_{n}c\left\{
					1+o_{p}(1)\right\} \\
					& =\frac{1}{2}a_{n}^{2}c^{\prime }\left( M_{n}^{\prime }\sum_{t=1}^{n}E\left[
					f_{t-1}(0)\tilde{x}_{t-1}\tilde{x}_{t-1}^{\prime }\right] M_{n}\right)
					c\left\{ 1+o_{p}(1)\right\} \\
					& \geq \frac{1}{4}a_{n}^{2}\underline{f}\underline{c}_{A(n,p)}C^{2}n^{2}%
					\text{ \quad (by Assumption L2)} \\
					& =O(a_{n}^{2}n^{2}\underline{c}_{A(n,p)}).
				\end{align*}%
				Finally, for $I_{3}$, for a large $n$,
				\begin{eqnarray*}
					Var(I_{3}) &=&Var\left( \sum_{t=1}^{n}\int_{0}^{\tilde{x}_{t-1}^{\prime
						}a_{n}M_{n}c}\left( \mathbf{1}(u_{t\tau }\leq s)-\mathbf{1}(u_{t\tau }\leq
					0)\right) ds\right) \\
					&\leq &E\left[ \left( \sum_{t=1}^{n}\int_{0}^{\tilde{x}_{t-1}^{\prime
						}a_{n}M_{n}c}\left( \mathbf{1}(u_{t\tau }\leq s)-\mathbf{1}(u_{t\tau }\leq
					0)\right) ds\right) ^{2}\right] \\
					&\leq &a_{n}^{2}\sum_{t=1}^{n}c^{\prime }M_{n}^{\prime }E\left[
					x_{t-1}x_{t-1}^{\prime }\right] M_{n}c+2a_{n}^{2}\sum_{t=2}^{n}%
					\sum_{k=1}^{t-1}E\left[ \left\vert x_{t-1}^{\prime }M_{n}c\right\vert
					\left\vert x_{k-1}^{\prime }M_{n}c\right\vert \right] \\
					&=&V_{3,1}+V_{3,2}.
				\end{eqnarray*}%
				By the proof of Lemma \ref{initial estimator}, we can easily show that the following bounds hold
				\begin{equation*}
					V_{3,1}\leq O(a_{n}^{2}\overline{c}_{A(n,p)}n^{2}),\text{ and }V_{3,2}\leq
					O(a_{n}^{2}\overline{c}_{A(n,p)}n^{3}).
				\end{equation*}%
				Based on the results above, we have
				\begin{equation*}
					I_{3}=O_{p}(\overline{c}_{A(n,p)}^{1/2}a_{n}n^{3/2}),
				\end{equation*}%
				which establishes the desired result.
			\end{proof}


			\begin{remark}
				\label{M star}Define $H^{-1}M_{n}=M_{n}^{\ast }$. The normalizing
				matrix then is
				\begin{equation*}
					\left(
					\begin{array}{cccc}
						\sqrt{n}I_{p_{z}} & 0 & 0 & 0 \\
						0 & \sqrt{n}I_{p_{1}} & 0 & 0 \\
						0 & -\sqrt{n}A_{1}^{\prime } & I_{p_{2}} & 0 \\
						0 & 0 & 0 & I_{p_{x}}%
					\end{array}%
					\right),
				\end{equation*}%
				and
				\begin{eqnarray*}
					a_{n}^{(1)}M_{n}^{\ast }c &=&\frac{p^{\alpha }}{n}\left(
					\begin{array}{cccc}
						\sqrt{n}I_{p_{z}} & 0 & 0 & 0 \\
						0 & \sqrt{n}I_{p_{1}} & 0 & 0 \\
						0 & -\sqrt{n}A_{1}^{\prime } & I_{p_{2}} & 0 \\
						0 & 0 & 0 & I_{p_{x}}%
					\end{array}%
					\right) c=\left(
					\begin{array}{cccc}
						\frac{p^{\alpha }}{\sqrt{n}}I_{p_{z}} & 0 & 0 & 0 \\
						0 & \frac{p^{\alpha }}{\sqrt{n}}I_{p_{1}} & 0 & 0 \\
						0 & -\frac{p^{\alpha }}{\sqrt{n}}A_{1}^{\prime } & \frac{p^{\alpha }}{n}%
						I_{p_{2}} & 0 \\
						0 & 0 & 0 & \frac{p^{\alpha }}{n}I_{p_{x}}%
					\end{array}%
					\right) c \\
					&=&(\frac{p^{\alpha }}{\sqrt{n}}(c_{1},...,c_{p_{z}}),\frac{p^{\alpha }}{%
						\sqrt{n}}(c_{p_{z}+1},...,c_{p_{z}+p_{1}}),{\ }\frac{p^{\alpha }}{\sqrt{n}}%
					(c_{1}^{\ast },...,c_{p_{2}}^{\ast })+O\left( \frac{p^{\alpha }}{n}\right) ,%
					\frac{p^{\alpha }}{n}(c_{p_{z}+p_{1}+p_{2}+1},...,c_{p})),
				\end{eqnarray*}%
				where
				\begin{equation*}
					(c_{1}^{\ast },...,c_{p_{2}}^{\ast })^{\prime }=A_{1}^{\prime
					}(c_{p_{z}+1},...,c_{p_{z}+p_{1}})^{\prime }.
				\end{equation*}%
				Therefore, the reduced rate for the 3rd block component is well accommodated. We
				define the convergence rates $\tilde{a}_{n,j}^{\ast }$ by
				\begin{equation*}
					\tilde{a}_{n,j}^{\ast }=\left\{
					\begin{array}{ccc}
						\frac{p^{\alpha }}{\sqrt{n}}=a_{n}^{(0)}, & \text{for }j=1,...,r+p_{2}, &  \\
						\frac{p^{\alpha }}{n}=a_{n}^{(1)}, & \text{ for }j=r+p_{2}+1,...,p. &
					\end{array}%
					\right.
				\end{equation*}
			\end{remark}

			\begin{proof}[Proof of Theorem \protect\ref{thm_consistency_mix}]
				Following the proof of Theorem \ref{thm_consistency}, it suffices to show
				that for any $\epsilon >0$ , there exists a sufficiently large $C$ such that%
				\begin{equation*}
					P\left\{ \inf_{||c||\leq C}Q_{n}(\beta _{0\tau }+a_{n}^{(1)}M_{n}^{\ast
					}c)>Q_{n}(\beta _{0\tau })\right\} \geq 1-\epsilon,
				\end{equation*}%
				where $M_{n}^{\ast }=Q^{-1}M_{n}$ as in Remark \ref{M star}.

				This inequality implies that with probability at least $1-\epsilon $, there
				is a local minimizer $\hat{\beta}_{\tau }^{\ast }$ in the shrinking ball $%
				\{\beta _{0\tau }+a_{n}^{(1)\ast }M_{n}^{\ast }c,||c||\leq C\}$ such that $||%
				\hat{\beta}_{\tau }^{\ast }-\beta _{0\tau }||=O_{p}(\tilde{a}_{n,j}^{\ast })$%
				, where $\tilde{a}_{n,j}^{\ast }$ is the $j$-th dominating rates from $%
				a_{n}^{(1)}M_{n}^{\ast }c$, i.e.,%
				\begin{equation*}
					\tilde{a}_{n,j}^{\ast }=\left\{
					\begin{array}{ccc}
						\frac{p^{\alpha }}{\sqrt{n}}=a_{n}^{(0)}, & \text{for }j=1,...,r+p_{2}\text{,%
						} &  \\
						\frac{p^{\alpha }}{n}=a_{n}^{(1)}, & \text{ for }j=r+p_{2}+1,...,p\text{.%
						} &
					\end{array}%
					\right.
				\end{equation*}%
				Then we obtain
				\begin{eqnarray*}
					&&Q_{n}(\beta _{0\tau }+a_{n}^{(1)}M_{n}^{\ast }c)-Q_{n}(\beta _{0\tau }) \\
					&\geq &\left( \sum_{t=1}^{n}\rho _{\tau }(u_{t\tau }-X_{t-1}^{\prime
					}a_{n}^{(1)}M_{n}^{\ast }c)-\sum_{t=1}^{n}\rho _{\tau }(u_{t\tau })\right)
					\text{ }+\sum_{j=1}^{q_{n}}\lambda _{n,j}(|\beta _{0\tau ,j}+\tilde{a}%
					_{n,j}^{\ast }c_{j}|-|\beta _{0\tau ,j}|) \\
					&=&d_{1}^{\ast }+d_{2}^{\ast }\text{.}
				\end{eqnarray*}

				For $d_{1}^{\ast }$, similarly to Theorem \ref{thm_consistency}, we have%
				\begin{eqnarray*}
					&&\sum_{t=1}^{n}\rho _{\tau }(u_{t\tau }-X_{t-1}^{\prime
					}a_{n}^{(1)}M_{n}^{\ast }c)-\sum_{t=1}^{n}\rho _{\tau }(u_{t\tau }) \\
					&=&-a_{n}^{(1)}\sum_{t=1}^{n}X_{t-1}^{\prime }M_{n}^{\ast }c\cdot \psi
					_{\tau }(u_{t\tau })+\sum_{t=1}^{n}\int_{0}^{X_{t-1}^{\prime }a_{n}^{(1)\ast
						}M_{n}^{\ast }c}\left( \mathbf{1}(u_{t\tau }\leq s)-\mathbf{1}(u_{t\tau
					}\leq 0)\right) ds.
				\end{eqnarray*}%
				Note that%
				\begin{eqnarray*}
					M_{n}^{\ast \prime }\sum_{t=1}^{n}E\left[ f_{t-1}(0)X_{t-1}X_{t-1}^{\prime }%
					\right] M_{n}^{\ast } &=&M_{n}^{\prime }\left( H^{-1}\right) ^{\prime
					}\sum_{t=1}^{n}E\left[ f_{t-1}(0)X_{t-1}X_{t-1}^{\prime }\right] H^{-1}M_{n}
					\\
					&=&M_{n}^{\prime }\sum_{t=1}^{n}E\left[ f_{t-1}(0)\left( H^{-1}\right)
					^{\prime }X_{t-1}X_{t-1}^{\prime }H^{-1}\right] M_{n} \\
					&=&M_{n}^{\prime }\sum_{t=1}^{n}E\left[ f_{t-1}(0)\tilde{x}_{t-1}\tilde{x}%
					_{t-1}^{\prime }\right] M_{n},
				\end{eqnarray*}%
				which is controlled by Assumptions $L2$ and $U2$. Thus, using the exactly same proof of
				Theorem \ref{thm_consistency}, the dominating order in $d_{1}^{\ast }$ is
				\begin{eqnarray*}
					&&\sum_{t=1}^{n}E\int_{0}^{x_{t-1}^{\prime }a_{n}^{(1)}M_{n}^{\ast }c}\left(
					\mathbf{1}(u_{t\tau }\leq s)-\mathbf{1}(u_{t\tau }\leq 0)\right) ds \\
					&=&\sum_{t=1}^{n}E\int_{0}^{x_{t-1}^{\prime }a_{n}^{(1)}M_{n}^{\ast
						}c}\left( f_{t-1}(0)\cdot s\right) ds\left\{ 1+o_{p}(1)\right\} \\
					&=&\frac{1}{2}\left( a_{n}^{(1)}\right) ^{2}c^{\prime }\left( M_{n}^{\ast
						\prime }\sum_{t=1}^{n}E\left[ f_{t-1}(0)X_{t-1}X_{t-1}^{\prime }\right]
					M_{n}^{\ast }\right) c\left\{ 1+o_{p}(1)\right\} \\
					&\geq &O_{p}(\left( a_{n}^{(1)}\right) ^{2}n^{2}\underline{c}%
					_{A(n,p)})=O_{p}\left( p^{2\alpha }\underline{c}_{A(n,p)}\right).
				\end{eqnarray*}%
				For $d_{2}^{\ast }$, the only differences with the proof of Theorem \ref%
				{thm_consistency} are the rate of divergence in $\lambda _{n,j}$'s for $%
				j=1,...,r+p_{2}$, and for $j=r+p_{2}+1,...,p$. From Corollary \ref%
				{corollary consistency mix},%
				\begin{equation*}
					\hat{\beta}_{\tau ,j}=%
					\begin{cases}
						O_{p}(a_{n}^{(0)})+\beta _{0\tau ,j}=o_{p}(1)+\beta _{0\tau ,j}, &
						j=1,...,r+p_{2}, \\
						O_{p}(a_{n}^{(1)})+\beta _{0\tau ,j}=o_{p}(1)+\beta _{0\tau ,j}, &
						j=r+p_{2}+1,...,p,
					\end{cases}%
				\end{equation*}%
				and clearly $a_{n}^{(0)}>a_{n}^{(1)}$ for any given $n$. Since $\beta
				_{0\tau ,j}\neq 0$ for $j=1,...,q_{n}$, we have%
				\begin{eqnarray*}
					\sum_{j=1}^{q_{n}}\lambda _{n,j}(|\beta _{0\tau ,j}+\tilde{a}_{n,j}^{\ast
					}c_{j}|-|\beta _{0\tau ,j}|) &\leq &\sum_{j=1}^{q_{n}}\lambda _{n,j}\tilde{a}%
					_{n,j}^{\ast }|c_{j}|=\lambda _{n}\sum_{j=1}^{q_{n}}\tilde{a}_{n,j}^{\ast
					}|c_{j}|\frac{1}{|\hat{\beta}_{\tau ,j}|^{\gamma }} \\
					&\leq &\lambda _{n}\max_{j}\left( \tilde{a}_{n,j}^{\ast }\right) \left(
					\sum_{j=1}^{q_{n}}\frac{1}{(o_{p}(1)+\beta _{0\tau ,j})^{2\gamma }}\right)
					^{1/2}||c|| \\
					&=&\lambda _{n}a_{n}^{(0)}q_{n}^{1/2}O_{p}(1) \\
					&=&O_{p}\left( \lambda _{n}(p^{\alpha }/\sqrt{n})p^{1/2}\right) \\
					&=&O_{p}\left( \lambda _{n}\frac{p^{\frac{1}{2}+\alpha }}{n^{1/2}}\right),
				\end{eqnarray*}%
				so is dominated by $d_{1}^{\ast }=O_{p}\left( p^{2\alpha }\underline{c}%
				_{A(n,p)}\right) $ under the condition \textbf{\ }%
				\begin{equation*}
					\lambda _{n}\frac{p^{\frac{1}{2}-\alpha }}{n^{1/2}\underline{c}_{A(n,p)}}%
					=\lambda _{n}\frac{n^{\left( \frac{1}{2}-\alpha \right) \zeta }}{n^{1/2}%
						\underline{c}_{A(n,p)}}=\lambda _{n}\frac{n^{\frac{1}{2}\zeta }}{n^{\frac{1}{%
								2}+\alpha \zeta }\underline{c}_{A(n,p)}}\rightarrow 0.
				\end{equation*}
			\end{proof}


			\begin{proof}[Proof of Theorem \protect\ref{thm_sparsity_mix}]
				Among $\hat{\beta}_{\tau }^{ALQR\ast }=\hat{\beta}_{\tau }^{\ast }=(\hat{%
					\beta}_{\tau }^{z\ast ^{\prime }},\hat{\beta}_{1\tau }^{c\ast ^{\prime }},%
				\hat{\beta}_{2\tau }^{c\ast ^{\prime }},\hat{\beta}_{\tau }^{x\ast ^{\prime
				}})^{\prime }$, we just to need to show ALQR\ sparsity for $\hat{\beta}%
				_{2\tau }^{c\ast }$, since other parts are shown by Theorem \ref%
				{thm_sparsity} and the existing proof for the $I$(0) case. Among $%
				j=r+1,...,r+p_{2}$, for any $j\notin \mathcal{A}_{0}$, $\left\vert \hat{\beta%
				}_{j,\tau }^{\ast }-\beta _{j,0\tau }\right\vert =\left\vert \hat{\beta}%
				_{j,\tau }^{\ast }\right\vert =O_{p}\left( \frac{p^{\alpha }}{\sqrt{n}}%
				\right) $, so following the proof of Theorem \ref{thm_sparsity},
				\begin{eqnarray}
					&&\Pr \left( j\in \hat{\mathcal{A}}_{n}\right)  \notag \\
					&\leq &\Pr \left( \left\vert \sum_{t=1}^{n}x_{t-1,j}\psi _{\tau }\left(
					u_{t\tau }-\delta _{\tau }^{\prime }x_{t-1}\right) -h_{n}\right\vert =\frac{%
						\lambda _{n}}{|\hat{\beta}_{j,\tau }^{\ast }|^{\gamma }}\right)  \notag \\
					&=&\Pr \left( \frac{1}{n^{\alpha \zeta +2}\overline{c}_{A(n,p)}^{1/2}}%
					\left\vert \sum_{t=1}^{n}x_{t-1,j}\psi _{\tau }\left( u_{t\tau }-\delta
					_{\tau }^{\prime }x_{t-1}\right) -h_{n}\right\vert =\frac{\lambda _{n}\frac{%
							\sqrt{n}}{p^{\alpha }}}{n^{\alpha \zeta +2}\overline{c}_{A(n,p)}^{1/2}}\frac{%
						1}{|\frac{\sqrt{n}}{p^{\alpha }}\hat{\beta}_{j,\tau }^{\ast }|^{\gamma }}%
					\right)  \notag \\
					&\longrightarrow &0,  \notag
				\end{eqnarray}%
				as long as
				\begin{equation*}
					\frac{\lambda _{n}\left( \frac{\sqrt{n}}{p^{\alpha }}\right) ^{\gamma }}{%
						n^{\alpha \zeta +2}\overline{c}_{A(n,p)}^{1/2}}=\frac{\lambda _{n}n^{\left(
							1/2-\alpha \zeta \right) \gamma }}{n^{\alpha \zeta +2}\overline{c}%
						_{A(n,p)}^{1/2}}\rightarrow \infty .
				\end{equation*}
			\end{proof}

	\section{Proof for Section \protect\ref{section_LimitTheory}} \label{appen_proof_section5}
		\begin{proof}[Proof of Theorem \protect\ref{thm_AsymDist_mix}]
		We first show the limit theory of ALQR\ estimator with I(0) predictors. For simplicity, we use $A$ and $B$ to
		represent $A_{(t,p)}$ and $B_{(t,p)}$, respectively. For an $m\times n$ real
		matrix $A$, we denote its Frobenius norm as $||A||_{F}$ $:=\sqrt{%
			trace(A^{\prime }A)}$. Recall that the I(0) part of our predictors is a linear process of the form
		of $z_{t}=\sum_{j=0}^{\infty }D_{zj}\epsilon _{t-j}$ with the conditions
		given in Section 2.2. In this proof, we use notation $x_{t}=\sum_{j=0}^{%
			\infty }D_{j}\epsilon _{t-j}$ for this I(0) predictors for simplicity.
		Moreover, for simple notation, we drop some superscript/subscript and write $%
		\hat{\beta}_{\tau }^{(0),ALQR\ast }$ as $\hat{\beta}$ unless we need some clarity. Also, without loss of
		generality, we set the first $q_{n}$ elements of $\hat{\beta}$ be non-zero.
		Let $\delta =\sqrt{n}(\beta -\beta _{0})$ and subscript $\mathcal{A}$ denote
		elements of the active set. Thus, the oracle (local) estimator is $\hat{%
			\delta}=\left( \hat{\delta}_{\mathcal{A}},0\right) $. The oracle property of
		the ALQR estimator in Section \ref{SEC:Oracle} implies that $\hat{\delta}$ is a minimizer
		of 
		\begin{equation*}
			\sum_{t=1}^{n}\rho _{\tau }\left( y_{t}-x_{t}^{\prime }\left( n^{-1/2}\delta
			+\beta _{0}\right) \right) +\sum_{j=1}^{p}\lambda _{n,j}\left\vert
			n^{-1/2}\delta +\beta _{0}\right\vert 
		\end{equation*}%
		with probability approaching 1. Define
		\begin{equation*}
			V_{j}(\delta )=-n^{-1/2}\sum_{t=1}^{n}\psi _{\tau }\left(
			y_{t}-x_{t}^{\prime }\left( n^{-1/2}\delta +\beta _{0}\right) \right)
			x_{t,j}+n^{-1/2}\lambda _{n,j}sgn\left( n^{-1/2}\delta _{j}+\beta
			_{0,j}\right) .
		\end{equation*}%
		Using vector notation, we have 
		\begin{equation*}
			V(\hat{\delta})-V(0)=-n^{-1/2}\sum_{t=1}^{n}\left[ \psi _{\tau }\left(
			y_{t}-x_{t}^{\prime }\left( n^{-1/2}\hat\delta +\beta _{0}\right) \right) -\psi
			_{\tau }\left( u_{t}\right) \right] x_{t}+n^{-1/2}\omega _{n}(\hat\delta ),
		\end{equation*}%
		where $\omega _{n}\text{$\left( \hat\delta \right) $}=\left( \lambda
		_{n,1}\left\vert \tilde{\beta}_{\tau ,1}\right\vert ^{-\gamma }sgn\left(
		n^{-1/2}\hat\delta _{1}+\beta _{0,1}\right) ,\ldots ,\lambda _{n,q}\left\vert 
		\tilde{\beta}_{\tau ,q}\right\vert ^{-\gamma }sgn\left( n^{-1/2}\hat\delta
		_{q}+\beta _{0,q}\right) ,0,\ldots ,0\right) .$
		
		Note that, except the penalty term $n^{-1/2}\omega _{n}(\hat\delta )$, 
		\begin{equation*}
			V(\hat{\delta})-V(0)=\left[ V(\hat{\delta})-V(0)-E\left[ V(\hat{\delta})%
			\right] +E\left[ V(0)\right] \right] +\left[ E\left[ V(\hat{\delta})\right]
			-E\left[ V(0)\right] +A\delta \right] -A\delta 
		\end{equation*}%
		and by rearranging 
		\begin{equation}
			A{\delta}-V(0)=\left[ V(\hat{\delta})-V(0)-E\left[ V(\hat{\delta})\right]
			+E\left[ V(0)\right] \right] +\left[ E\left[ V(\hat{\delta})\right] -E\left[
			V(0)\right] +A\delta \right] -V(\hat{\delta}).  \label{arrange0}
		\end{equation}
		
		Define the weighted norm $||\cdot ||_{c}$ by $||A||_{c}=||cA||$ where $c$ is
		an arbitrary $l\times p$ matrix with $||c||\leq \underline{c}_{B}^{-1/2}L_{c}
		$ for a large constant $L_{c}<\infty $. Since $l$ is fixed, without loss of
		generality we assume that $l=1$. \bigskip 
		
		Using this weighted norm $||\cdot ||_{c}$, we will show 
		\begin{align}
			\left\Vert A\delta -V(0)\right\Vert _{c} 
			\leq \sup_{||\hat{\delta}||\leq \sqrt{p}L} & \Big[ \left\Vert V(\hat{\delta})-V(0)-E\left[ V(\hat{\delta}) \right]  +E\left[ V(0)\right] \right\Vert _{c} \notag \\
			& + \left\Vert E\left[ V(\hat{\delta})\right] -E\left[ V(0)\right] +A\delta \right\Vert_{c}
			+\left\Vert V\left( \hat{\delta}\right) \right\Vert _{c} \Big]=o_{p}(1),
			\label{goal0}
		\end{align}%
		for a large constant $L$. 
		
		We denote the terms on the RHS of \eqref{goal0} as:
		\begin{align*}
			\mbox{(CA.6)}&:=\sup_{\Vert\hat{\delta}\Vert \leq \sqrt{p}L}\left\Vert V(\hat{\delta})-V(0)-E\left[
			V(\hat{\delta})\right] +E\left[ V(0)\right] \right\Vert_{c}\\
			\mbox{(CA.7)}&:= \sup_{||\hat{\delta}||\leq \sqrt{p}L}\left\Vert E\left[ V(\hat{\delta})
			\right] -E\left[ V(0)\right] +A\delta \right\Vert_{c} \\
			\mbox{(CA.8)}&:=\sup_{||\hat{\delta}||\leq \sqrt{p}L}\left\Vert V\left( \hat{\Delta}\right)
			\right\Vert_{c}.
		\end{align*}
		
		For (CA.6), we need to show that for any large constant $L<\infty $, 
		\begin{equation*}
			\sup_{||\hat{\delta}||\leq \sqrt{p}L}\left\Vert V(\hat{\delta})-V(0)-E\left[ V(\hat{%
				\delta})-V(0)\right] \right\Vert_{c}=o_{p}(1).
		\end{equation*}
		
		First, we write $a_{t}\equiv cx_{t}=a_{t}^{+}-a_{t}^{-}$ where $%
		a_{t}^{+}=max\{a_{t},0\}$ and $a_{t}^{-}=max\{-a_{t},0\}$. By Minkowski's
		inequality, we obtain 
		\begin{align}
			\sup_{||\hat{\delta}||\leq \sqrt{p}L} & \left\Vert V(\hat{\delta})-V(0)-E\left[ V(\hat{%
				\delta})-V(0)\right] \right\Vert_{c}  \notag \\
			&\leq \sup_{||\hat{\delta}||\leq \sqrt{p}L}\left\vert V^{+}(\hat{\delta}%
			)-V^{+}(0)-E\left[ V^{+}(\hat{\delta})-V^{+}(0)\right] \right\vert  \notag \\
			&+\sup_{||\hat{\delta}||\leq \sqrt{p}L}\left\vert V^{-}(\hat{\delta}%
			)-V^{-}(0)-E\left[ V^{-}(\hat{\delta})-V^{-}(0)\right] \right\vert ,
			\label{eqn::A6}
		\end{align}%
		where $V^{+}(\hat{\delta})\equiv n^{-1/2}\sum_{t=1}^{n}\psi _{\tau }\left(
		y_{t}-x_{t-1}^{\prime }\left( n^{-1/2}\hat{\delta}+\beta _{0}\right) \right)
		a_{t-1}^{+}$, and $V^{-}(\hat{\delta})$ is analogously defined. Thus, it
		suffices to show that each term on the right hand side of (\ref{eqn::A6}) is 
		$o_{p}(1)$.
		
		We show the first term of (\ref{eqn::A6}) is $o_p(1)$. Let $\mathbf{D}
		\equiv \{\hat{\delta} \in \mathcal{R}^p : ||\hat\delta||\leq \sqrt{p}L\}$
		for some finite constant $L$. Let $|t|_{\infty}$ denote the maximum of the
		absolute values of the coordinates of $t$. We select $N_1 = (2n^2)^p$ grid
		points, $\hat{\delta}_1,...,\hat{\delta}_{N_1}$ and cover $\mathbf{D}$ by
		cubes $\mathbf{D}_s = \left\{ \hat{\delta} \in \mathcal{R}^p: |\hat{\delta}
		- \hat{\delta}_s|_{\infty} \leq \delta_{\epsilon n} \right\}$ with sides of
		length $\delta_{\epsilon n}$ where $\delta_{\epsilon n} = Lp^{1/2}/n^2$.
		Since $\psi_{\tau}(\cdot)$ is monotone, by the Minkowski's inequality we
		obtain that 
		\begin{eqnarray}
			&& \sup_{||\hat\delta||\leq \sqrt{p}L} \left|V^+(\hat\delta)-V^+(0)-E\left[%
			V^+(\hat\delta)-V^+(0)\right]\right|  \notag \\
			&\leq& \max_{1 \leq s \leq N_1} \left|V^+(\hat\delta)-V^+(0)-E\left[%
			V^+(\hat\delta)-V^+(0)\right]\right|  \notag \\
			&& + \max_{1 \leq s \leq N_1} \left| n^{-1/2} \sum_{t=1}^{n} E \left[
			\psi_{st}(\delta_{\epsilon n}) a_{t-1}^+\right] - E\left[ \psi_{st}(-%
			\delta_{\epsilon n} )a_{t-1}^+ \right] \right|  \notag \\
			&& + \max_{1 \leq s \leq N_1} \left| n^{-1/2} \sum_{t=1}^{n} \left[ \left[%
			\psi_{st}(\delta_{\epsilon n})-\psi_{st}(0) \right]a_{t-1}^+ - E\left\{ %
			\left[\psi_{st}(\delta_{\epsilon n})-\psi_{st}(0) \right]a_{t-1}^+ \right\}%
			\right] \right|  \notag \\
			&\equiv& I_1 + I_2 + I_3,  \notag
		\end{eqnarray}
		where $\psi_{st}(\delta) = \psi_{\tau}\left(u_{t\tau} - n^{-1/2}\hat{\delta}%
		_s^{\prime }x_{t-1} +n^{-1/2}\delta ||x_{t-1}||\right)$.
		
		For $I_{1}$, note that 
		\begin{eqnarray*}
			&&V^{+}(\hat{\delta})-V^{+}(0)-E\left[ V^{+}(\hat{\delta})-V^{+}(0)\right] \\
			&=&n^{-1/2}\sum_{t=1}^{n}\left[ \psi _{\tau }\left( y_{t}-x_{t-1}^{\prime
			}\left( n^{-1/2}\hat{\delta}+\beta _{0}\right) \right) a_{t-1}^{+}-\psi
			_{\tau }\left( y_{t}-x_{t-1}^{\prime }\beta _{0}\right) a_{t-1}^{+}\right] \\
			&&-n^{-1/2}\sum_{t=1}^{n}E\left[ \psi _{\tau }\left( y_{t}-x_{t-1}^{\prime
			}\left( n^{-1/2}\hat{\delta}+\beta _{0}\right) \right) a_{t-1}^{+}-\psi
			_{\tau }\left( y_{t}-x_{t-1}^{\prime }\beta _{0}\right) a_{t-1}^{+}\right] \\
			&=&n^{-1}\sum_{t=1}^{n}\eta _{ts} \\
			&=&n^{-1}\sum_{t=1}^{n}\eta _{ts}\mathbf{1}\left\{ a_{t-1}^{+}\leq
			e_{1n}\right\} +n^{-1}\sum_{t=1}^{n}\eta _{ts}\mathbf{1}\left\{
			a_{t-1}^{+}>e_{1n}\right\} \\
			&\equiv &D_{1s}+D_{2s},
		\end{eqnarray*}%
		where $\eta _{ts}\equiv n^{1/2}[\eta _{ts,0}-E(\eta _{ts,0})]$, $\eta
		_{ts,0}=[\psi _{\tau }(u_{t\tau }-n^{-1/2}\hat{\delta}_{s}^{\prime
		}x_{t-1})-\psi _{\tau }(u_{t\tau })]a_{t-1}^{+}$, and $e_{1n}=(np^{4}%
		\underline{c}_{B}^{-4})^{1/8}$. To prove that $I_{1}=o_{p}(1)$, it suffices
		to show 
		\begin{equation*}
			\max_{1\leq s\leq N_{1}}\left\vert D_{ks}\right\vert =o_{p}(1)\quad \text{%
				for k$=1$ and $2$}.
		\end{equation*}
		
		For the I(0) predictors $x_{t}=\sum_{j=0}^{\infty }D_{j}\epsilon _{t-j}$, we
		have $D_{j}=O\left( e^{-vj}\right) $ with $v>0$ along with the conditions
		(1), (2), (5) of \citet{withers1981conditions}: $\left\{ \epsilon _{j}\right\} _{j=-\infty
		}^{\infty }$ are independent r.v's with characteristic functions $\left\{
		\phi _{j}\right\} $ such that%
		\begin{equation}
			\text{(1):  \quad}(2\pi \text{ })^{-1}\max_{j}\int \left\vert \phi
			_{j}(t)\right\vert dt<\infty ,  \label{withers1}
		\end{equation}%
		\begin{equation}
			\text{(2): \quad}\max_{j}E\left\vert \epsilon _{j}\right\vert ^{\delta }<\infty 
			\text{ for some }\delta >0,  \label{withers2}
		\end{equation}%
		\begin{equation}
			\text{(5): \quad}\sup_{m,s,k\geq 1}\sup_{\alpha ,\beta ,v}\max_{t}\left\vert 
			\frac{\partial }{\partial v_{t}}P\left( W+v\in \cup _{1}^{s}D_{j}\right)
			\right\vert <\infty  , \label{withers5}
		\end{equation}%
		where 
		\begin{equation*}
			D_{j}=\mathbb{X}_{t=k}^{k+m-1}\left( \alpha _{jt},\beta _{jt}\right) \text{, 
			}v=(v_{k},...,v_{k+m-1})\text{, }W=(W_{k},...,W_{k+m-1})\,
		\end{equation*}%
		with $W_{t}=\sum_{j=0}^{t}D_{j}\epsilon _{t-1-j}$ and $\mathbb{X}$
		represents a product space.
		
		As a result, $x_{t}$ is strong mixing with $\alpha (j)=O(e^{-v\lambda j})$
		with another constant $\lambda >0$ by Corollary 2 of \citet{withers1981conditions}. By the
		invariance property of the strong mixing processes, $\frac{\eta _{ts}}{%
			n^{1/2}}=[\eta _{ts,0}-E(\eta _{ts,0})]$ is also strong mixing with $\alpha
		(j)=O(e^{-v\lambda j})$, and 
		\begin{equation*}
			\frac{\eta _{ts}}{n^{1/2}}\mathbf{1}\left\{ a_{t-1}^{+}\leq e_{1n}\right\}
			=[\eta _{ts,0}-E(\eta _{ts,0})]\mathbf{1}\left\{ a_{t-1}^{+}\leq
			e_{1n}\right\} \leq 2e_{1n}\text{ for all }t\text{ and }s\text{.}
		\end{equation*}
		
		From Theorem 2, Equation (2.3) of \citet{merlevede2009bernstein}, there is a
		constant $C_{3}$ depending only on $\widetilde{c}$ such that, for all $n\geq
		2,$%
		\begin{equation*}
			P\left( \left\vert \sum_{t=1}^{n}X_{t}\right\vert \geq \eta \right) \leq
			\exp \left(-\frac{C_{3}\eta ^{2}}{v^{2}n+M^{2}+\eta M\left( \log n\right) ^{2}}\right).
		\end{equation*}%
		with $v^{2}=\sup_{i>0}\left( Var(X_{i})+2\sum_{j>i}\left\vert Cov\left(
		X_{i},X_{j}\right) \right\vert \right) =Var(X_{0})+2\sum_{j=1}^{\infty
		}\left\vert Cov\left( X_{j},X_{0}\right) \right\vert ~$(using covariance
		stationarity).
		
		In our case, 
		\begin{equation*}
			P\left( \left\vert \sum_{t=1}^{n}\frac{\eta _{ts}}{n^{1/2}}\mathbf{1}\left\{
			a_{t-1}^{+}\leq e_{1n}\right\} \right\vert \geq \varepsilon n^{1/2}\right) ,
		\end{equation*}%
		so $X_{t}=\frac{\eta _{ts}}{n^{1/2}}\mathbf{1}\left\{ a_{t-1}^{+}\leq
		e_{1n}\right\} $, $\eta =\varepsilon n^{1/2}$, and $M=2e_{1n}$, thereby
		providing 
		\begin{equation*}
			P\left( \left\vert \sum_{t=1}^{n}\frac{\eta _{ts}}{n^{1/2}}\mathbf{1}\left\{
			a_{t-1}^{+}\leq e_{1n}\right\} \right\vert \geq \varepsilon n^{1/2}\right)
			\leq \exp \left(-\frac{C_{3}\varepsilon ^{2}n}{v^{2}n+4e_{1n}^{2}+2e_{1n}%
				\varepsilon n^{1/2}\left( \log n\right) ^{2}}\right).
		\end{equation*}
		
		Thus, it suffices to show $v^{2}n$ is slower than $n$.
		
		Note that 
		\begin{eqnarray*}
			&&Var\left( \eta _{ts}\mathbf{1}\left\{ a_{t-1}^{+}\leq e_{1n}\right\}
			\right)  \\
			&=&E\left[ \eta _{ts}^{2}\mathbf{1}\left\{ a_{t-1}^{+}\leq e_{1n}\right\} %
			\right] -E\left[ \eta _{ts}\mathbf{1}\left\{ a_{t-1}^{+}\leq e_{1n}\right\} %
			\right] ^{2} \\
			&\leq &E\left[ \eta _{ts}^{2}\mathbf{1}\left\{ a_{t-1}^{+}\leq
			e_{1n}\right\} \right]  \\
			&=&E\left[ n[\eta _{ts,0}-E(\eta _{ts,0})]^{2}\mathbf{1}\left\{
			a_{t-1}^{+}\leq e_{1n}\right\} \right]  \\
			&\leq &nE\left[ [\eta _{ts,0}]^{2}\right] =nE\left[ [\left( \psi _{\tau
			}(u_{t\tau }-n^{-1/2}\hat{\delta}_{s}^{\prime }x_{t-1})-\psi _{\tau
			}(u_{t\tau })\right) \left( a_{t-1}^{+}\right) ^{2}\right]  \\
			&=&nE\left[ [E\left[ \left( \psi _{\tau }(u_{t\tau }-n^{-1/2}\hat{\delta}%
			_{s}^{\prime }x_{t-1})-\psi _{\tau }(u_{t\tau })\right) |x_{t-1}\right]
			\left( a_{t-1}^{+}\right) ^{2}\right]  \\
			&=&nE\left[ \left( F_{u}\left( n^{-1/2}\hat{\delta}_{s}^{\prime
			}x_{t-1}|x_{t-1}\right) -F_{u}\left( 0|x_{t-1}\right) \right) \left(
			a_{t-1}^{+}\right) ^{2}\right]  \\
			&=&nE\left[ \left( n^{-1/2}\hat{\delta}_{s}^{\prime }x_{t-1}f_{u}\left(
			0|x_{t-1}\right) +o_{p}(n^{-1/2}\right) \left( a_{t-1}^{+}\right) ^{2}\right]
			=Cn^{1/2}p^{1/2}E[\left( a_{t-1}^{+}\right) ^{2}],
		\end{eqnarray*}%
		where we used the fact $\left\Vert \hat{\delta}_{s}^{\prime
		}x_{t-1}\right\Vert \leq O_{p}(p^{1/2})$ (see (\ref{eqn_delta_x}) below). 
	Therefore, we have $Var\left( X_{t}\right) =Var\left( \frac{\eta _{ts}}{n}\mathbf{1}%
		\left\{ a_{t-1}^{+}\leq e_{1n}\right\} \right) =\frac{1}{n^{2}}Var\left(
		\eta _{ts}\mathbf{1}\left\{ a_{t-1}^{+}\leq e_{1n}\right\} \right) \leq
		Cn^{-3/2}\underline{c}_{B}^{-1}p^{3/2}.$ 
		\begin{equation}
			Var\left( \frac{\eta _{ts}}{n^{1/2}}\mathbf{1}\left\{ a_{t-1}^{+}\leq
			e_{1n}\right\} \right) \leq Cn^{-1/2}\underline{c}_{B}^{-1}p^{3/2}.
			\label{varx}
		\end{equation}%
		It remains to show the covariance terms $\sum_{j=1}^{\infty }\left\vert
		Cov\left( X_{j},X_{0}\right) \right\vert $ is same or smaller order than
		the term in (\ref{varx}). By the inequality of the strong mixing processes, see
		e.g., Corollary 14.3 of \citet{davidson1994stochastic}, for $\frac{1}{p}+\frac{1}{q}+\frac{1%
		}{r}=1$, 
		\begin{equation*}
			\left\vert Cov\left( X_{j},X_{0}\right) \right\vert \preceq \alpha
			(j)^{1/r}\left\Vert X_{j}\right\Vert _{p}\left\Vert X_{0}\right\Vert _{r}.
		\end{equation*}%
		where $\preceq $ is $\leq $ up to a fixed constant term. Letting $%
		p=q=2+\zeta ,~$\ we have $r=\frac{2+\zeta }{\zeta }$. Note that 
		\begin{equation*}
			\left\Vert X_{j}\right\Vert _{2+\zeta }=\left( E\left[ \left\vert
			X_{j}\right\vert ^{2+\zeta }\right] \right) ^{\frac{1}{2+\zeta }}
		\end{equation*}%
		and 
		\begin{eqnarray*}
			&&E\left[ \left\vert X_{j}\right\vert ^{2+\zeta }\right]  \\
			&=&E\left[ \left\vert \frac{\eta _{js}}{n^{1/2}}\mathbf{1}\left\{
			a_{j-1}^{+}\leq e_{1n}\right\} \right\vert ^{2+\zeta }\right] =E\left[
			\left\vert [\eta _{ts,0}-E(\eta _{ts,0})]\mathbf{1}\left\{ a_{t-1}^{+}\leq
			e_{1n}\right\} \right\vert ^{2+\zeta }\right]  \\
			&=&E\Big[ \left\vert \left\{ [\psi _{\tau }(u_{t\tau }-n^{-1/2}\hat{\delta}%
			_{s}^{\prime }x_{t-1})-\psi _{\tau }(u_{t\tau })]a_{t-1}^{+}   -E\left[ [\psi
			_{\tau }(u_{t\tau }-n^{-1/2}\hat{\delta}_{s}^{\prime }x_{t-1})-\psi _{\tau
			}(u_{t\tau })]a_{t-1}^{+}\right] \right\} \right\vert ^{2+\zeta }\\
			&& \times \mathbf{1}%
			\left\{ a_{j-1}^{+}\leq e_{1n}\right\} \Big]  \\
			&\leq &E\left[ \left\vert \psi _{\tau }(u_{t\tau }-n^{-1/2}\hat{\delta}%
			_{s}^{\prime }x_{t-1})-\psi _{\tau }(u_{t\tau })\right\vert \left(
			a_{t-1}^{+}\right) ^{2+\zeta }\right]  \\
			&=&E\left[ E\left[ \left\vert \psi _{\tau }(u_{t\tau }-n^{-1/2}\hat{\delta}%
			_{s}^{\prime }x_{t-1})-\psi _{\tau }(u_{t\tau })\right\vert |x_{t-1}\right]
			\left( a_{t-1}^{+}\right) ^{2+\zeta }\right] 
		\end{eqnarray*}%
		From the same argument to derive (\ref{varx}), we have
		\[E\left[ \left\vert \psi
		_{\tau }(u_{t\tau }-n^{-1/2}\hat{\delta}_{s}^{\prime }x_{t-1})-\psi _{\tau
		}(u_{t\tau })\right\vert |x_{t-1}\right] =n^{-1/2}\hat{\delta}_{s}^{\prime
		}x_{t-1}+o_{p}\left( n^{-1/2}\right).
		\]
		Therefore, we get
		\begin{equation*}
			E\left[ \left\vert X_{j}\right\vert ^{2+\zeta }\right] \preceq
			n^{-1/2}p^{1/2}E\left[ \left( a_{t-1}^{+}\right) ^{2+\zeta }\right]
		\end{equation*}%
		and 
		\begin{equation*}
			\left\Vert X_{j}\right\Vert _{2+\zeta }\preceq n^{-\frac{1}{2\left(
					2+\zeta \right) }}p^{\frac{1}{2\left( 2+\zeta \right) }}\left\Vert
			a_{t-1}^{+}\right\Vert _{2+\zeta }.
		\end{equation*}%
		The stationarity assumption implies that $\left\Vert X_{0}\right\Vert _{2+\zeta }$
		is the same order. Therefore, 
		\begin{equation*}
			\left\vert Cov\left( X_{j},X_{0}\right) \right\vert \preceq \alpha
			(j)^{1/r}n^{-\frac{1}{\left( 2+\zeta \right) }}p^{\frac{1}{\left( 2+\zeta
					\right) }}\left\Vert a_{t-1}^{+}\right\Vert _{2+\zeta }^{2}
		\end{equation*}%
		and
		\begin{equation*}
			\sum_{j=1}^{\infty }\left\vert Cov\left( X_{j},X_{0}\right) \right\vert
			\preceq Cn^{-\frac{1}{\left( 2+\zeta \right) }}\underline{c}_{B}^{-1}p^{1+%
				\frac{1}{\left( 2+\zeta \right) }}.
		\end{equation*}%
		In other words, the leading term of $v^{2}=\sup_{i>0}\left(
		Var(X_{i})+2\sum_{j>i}\left\vert Cov\left( X_{i},X_{j}\right) \right\vert
		\right) $ is $Var\left( \frac{\eta _{ts}}{n^{1/2}}\mathbf{1}\left\{
		a_{t-1}^{+}\leq e_{1n}\right\} \right) $, which is $Cn^{-1/2}\underline{c}%
		_{B}^{-1}p^{3/2}$. Therefore, we conclude that 
		\begin{equation*}
			P\left( \left\vert \sum_{t=1}^{n}\frac{\eta _{ts}}{n^{1/2}}\mathbf{1}\left\{
			a_{t-1}^{+}\leq e_{1n}\right\} \right\vert \geq \varepsilon n^{1/2}\right)
			\leq \exp (-\frac{C_{3}\varepsilon ^{2}n}{Cn^{-1/2}\underline{c}%
				_{B}^{-1}p^{3/2}+4e_{1n}^{2}+2e_{1n}\varepsilon n^{1/2}\left( \log n\right)
				^{2}})\rightarrow 0
		\end{equation*}%
		using the same argument from \citet[p.~53]{lu2015jackknife}. Note that, by the
		weak dependence assumption (asymptotic independence), we achieve the same
		order of magnitudes (in terms of $n$) in the exponent of the Bernstein's
		inequality.
		
		The proofs of the other terms $D_{2}$, $I_{2}$ and $I_{3}$ do
		not rely on the independence assumption, such that the same argument of \citet{lu2015jackknife} can carry over under our additional assumptions. Thus the proofs are omitted.
		
		For (CA.7), we need to show: 
		\begin{equation*}
			\sup_{||\hat{\delta}||\leq \sqrt{p}L}||E[V(\delta )]-E[V(0)]+A\delta
			||_{c}=o(1).
		\end{equation*}
		
		By Assumption 4.1 and 4.6, 
		\allowdisplaybreaks
		\begin{align*}
			&\sup_{||\hat{\delta}||\leq \sqrt{p}L}||E[V(\delta )]-E[V(0)]+A\delta ||_{c}
			\\
			&=\sup_{||\hat{\delta}||\leq \sqrt{p}L}||n^{-1/2}\sum_{t=1}^{n}E\left[ \psi
			_{\tau }(u_{t}-n^{-1/2}\delta ^{\prime }x_{t-1})x_{t-1}\right]
			-n^{-1/2}\sum_{t=1}^{n}E\left[ \psi _{\tau }(u_{t})x_{t-1}\right] +A\delta
			||_{c}    \\
			&=\sup_{||\hat{\delta}||\leq \sqrt{p}L}||n^{-1/2}\sum_{t=1}^{n}E\left[
			\left( \tau -\mathbf{1}(u_{t}-n^{-1/2}\delta ^{\prime }x_{t-1}\leq 0)\right)
			x_{t-1}-\left( \tau -\mathbf{1}(u_{t}\leq 0)\right) x_{t-1}\right] +A\delta
			||_{c}    \\
			&=\sup_{||\hat{\delta}||\leq \sqrt{p}L}||n^{-1/2}\sum_{t=1}^{n}E\left[
			\left( -\mathbf{1}(u_{t}-n^{-1/2}\delta ^{\prime }x_{t-1}\leq 0)+\mathbf{1}%
			(u_{t}\leq 0)\right) x_{t-1}\right] +A\delta ||_{c}    \\
			&=\sup_{||\hat{\delta}||\leq \sqrt{p}L}||n^{-1/2}\sum_{t=1}^{n}E\left[
			F(-u_{t}-n^{-1/2}\delta ^{\prime
			}x_{t-1}|x_{t-1})x_{t-1}-F(u_{t}|x_{t-1})x_{t-1}\right] -A\delta ||_{c} 
			\\
			&=\sup_{||\hat{\delta}||\leq \sqrt{p}L}||n^{-1/2}\sum_{t=1}^{n}E\left[
			n^{-1/2}\delta ^{\prime }x_{t-1}\left( \int_{0}^{1}f(-u_{t}+sn^{-1/2}\delta
			^{\prime }x_{t-1}|x_{t-1})ds\right) x_{t-1}\right] -E\left[
			f(-u_{t}|x_{t-1})x_{t-1}x_{t-1}^{\prime }\right] \delta ||_{c}    \\
			&=\sup_{||\hat{\delta}||\leq \sqrt{p}L}||n^{-1}\sum_{t=1}^{n}E\left[ \left(
			\int_{0}^{1}\left[ f(-u_{t}+sn^{-1/2}\delta ^{\prime
			}x_{t-1}|x_{t-1})-f(-u_{t}|x_{t-1})\right] ds\right) x_{t-1}x_{t-1}^{\prime
			}\delta \right] ||_{c}\quad (\text{stationary $x_{t-1}$})    \\
			&\leq C\sup_{||\hat{\delta}||\leq \sqrt{p}L}n^{-3/2}\sum_{t=1}^{n}E||\delta
			^{\prime }x_{t-1}x_{t-1}x_{t-1}^{\prime }\delta ||_{c}\quad (\text{Taylor
				expansion})    \\
			&\leq Cn^{-3/2}||c||\sup_{||\hat{\delta}||\leq \sqrt{p}L}\sum_{t=1}^{n}%
			\left( E||\delta ^{\prime }x_{t-1}||_F^{2}\right) ^{1/2}\left(
			E||x_{t-1}x_{t-1}^{\prime }\delta||_F^2\right) ^{1/2}\quad (\text{by Cauchy-Schwarz
				inequality})    \\
			&\leq Cn^{-3/2}\underline{c}_{B}^{-1/2}L_{c}\sup_{||\hat{\delta}||\leq 
				\sqrt{p}L}\sum_{t=1}^{n}\left( E||\delta ^{\prime }x_{t-1}||_F^{2}\right)
			^{1/2}\left( E||x_{t-1}x_{t-1}^{\prime }\delta||_F^2\right) ^{1/2}\quad (\text{since $%
				||c||\leq \underline{c}_{B}^{1/2}L_{c}$})    \\
			&=Cn^{-3/2}\underline{c}_{B}^{-1/2}L_{c}\sum_{t=1}^{n}O(\overline{c}%
			_{A}^{1/2}p^{1/2})O(p^{3/2})    \\
			&=O(n^{-1/2}\underline{c}_{B}^{-1/2}\overline{c}_{A}^{1/2}p^{2})=o_{p}(1), 
		\end{align*}%
		where the second last equality uses the facts that 
		\begin{eqnarray}
			E||\delta ^{\prime }x_{t-1}||_F^{2} &\leq &E||x_{t-1}||^{2}||\delta
			||_F^{2}\quad   \notag \\
			&=&\lambda _{\max }\left( E[x_{t-1}x_{t-1}^{\prime }]\right) ||\delta ||_F^{2}
			\notag \\
			&\leq &O(\overline{c}_{A}p)\quad (\text{Assumption \ref{assumption L2}})  \label{eqn_delta_x}
		\end{eqnarray}%
		and 
		\begin{eqnarray}
			E||x_{t-1}x_{t-1}^{\prime }\delta ||_F^{2} &\leq &E||x_{t-1}x_{t-1}^{\prime
			}||_F^{2}||\delta ||_F^{2}  \notag \\
			&=&E\left[ \left( \sum_{j=1}^{p}x_{t-1,j}^{2}\right) ^{2}||\delta ||_F^{2}%
			\right]   \notag \\
			&\leq &O(p^{2})O(p)=O(p^{3}).  \notag
		\end{eqnarray}
		
		\bigskip
		
		Next, we show (CA.8): 
		\begin{eqnarray}
			|| V(\hat{\delta}) ||_c = o_p(1).  \notag
		\end{eqnarray}
		Note that
		\begin{eqnarray}
			||V(\hat{\delta})||_{c} &=&\lVert n^{-1/2}\sum_{t=1}^{n}\psi _{\tau
			}(u_{t}-n^{-1/2}\delta ^{\prime }x_{t-1})x_{t-1}\rVert _{c}  \notag \\
			&=&\lVert n^{-1/2}\sum_{t=1}^{n}\psi _{\tau }(y_{t}-\hat{\beta}^{\prime
			}x_{t-1})x_{t-1}\rVert _{c}  \notag \\
			&=&\left\vert n^{-1/2}\sum_{t=1}^{n}\psi _{\tau }(y_{t}-\hat{\beta}^{\prime
			}x_{t-1})cx_{t-1}\right\vert   \notag \\
			&\leq &n^{-1/2}\sum_{t=1}^{n}\left\vert cx_{t-1}\right\vert \mathbf{1}(y_{t}-%
			\hat{\beta}^{\prime }x_{t-1}=0)\quad (\text{by the proof of Lemma A2 in
				\citet{ruppert1980trimmed}})  \notag \\
			&\leq &n^{-1/2}\max_{1\leq t\leq n}\left\vert cx_{t-1}\right\vert p  \notag
			\\
			&=&o_{p}(1),  \notag
		\end{eqnarray}%
		where we use the fact that 
		\begin{eqnarray}
			P\left( \max_{1\leq t\leq n}\left\vert cx_{t-1}\right\vert \geq
			n^{1/2}p^{-1}\right)  &\leq &nP\left( \left\vert cx_{t-1}\right\vert \geq
			n^{1/2}p^{-1}\right) \quad (\text{by Boole's inequality})  \notag \\
			&\leq &n\frac{E\left\vert cx_{t-1}\right\vert ^{8}}{n^{4}p^{-8}}\quad (\text{%
				by Markov's inequality})  \notag \\
			&\leq &n^{-3}p^{8}\lVert c\rVert ^{8}E\lVert x_{t-1}\rVert ^{8}  \notag \\
			&=&O(n^{-3}p^{8}\underline{c}_{B}^{-4}p^{4})\quad (\text{since $\sup_{j\geq
					1}E(x_{t-1,j}^{8})\leq c_{x}$ for some $c_{x}\leq \infty $})  \notag \\
			&=&O(n^{-3}p^{12}\underline{c}_{B}^{-4})  \notag \\
			&=&o(1),  \notag
		\end{eqnarray}%
		by the assumption that $n^{-3}p^{12}\underline{c}_{B}^{-4}\rightarrow 0$ as $%
		n\rightarrow \infty $.
		
		From (\ref{goal0}), the limit theory of $A\hat{\delta}$ follows $
		V(0)$ because the penalty term $n^{-1/2}\omega _{n}(\delta )$ for the
		non-zero $q_{n}$ element is asymptotically negligible under our rate
		conditions. Note that $V(0):=n^{-1/2}\sum_{t=1}^{n}\psi _{\tau }\left(
		u_{t\tau }\right) x_{t}$, then under our strong mixing condition of Assumption \ref{5.2}, Proposition B.2 of the supplement of \citet{li2020uniform} provides: 
		\begin{equation*}
			\left\Vert V(0)-\tilde{S}_{n}\right\Vert _{F}=o_{p}(1),
		\end{equation*}%
		where $\tilde{S}_{n}$ is a $q_{n}$-dimensional random vector with
		distribution $N\left( 0,B\right) $. Therefore, $%
		\hat{\delta}:=\sqrt{n}(\hat{\beta}-\beta _{0})$ converges to a $q_{n}$%
		-dimensional random vector with distribution $N\left(
		0,A^{-1}BA^{-1}\right) $ in $\left\Vert \cdot
		\right\Vert _{F}$-norm, implying Theorem \ref{thm_AsymDist_mix}-(i), from the relation $\left\Vert M
		\right\Vert \leq
		\left\Vert M  \right\Vert _{F} \leq \sqrt{r}\left\Vert M
		\right\Vert$ for a matrix $M$ of rank $r$.
	
		For the proof of Theorem  \ref{thm_AsymDist_mix}-(ii), following the argument in Section \ref{oracle mr}, we
		have 
		\begin{equation*}
			R\left[ \hat{\beta}_{\tau }^{(1),QR\ast }(1)-\beta _{0\tau }^{(1)}(1)\right]
			=R\left[ \left( \tilde{\beta}_{\tau }^{(1),QR}(1)-\tilde{\beta}_{0\tau
			}^{(1)}(1)\right) -A_{1}^{\prime }\left( \hat{\beta}_{\tau
			}^{(1),QR}(1)-\beta _{0\tau }^{(1)}(1)\right) \right], 
		\end{equation*}%
		so that 
		\begin{eqnarray*}
			\sqrt{n}R\left[ \hat{\beta}_{\tau }^{(1),QR\ast }(1)-\beta _{0\tau }^{(1)}(1)%
			\right]  &=&-\sqrt{n}RA_{1}^{\prime }\left( \hat{\beta}_{\tau
			}^{(1),QR}(1)-\beta _{0\tau }^{(1)}(1)\right) +\frac{1}{\sqrt{n}}nR\left( 
			\tilde{\beta}_{\tau }^{(1),QR}(1)-\tilde{\beta}_{0\tau }^{(1)}(1)\right)  \\
			&=&-\sqrt{n}RA_{1}^{\prime }\left( \hat{\beta}_{\tau }^{(1),QR}(1)-\beta
			_{0\tau }^{(1)}(1)\right) +O_{p}\left( \frac{1}{\sqrt{n}}\right)  \\
			&\Longrightarrow &N\left( 0,RA_{1}^{\prime
			}A^{-1}BA^{-1}A_{1}R^{\prime }\right) \text{.}
		\end{eqnarray*}%
		Finally, for the fixed dimension of $p_{x}<\infty $, 
		Theorem  \ref{thm_AsymDist_mix}-(iii) follows from \citet{lee2016predictive}. Unlike the cointegrated parts of
		the system, this non-cointegrated local unit root regressors have their own
		limit theory from the block-wise diagonal structure of $H$ and $H^{-1}$
		matrix transformations given in Section \ref{oracle mr}.
						
	\end{proof}

		\newpage
		
			\section{Additional Tables and Figures}

			\begin{table}[tbhp]
				\caption{Prediction Results of Stock Returns: 12 one-period-ahead forecasts, GIC}
				\label{table_stockreturns_FPE_and_R2_GIC}\centering
				\begin{tabular}{l rrrrr}
					\hline
					& \multicolumn{5}{c}{Quantile ($\tau$)} \\
					\cline{2-6}
					& \multicolumn{1}{c}{0.05} & \multicolumn{1}{c}{0.1} & \multicolumn{1}{c}{0.5} & \multicolumn{1}{c}{0.9} & \multicolumn{1}{c}{0.95} \\
					\hline
					& \multicolumn{5}{c}{\underline{Final Prediction Error (FPE)}}\\
					QR & 0.0047 & 0.0099 & 0.0134 & 0.0078 & 0.0060 \\
					LASSO & \textbf{0.0045} & \textbf{0.0083} & 0.0123 & \textbf{0.0032} & \textbf{0.0020} \\
					ALQR  & \textbf{0.0045} & \textbf{0.0083} & \textbf{0.0122} & \textbf{0.0032} & \textbf{0.0020} \\
					QUANT  & 0.0046 & \textbf{0.0083} & 0.0124 & 0.0055 & 0.0029 \\
					\\
					& \multicolumn{5}{c}{\underline{Out-of-Sample $R^2$}}\\
					QR & -0.0147 & -0.1930 & -0.0831 & -0.4048 & -1.0666 \\
					LASSO & 0.0219 & \textbf{-0.0009} & 0.0094 & 0.4219 & \textbf{0.3233 }\\
					ALQR  & \textbf{0.0223} & -0.0029 & \textbf{0.0167} & \textbf{0.4288} & 0.3211 \\
					\\
					& \multicolumn{5}{c}{\underline{Average \# of Selected Predictors}}\\
					LASSO & 11.00 & 10.00 & 12.00 & 11.42 & 10.58 \\
					ALQR & 10.00 & 7.92 & 10.00 & 6.33 & 9.25 \\
					\\
					& \multicolumn{5}{c}{\underline{Tuning Parameter ($\lambda$) by GIC}}\\
					LASSO $(\times10^{-4})$ & 82.07 & 225.20 & 0.04 & 49.17 & 27.66 \\
					ALQR $(\times10^{-7})$& 5.26 & 85.12 & 1.46 & 197.67 & 0.88 \\
					\hline
				\end{tabular}%
			\end{table}



			\begin{table}[tbhp]
				\caption{Prediction Results of Stock Returns: 24 one-period-ahead forecasts, GIC}
				\label{table_stockreturns_FPE_and_R2_GIC_24step}\centering
				\begin{tabular}{l rrrrr}
					\hline
					& \multicolumn{5}{c}{Quantile ($\tau$)} \\
					\cline{2-6}
					& \multicolumn{1}{c}{0.05} & \multicolumn{1}{c}{0.1} & \multicolumn{1}{c}{0.5} & \multicolumn{1}{c}{0.9} & \multicolumn{1}{c}{0.95} \\
					\hline
					& \multicolumn{5}{c}{\underline{Final Prediction Error (FPE)}}\\
					QR & \textbf{0.0045} & 0.0102 & 0.0147 & 0.0067 & 0.0050 \\
					LASSO & 0.0058 & 0.0101 & \textbf{0.0140} & \textbf{0.0045} & \textbf{0.0025} \\
					ALQR & 0.0058 & \textbf{0.0097} & \textbf{0.0140} & \textbf{0.0045} & \textbf{0.0025} \\
					QUANT & 0.0055 & 0.0098 & 0.0148 & 0.0063 & 0.0036 \\
					\\
					& \multicolumn{5}{c}{\underline{Out-of-Sample $R^2$}}\\
					QR & \textbf{0.1751} & -0.0388 & 0.0100 & -0.0745 & -0.3938 \\
					LASSO & -0.0590 & -0.0245 & 0.0573 & 0.2747 & 0.2881 \\
					ALQR & -0.0526 & \textbf{0.0103} & \textbf{0.0577} & \textbf{0.2798} & \textbf{0.3088} \\
					\\
					& \multicolumn{5}{c}{\underline{Average \# of Selected Predictors}}\\
					LASSO & 10.88 & 9.63 & 12.00 & 10.67 & 10.71 \\
					ALQR & 9.96 & 8.88 & 10.63 & 10.25 & 9.67 \\
					\\
					& \multicolumn{5}{c}{\underline{Tuning Parameter ($\lambda$) by GIC}}\\
					LASSO $(\times10^{-4})$ &66.33 & 174.85 & 0.30 & 197.25 & 14.76 \\
					ALQR $(\times10^{-7})$& 3.74 & 24.70 & 0.65 & 1.55 & 0.59 \\
					\hline\hline
				\end{tabular}%
			\end{table}


			\begin{figure}
					\caption{The Correlation Heat Map of the Persistent Predictors}
				\label{fig_heatmap}\centering
				\begin{threeparttable}

						\includegraphics[scale=0.6]{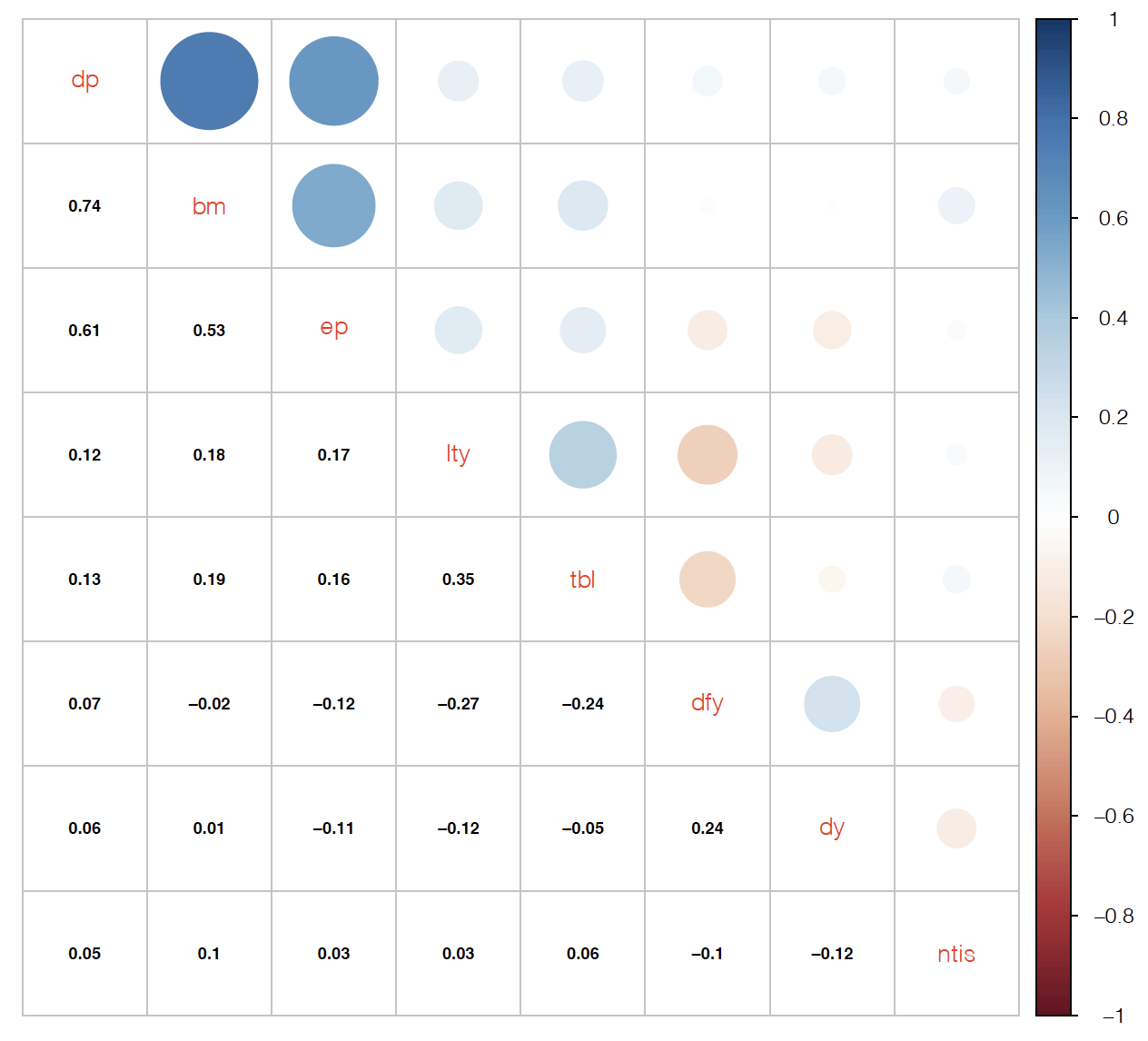}

					\begin{tablenotes}[para,flushleft]
						\item {\footnotesize Note: The correlation heat map is based on 816 monthly observations ranging from January 1952 to December 2019. All correlations are computed using the first difference of predictors. The full predictor names are defined in Table \ref{tb:v_names}.}
					\end{tablenotes}

				\end{threeparttable}
			\end{figure}


		\begin{table}[tbhp]
			\caption{Simulation Results: Scenario 1 with Dependent Predictors}
			\label{table_more_simulation_S1_rho010}\centering
			\begin{tabular}{l rrrrr}
				\hline
				& \multicolumn{5}{c}{Quantile ($\tau$)} \\
				\cline{2-6}
				& \multicolumn{1}{c}{0.05} & \multicolumn{1}{c}{0.1} & \multicolumn{1}{c}{0.5} & \multicolumn{1}{c}{0.9} & \multicolumn{1}{c}{0.95} \\
				\hline
				\\
				\multicolumn{5}{l}{\underline{Dependency Rate: $\rho=0.1$}} \\
				\\
				& \multicolumn{5}{c}{\underline{Final Prediction Error (FPE)}}\\
				QR    &   0.1076 & 0.1809 & 0.4072 & 0.1812 & 0.1081 \\ 
				LASSO &   0.1073 & 0.1811 & 0.4084 & 0.1809 & 0.1077 \\ 
				ALQR  &   \textbf{0.1052} & \textbf{0.1785} & \textbf{0.4048} & \textbf{0.1794} & \textbf{0.1060} \\ 
				RIDGE &   0.2288 & 0.3050 & 0.4053 & 0.3031 & 0.2259 \\ 
				QUANT &   2.2753 & 4.2477 & 11.4435 & 4.3347 & 2.3361 \\ 
				\\
				& \multicolumn{5}{c}{\underline{Out-of-Sample $R^2$}}\\
				QR    &   0.9527 & 0.9574 & 0.9644 & 0.9582 & 0.9537 \\ 
				LASSO &   0.9528 & 0.9574 & 0.9643 & 0.9583 & 0.9539 \\ 
				ALQR  &   \textbf{0.9537} & \textbf{0.9580} & \textbf{0.9646} & \textbf{0.9586} & \textbf{0.9546} \\ 
				RIDGE &   0.8994 & 0.9282 & 0.9646 & 0.9301 & 0.9033 \\ 
				\\
				& \multicolumn{5}{c}{\underline{Average \# of Selected Predictors}}\\
				LASSO & 10.42 & 10.15 & 9.67 & 10.14 & 10.47 \\ 
				ALQR  &  6.56 & 6.35  & 6.38 & 6.35 & 6.50 \\ 
				\\
				& \multicolumn{5}{c}{\underline{Tuning Parameter ($\lambda$)}}\\
				LASSO $(\times10^{0})$  & 8.57 & 15.18 & 37.39 & 15.33 & 8.43 \\ 
				ALQR $(\times10^{-4})$  & 1.68 & 2.39 & 3.81 & 2.43 & 1.73 \\ 
				RIDGE $(\times10^{-5})$ &   0.8 & 1.7 & 1.6 & 1.7 & 1.0 \\ 
				\hline
			\end{tabular}%
		\end{table}

		\begin{table}[tbhp]
			\caption{Simulation Results: Scenario 1 with Dependent Predictors}
			\label{table_more_simulation_S1_rho050}\centering
			\begin{tabular}{l rrrrr}
				\hline
				& \multicolumn{5}{c}{Quantile ($\tau$)} \\
				\cline{2-6}
				& \multicolumn{1}{c}{0.05} & \multicolumn{1}{c}{0.1} & \multicolumn{1}{c}{0.5} & \multicolumn{1}{c}{0.9} & \multicolumn{1}{c}{0.95} \\
				\hline
				\\
				\multicolumn{5}{l}{\underline{Dependency Rate: $\rho=0.5$}} \\
				\\
				& \multicolumn{5}{c}{\underline{Final Prediction Error (FPE)}}\\
				QR    &   0.1064 & 0.1800 & 0.4074 & 0.1805 & 0.1074 \\ 
				LASSO &     0.1069 & 0.1808 & 0.4094 & 0.1810 & 0.1078 \\ 
				ALQR  &     \textbf{0.1053} & \textbf{0.1785} & \textbf{0.4055} & \textbf{0.1790} & \textbf{0.1060} \\ 
				RIDGE &     0.2234 & 0.3046 & 0.4057 & 0.3142 & 0.2296 \\ 
				QUANT &     2.5896 & 4.6670 & 11.7537 & 4.5306 & 2.4595 \\ 
				\\
				& \multicolumn{5}{c}{\underline{Out-of-Sample $R^2$}}\\
				QR    &   0.9589 & 0.9614 & 0.9653 & 0.9601 & 0.9563 \\ 
				LASSO &     0.9587 & 0.9613 & 0.9652 & 0.9600 & 0.9562 \\ 
				ALQR  &     \textbf{0.9593} & \textbf{0.9617} & \textbf{0.9655} & \textbf{0.9605} & \textbf{0.9569} \\ 
				RIDGE &     0.9137 & 0.9347 & \textbf{0.9655} & 0.9306 & 0.9067 \\ 
				\\
				& \multicolumn{5}{c}{\underline{Average \# of Selected Predictors}}\\
				LASSO & 9.90 & 9.67 & 9.34 & 9.78 & 10.00 \\ 
				ALQR  &   6.46 & 6.32 & 6.33 & 6.29 & 6.46 \\ 
				\\
				& \multicolumn{5}{c}{\underline{Tuning Parameter ($\lambda$)}}\\
				LASSO $(\times10^{0})$  & 9.82 & 17.20 & 37.28 & 16.12 & 9.37 \\ 
				ALQR $(\times10^{-4})$  &   1.50 & 2.08 & 3.25 & 2.09 & 1.53 \\ 
				RIDGE $(\times10^{-5})$ &   0.9 & 1.8 & 2.2 & 1.6 & 0.8 \\ 
				\hline
			\end{tabular}%
		\end{table}

		\begin{table}[tbhp]
			\caption{Simulation Results: Scenario 1 with Dependent Predictors}
			\label{table_more_simulation_S1_rho090}\centering
			\begin{tabular}{l rrrrr}
				\hline
				& \multicolumn{5}{c}{Quantile ($\tau$)} \\
				\cline{2-6}
				& \multicolumn{1}{c}{0.05} & \multicolumn{1}{c}{0.1} & \multicolumn{1}{c}{0.5} & \multicolumn{1}{c}{0.9} & \multicolumn{1}{c}{0.95} \\
				\hline
				\\
				\multicolumn{5}{l}{\underline{Dependency Rate: $\rho=0.9$}} \\
				\\
				& \multicolumn{5}{c}{\underline{Final Prediction Error (FPE)}}\\
				QR    &   0.1067 & 0.1804 & 0.4077 & 0.1809 & 0.1083 \\ 
				LASSO &     0.1060 & 0.1800 & 0.4095 & 0.1810 & 0.1075 \\ 
				ALQR  &     \textbf{0.1052} & \textbf{0.1783} & \textbf{0.4055} & \textbf{0.1787} & \textbf{0.1065} \\ 
				RIDGE &     0.2157 & 0.2955 & 0.4058 & 0.2990 & 0.2193 \\ 
				QUANT &     2.2630 & 4.1512 & 11.0832 & 4.3207 & 2.4073 \\ 
				\\
				& \multicolumn{5}{c}{\underline{Out-of-Sample $R^2$}}\\
				QR    &   0.9528 & 0.9565 & 0.9632 & 0.9581 & 0.9550 \\ 
				LASSO &     0.9532 & 0.9566 & 0.9630 & 0.9581 & 0.9553 \\ 
				ALQR  &     \textbf{0.9535} & \textbf{0.9570} & \textbf{0.9634} & \textbf{0.9586} & \textbf{0.9558} \\ 
				RIDGE &     0.9047 & 0.9288 & \textbf{0.9634} & 0.9308 & 0.9089 \\ 
				\\
				& \multicolumn{5}{c}{\underline{Average \# of Selected Predictors}}\\
				LASSO & 8.33 & 7.74 & 7.17 & 7.71 & 8.26 \\ 
				ALQR  &   6.17 & 6.01 & 6.07 & 5.97 & 6.13 \\ 
				\\
				& \multicolumn{5}{c}{\underline{Tuning Parameter ($\lambda$)}}\\
				LASSO $(\times10^{0})$  & 13.09 & 23.81 & 52.67 & 23.93 & 13.39 \\ 
				ALQR $(\times10^{-4})$  &   1.36 & 1.67 & 2.11 & 1.57 & 1.38 \\ 
				RIDGE $(\times10^{-5})$ &   0.6 & 1.3 & 1.5 & 1.2 & 0.9 \\
				\hline
			\end{tabular}%
		\end{table}

\begin{table}[tbhp]
	\caption{Simulation Results: Scenario 2 with Dependent Predictors}
	\label{table_more_simulation_S2_rho010}\centering
	\begin{tabular}{l rrrrr}
		\hline
		& \multicolumn{5}{c}{Quantile ($\tau$)} \\
		\cline{2-6}
		& \multicolumn{1}{c}{0.05} & \multicolumn{1}{c}{0.1} & \multicolumn{1}{c}{0.5} & \multicolumn{1}{c}{0.9} & \multicolumn{1}{c}{0.95} \\
		\hline
		\\
		\multicolumn{5}{l}{\underline{Dependency Rate: $\rho=0.1$}} \\
		\\
		& \multicolumn{5}{c}{\underline{Final Prediction Error (FPE)}}\\
		QR    &   0.1075 & 0.1809 & 0.4074 & 0.1811 & 0.1081 \\ 
		LASSO &     0.1074 & 0.1810 & 0.4087 & 0.1807 & 0.1080 \\ 
		ALQR  &     \textbf{0.1067} & \textbf{0.1794} & 0.4068 & \textbf{0.1799} & \textbf{0.1068} \\ 
		RIDGE &     0.2363 & 0.3123 & \textbf{0.4057} & 0.3123 & 0.2349 \\ 
		QUANT &     2.3787 & 4.4253 & 11.8534 & 4.4989 & 2.4148 \\ 
		\\
		& \multicolumn{5}{c}{\underline{Out-of-Sample $R^2$}}\\
		QR    &   0.9548 & 0.9591 & 0.9656 & 0.9597 & 0.9552 \\ 
		LASSO &     0.9549 & 0.9591 & 0.9655 & 0.9598 & 0.9553 \\ 
		ALQR  &     \textbf{0.9552} & \textbf{0.9595 }& 0.9657 & \textbf{0.9600} & \textbf{0.9558} \\ 
		RIDGE &     0.9007 & 0.9294 & \textbf{0.9658} & 0.9306 & 0.9027 \\ 
		\\
		& \multicolumn{5}{c}{\underline{Average \# of Selected Predictors}}\\
		LASSO & 10.82 & 10.67 & 10.35 & 10.66 & 10.86 \\ 
		ALQR  &   8.44 & 8.33 & 8.31 & 8.34 & 8.45 \\ 
		\\
		& \multicolumn{5}{c}{\underline{Tuning Parameter ($\lambda$)}}\\
		LASSO $(\times10^{0})$  & 6.95 & 12.49 & 30.84 & 12.25 & 6.89 \\ 
		ALQR $(\times10^{-4})$  &    1.23 & 1.61 & 2.52 & 1.62 & 1.18 \\ 
		RIDGE $(\times10^{-5})$ &    0.7 & 1.7 & 1.6 & 1.4 & 0.6 \\ 
		\hline
	\end{tabular}%
\end{table}

\begin{table}[tbhp]
	\caption{Simulation Results: Scenario 2 with Dependent Predictors}
	\label{table_more_simulation_S2_rho050}\centering
	\begin{tabular}{l rrrrr}
		\hline
		& \multicolumn{5}{c}{Quantile ($\tau$)} \\
		\cline{2-6}
		& \multicolumn{1}{c}{0.05} & \multicolumn{1}{c}{0.1} & \multicolumn{1}{c}{0.5} & \multicolumn{1}{c}{0.9} & \multicolumn{1}{c}{0.95} \\
		\hline
		\\
		\multicolumn{5}{l}{\underline{Dependency Rate: $\rho=0.5$}} \\
		\\
		& \multicolumn{5}{c}{\underline{Final Prediction Error (FPE)}}\\
		QR    &   0.1072 & 0.1805 & 0.4081 & 0.1813 & 0.1082 \\ 
		LASSO &     0.1073 & 0.1811 & 0.4099 & 0.1815 & 0.1085 \\ 
		ALQR  &     \textbf{0.1068} & \textbf{0.1799} & 0.4070 & \textbf{0.1805} & \textbf{0.1072} \\ 
		RIDGE &     0.2484 & 0.3313 & \textbf{0.4063} & 0.3406 & 0.2554 \\ 
		QUANT &     2.7837 & 5.0151 & 12.9604 & 4.9791 & 2.6430 \\ 
		\\
		& \multicolumn{5}{c}{\underline{Out-of-Sample $R^2$}}\\
		QR    &   0.9615 & 0.9640 & 0.9685 & 0.9636 & 0.9591 \\ 
		LASSO &     0.9614 & 0.9639 & 0.9684 & 0.9635 & 0.9589 \\ 
		ALQR  &     \textbf{0.9616} & \textbf{0.9641} & \textbf{0.9686} & \textbf{0.9637} & \textbf{0.9594} \\ 
		RIDGE &     0.9108 & 0.9339 & \textbf{0.9686} & 0.9316 & 0.9034 \\ 
		\\
		& \multicolumn{5}{c}{\underline{Average \# of Selected Predictors}}\\
		LASSO & 10.58 & 10.45 & 10.27 & 10.54 & 10.66 \\ 
		ALQR  &   8.33 & 8.32 & 8.32 & 8.27 & 8.39 \\ 
		\\
		& \multicolumn{5}{c}{\underline{Tuning Parameter ($\lambda$)}}\\
		LASSO $(\times10^{0})$  & 7.23 & 12.75 & 28.74 & 12.08 & 6.94 \\ 
		ALQR $(\times10^{-4})$  &    1.19 & 1.41 & 2.07 & 1.44 & 1.05 \\ 
		RIDGE $(\times10^{-5})$ &    0.9 & 1.5 & 1.7 & 1.4 & 0.8 \\ 
		\hline
	\end{tabular}%
\end{table}

\begin{table}[tbhp]
	\caption{Simulation Results: Scenario 2 with Dependent Predictors}
	\label{table_more_simulation_S2_rho090}\centering
	\begin{tabular}{l rrrrr}
		\hline
		& \multicolumn{5}{c}{Quantile ($\tau$)} \\
		\cline{2-6}
		& \multicolumn{1}{c}{0.05} & \multicolumn{1}{c}{0.1} & \multicolumn{1}{c}{0.5} & \multicolumn{1}{c}{0.9} & \multicolumn{1}{c}{0.95} \\
		\hline
		\\
		\multicolumn{5}{l}{\underline{Dependency Rate: $\rho=0.9$}} \\
		\\
		& \multicolumn{5}{c}{\underline{Final Prediction Error (FPE)}}\\
		QR    &   0.1065 & 0.1803 & 0.4074 & 0.1808 & 0.1082 \\ 
		LASSO &     0.1065 & 0.1804 & 0.4103 & 0.1817 & 0.1078 \\ 
		ALQR  &     \textbf{0.1063} & \textbf{0.1797} & 0.4065 & \textbf{0.1798} & \textbf{0.1077} \\ 
		RIDGE &     0.2571 & 0.3418 & \textbf{0.4058} & 0.3469 & 0.2593 \\ 
		QUANT &     2.7380 & 5.0229 & 13.4626 & 5.2381 & 2.8857 \\ 
		\\
		& \multicolumn{5}{c}{\underline{Out-of-Sample $R^2$}}\\
		QR    &   0.9611 & 0.9641 & 0.9697 & 0.9655 & 0.9625 \\ 
		LASSO &     0.9611 & 0.9641 & 0.9695 & 0.9653 & \textbf{0.9627}\\ 
		ALQR  &     \textbf{0.9612} & \textbf{0.9642} & 0.9698 & \textbf{0.9657} & \textbf{0.9627} \\ 
		RIDGE &     0.9061 & 0.9320 & \textbf{0.9699} & 0.9338 & 0.9101 \\ 
		\\
		& \multicolumn{5}{c}{\underline{Average \# of Selected Predictors}}\\
		LASSO & 9.55 & 9.14 & 8.62 & 9.13 & 9.50 \\ 
		ALQR  &   8.04 & 7.98 & 8.05 & 7.94 & 8.05 \\ 
		\\
		& \multicolumn{5}{c}{\underline{Tuning Parameter ($\lambda$)}}\\
		LASSO $(\times10^{0})$  & 9.47 & 17.32 & 40.78 & 17.24 & 9.63 \\ 
		ALQR $(\times10^{-4})$  &    1.06 & 1.32 & 1.52 & 1.21 & 0.98 \\ 
		RIDGE $(\times10^{-5})$ &    0.6 & 1.3 & 1.4 & 1.2 & 0.6 \\ 
		\hline
	\end{tabular}%
\end{table}

		\begin{table}[hp]
			\caption{Simulation Results: Scenario 3 with Dependent Predictors}
			\label{table_more_simulation_S3_rho010}\centering
			\begin{tabular}{l rrrrr}
				\hline
				& \multicolumn{5}{c}{Quantile ($\tau$)} \\
				\cline{2-6}
				& \multicolumn{1}{c}{0.05} & \multicolumn{1}{c}{0.1} & \multicolumn{1}{c}{0.5} & \multicolumn{1}{c}{0.9} & \multicolumn{1}{c}{0.95} \\
				\hline
				\\
				\multicolumn{5}{l}{\underline{Dependency Rate: $\rho=0.1$}} \\
				\\
				& \multicolumn{5}{c}{\underline{Final Prediction Error (FPE)}}\\
				QR    &   \textbf{0.1076} & \textbf{0.1808} & 0.4074 & 0.1811 & 0.1082 \\ 
				LASSO &     \textbf{0.1076} & \textbf{0.1808} & 0.4074 & \textbf{0.1810} & \textbf{0.1081} \\ 
				ALQR  &     0.1079 & 0.1812 & 0.4074 & 0.1814 & 0.1086 \\ 
				RIDGE &     0.2394 & 0.3157 & \textbf{0.4057} & 0.3122 & 0.2347 \\ 
				QUANT &     2.3833 & 4.4624 & 11.9144 & 4.5103 & 2.4318 \\ 
				\\
				& \multicolumn{5}{c}{\underline{Out-of-Sample $R^2$}}\\
				QR    &   \textbf{0.9549} & \textbf{0.9595} & 0.9658 & \textbf{0.9599} & \textbf{0.9555} \\ 
				LASSO &     \textbf{0.9549} & \textbf{0.9595} & 0.9658 & \textbf{0.9599} & \textbf{0.9555} \\ 
				ALQR  &     0.9547 & 0.9594 & 0.9658 & 0.9598 & 0.9553 \\ 
				RIDGE &     0.8996 & 0.9292 & \textbf{0.9660} & 0.9308 & 0.9035 \\ 
				\\
				& \multicolumn{5}{c}{\underline{Average \# of Selected Predictors}}\\
				LASSO & 12.00 & 12.00 & 12.00 & 12.00 & 12.00 \\ 
				ALQR  &   11.83 & 11.92 & 12.00 & 11.93 & 11.82 \\ 
				\\
				& \multicolumn{5}{c}{\underline{Tuning Parameter ($\lambda$)}}\\
				LASSO $(\times10^{-1})$  & 1.67 & 1.82 & 1.92 & 1.80 & 1.73 \\ 
				ALQR $(\times10^{-5})$  &   1.1 & 0.9 & 0.2 & 0.8 & 1.0 \\ 
				RIDGE $(\times10^{-5})$ &   0.6 & 1.1 & 1.0 & 1.1 & 0.6 \\ 
				\hline
			\end{tabular}%
		\end{table}

		\begin{table}[tbhp]
			\caption{Simulation Results: Scenario 3 with Dependent Predictors}
			\label{table_more_simulation_S3_rho050}\centering
			\begin{tabular}{l rrrrr}
				\hline
				& \multicolumn{5}{c}{Quantile ($\tau$)} \\
				\cline{2-6}
				& \multicolumn{1}{c}{0.05} & \multicolumn{1}{c}{0.1} & \multicolumn{1}{c}{0.5} & \multicolumn{1}{c}{0.9} & \multicolumn{1}{c}{0.95} \\
				\hline
				\\
				\multicolumn{5}{l}{\underline{Dependency Rate: $\rho=0.5$}} \\
				\\
				& \multicolumn{5}{c}{\underline{Final Prediction Error (FPE)}}\\
				QR    &   0.1065 & 0.1799 & 0.4076 & \textbf{0.1804} & 0.1074 \\ 
				LASSO &     \textbf{0.1064} & \textbf{0.1798} & 0.4076 & 0.1805 & \textbf{0.1073} \\ 
				ALQR  &     0.1070 & 0.1805 & 0.4077 & 0.1810 & 0.1077 \\ 
				RIDGE &     0.2514 & 0.3381 & \textbf{0.4059} & 0.3487 & 0.2597 \\ 
				QUANT &     2.8938 & 5.2864 & 13.4026 & 5.2607 & 2.8399 \\
				\\
				& \multicolumn{5}{c}{\underline{Out-of-Sample $R^2$}}\\
				QR    &   \textbf{0.9632} & \textbf{0.9660} & 0.9696 & \textbf{0.9657} & \textbf{0.9622} \\ 
				LASSO &     \textbf{0.9632} & \textbf{0.9660} & 0.9696 & \textbf{0.9657} & \textbf{0.9622} \\ 
				ALQR  &     0.9630 & 0.9659 & 0.9696 & 0.9656 & 0.9621 \\ 
				RIDGE &     0.9131 & 0.9360 & \textbf{0.9697} & 0.9337 & 0.9086 \\ 
				\\
				& \multicolumn{5}{c}{\underline{Average \# of Selected Predictors}}\\
				LASSO & 12.00 & 12.00 & 12.00 & 12.00 & 12.00 \\ 
				ALQR  &   11.78 & 11.90 & 12.00 & 11.91 & 11.76 \\ 
				\\
				& \multicolumn{5}{c}{\underline{Tuning Parameter ($\lambda$)}}\\
				LASSO $(\times10^{-1})$  & 1.47 & 1.50 & 1.68 & 1.60 & 1.57 \\ 
				ALQR $(\times10^{-5})$  &   1.2 & 1.0 & 0.2 & 1.0 & 1.2 \\ 
				RIDGE $(\times10^{-5})$ &   0.6 & 1.0 & 1.2 & 1.1 & 0.6 \\ 
				\hline
			\end{tabular}%
		\end{table}

		\begin{table}[tbhp]
			\caption{Simulation Results: Scenario 3 with Dependent Predictors}
			\label{table_more_simulation_S3_rho090}\centering
			\begin{tabular}{l rrrrr}
				\hline
				& \multicolumn{5}{c}{Quantile ($\tau$)} \\
				\cline{2-6}
				& \multicolumn{1}{c}{0.05} & \multicolumn{1}{c}{0.1} & \multicolumn{1}{c}{0.5} & \multicolumn{1}{c}{0.9} & \multicolumn{1}{c}{0.95} \\
				\hline
				\\
				\multicolumn{5}{l}{\underline{Dependency Rate: $\rho=0.9$}} \\
				\\
				& \multicolumn{5}{c}{\underline{Final Prediction Error (FPE)}}\\
				QR    &   \textbf{0.1069} & \textbf{0.1803} & 0.4079 & \textbf{0.1810} & \textbf{0.1083} \\ 
				LASSO &     \textbf{0.1069} & \textbf{0.1803} & 0.4079 & \textbf{0.1810} & \textbf{0.1083} \\ 
				ALQR  &     0.1075 & 0.1818 & 0.4084 & 0.1823 & 0.1089 \\ 
				RIDGE &     0.2638 & 0.3498 & \textbf{0.4064} & 0.3538 & 0.2660 \\ 
				QUANT &     2.8993 & 5.1589 & 13.7720 & 5.3428 & 2.9630 \\ 
				\\
				& \multicolumn{5}{c}{\underline{Out-of-Sample $R^2$}}\\
				QR    &   \textbf{0.9631} & \textbf{0.9651} & 0.9704 & \textbf{0.9661} & \textbf{0.9634} \\ 
				LASSO &     \textbf{0.9631} & \textbf{0.9651} & 0.9704 & \textbf{0.9661} & \textbf{0.9634} \\ 
				ALQR  &     0.9629 & 0.9648 & 0.9703 & 0.9659 & 0.9632 \\ 
				RIDGE &     0.9090 & 0.9322 & \textbf{0.9705} & 0.9338 & 0.9102 \\ 
				\\
				& \multicolumn{5}{c}{\underline{Average \# of Selected Predictors}}\\
				LASSO & 11.95 & 11.99 & 12.00 & 11.98 & 11.96 \\ 
				ALQR  &   11.01 & 11.28 & 11.74 & 11.31 & 11.07 \\ 
				\\
				& \multicolumn{5}{c}{\underline{Tuning Parameter ($\lambda$)}}\\
				LASSO $(\times10^{-1})$  & 1.51 & 1.25 & 1.02 & 1.20 & 1.49 \\ 
				ALQR $(\times10^{-5})$  &   2.7 & 3.4 & 2.1 & 3.2 & 2.6 \\ 
				RIDGE $(\times10^{-5})$ &   0.6 & 1.0 & 1.0 & 1.1 & 0.5 \\ 
				\hline
			\end{tabular}%
		\end{table}

\begin{landscape}
\begin{table}
	{}	\caption{The Johansen Cointegration Test}
	\label{table_Johansen}\centering
	\begin{threeparttable}
\begin{tabular}{p{0.1\linewidth} | cccccccccccc c}
	\hline \hline
	Hypothesised number of CE(s) &\multicolumn{12}{c}{Trace statistic}& 0.05 critical value \\
	\hline
	& (1) & (2) & (3) & (4) & (5) & (6) & (7) & (8) & (9) & (10) & (11) & (12) \\
	\hline
	$r=0$ & 2.21 & 2.40 & 2.36 & 2.25 & 2.26 & 2.29 & 2.34 & 2.42 & 2.44 & 2.43 & 2.47 & 2.27 & 157.11 \\
	$r\leq1$ & 10.25 & 10.66 & 11.18 & 10.84 & 10.52 & 10.41 & 10.66 & 10.91 & 11.00 & 10.39 & 10.04 & 10.36 & 124.25 \\
	$r\leq2$ & 20.70 & 20.51 & 21.52 & 21.23 & 21.11 & 20.64 & 20.79 & 21.17 & 20.75 & 20.21 & 19.84 & 20.57 & 90.39 \\
	$r\leq3$ & 41.47 & 41.45 & 41.99 & 41.25 & 40.85 & 40.48 & 41.01 & 41.81 & 41.28 & 40.87 & 40.36 & 40.96 & 70.6 \\
	$r\leq4$ & 83.53 & 83.44 & 83.91 & 83.11 & 82.61 & 82.17 & 82.42 & 82.91 & 82.16 & 81.13 & 80.63 & 81.38 & 48.28 \\
	$r\leq5$ & 143.67 & 142.99 & 143.71 & 142.78 & 142.42 & 142.09 & 142.33 & 142.95 & 142.65 & 141.95 & 141.40 & 141.37 & 31.52 \\
	$r\leq6$ & 221.96 & 219.42 & 219.94 & 219.65 & 219.20 & 218.94 & 218.71 & 217.84 & 218.05 & 216.82 & 216.47 & 216.03 & 17.95 \\
	$r\leq7$ & 355.85 & 353.34 & 353.44 & 352.96 & 352.51 & 351.99 & 352.06 & 351.59 & 356.59 & 356.55 & 357.69 & 352.80 & 8.18 \\
	\hline\\ 
	\hline
	Hypothesised number of CE(s) & \multicolumn{12}{c}{Maximum eigenvalue statistic}  & 0.05 critical value \\
	\hline
	& (1) & (2) & (3) & (4) & (5) & (6) & (7) & (8) & (9) & (10) & (11) & (12) \\
	\hline
	$r\leq0$  & 2.21 & 2.40 & 2.36 & 2.25 & 2.26 & 2.29 & 2.34 & 2.42 & 2.44 & 2.43 & 2.47 & 2.27 & 51.07 \\
	$r\leq1$  & 8.03 & 8.26 & 8.81 & 8.59 & 8.26 & 8.12 & 8.33 & 8.49 & 8.56 & 7.96 & 7.57 & 8.09 & 44.91 \\
	$r\leq2$  & 10.45 & 9.84 & 10.34 & 10.39 & 10.59 & 10.23 & 10.12 & 10.26 & 9.75 & 9.82 & 9.80 & 10.21 & 39.43 \\
	$r\leq3$  & 20.77 & 20.94 & 20.47 & 20.02 & 19.74 & 19.84 & 20.23 & 20.64 & 20.53 & 20.65 & 20.53 & 20.38 & 33.32 \\
	$r\leq4$  & 42.06 & 41.99 & 41.92 & 41.85 & 41.76 & 41.68 & 41.41 & 41.10 & 40.89 & 40.27 & 40.27 & 40.43 & 27.14 \\
	$r\leq5$  & 60.14 & 59.55 & 59.80 & 59.68 & 59.80 & 59.92 & 59.91 & 60.04 & 60.48 & 60.81 & 60.76 & 59.99 & 21.07 \\
	$r\leq6$  & 78.29 & 76.43 & 76.23 & 76.87 & 76.78 & 76.85 & 76.38 & 74.89 & 75.40 & 74.87 & 75.07 & 74.66 & 14.9 \\
	$r\leq7$  & 133.90 & 133.92 & 133.50 & 133.31 & 133.31 & 133.05 & 133.36 & 133.75 & 138.55 & 139.74 & 141.23 & 136.77 & 8.18 \\
	\hline\hline
\end{tabular}
	\begin{tablenotes}[para,flushleft]
	\item {\footnotesize Notes: 1. $r$ indicates the number of cointegrating vectors. 2. From Column (1) to Column (12), each column states the value of test statistic using data based on the rolling fixed window scheme for 12 one-step-ahead forecasts, respectively.}
\end{tablenotes}
\end{threeparttable}
\end{table}
\end{landscape}

		\end{appendix}

		\bibliographystyle{chicago}
		\bibliography{FLS-ALASSOQR}


	\end{document}